\documentclass[12pt]{article} 
\usepackage{amsmath} 
\usepackage{amssymb,amsfonts,euscript} 
\usepackage{hyperref} 
\renewcommand{\baselinestretch}{1.2} 
\newcommand\nn{\nonumber}

\newcommand{\D}{{\cal D}}

\newcommand{\M}{{\cal M}}

\newcommand{\cL}{\cal{L}}

\newcommand{\Zint}{\mathbb{Z}}

\newcommand{\bz}{\bar{z}}

\def\a{\alpha} 
\def\be{\beta} 
\def\l{\lambda} 
\def\m{\mu} 
\def\n{\nu} 
\def\r{\rho} 
\def\p{\pi}
\def\de{\delta} 
\def\k{\kappa}

\def\hB{\hat{B}}

\def\hg{\hat{g}}

\def\hb{\hat{B}}
\def\hh{\hat{H}}
\def\ha{\hat{A}}
 
\def\part{\partial}

\def\thickone{{\rm 1\mskip-4.5mu l}}

\newcommand	{\vshalf}	{\vspace{-.35in}}
\newcommand	{\vsvs}		{\vspace{-.3 in}}
\newcommand	{\vsvsvs}	{\vspace{-.2 in}}
\renewcommand{\=}{\hspace{-.03in}=\hspace{-.02in}}
\newcommand{\equivs}{\hspace{-.03in}\equiv\hspace{-.02in}}

\newcommand{\sa}{\mathop{\vtop{\ialign{##\crcr 
$\hfil\displaystyle{\longrightarrow}\hfil$\crcr\noalign{\kern-1pt\nointerlineskip} 
\hspace{.12in}$^\sigma$\hskip6pt\crcr\noalign{\kern3pt}}}}} 
\newcommand{\slra}{\mathop{\vtop{\ialign{##\crcr 
$\hfil\displaystyle{\longleftrightarrow}\hfil$\crcr\noalign{\kern-1pt\nointerlineskip} 
\hspace{.12in}$^\sigma$\hskip6pt\crcr\noalign{\kern3pt}}}}} 
\newcommand{\sat}{\mathop{\vtop{\ialign{##\crcr 
$\hfil\displaystyle{\longrightarrow}\hfil$\crcr\noalign{\kern-1pt\nointerlineskip} 
\hspace{.12in}$^\sigma$\hskip6pt\crcr\noalign{\kern3pt}}}}} 
\newcommand{\pa}{\mathop{\vtop{\ialign{##\crcr 
$\hfil\displaystyle{\oplus}\hfil$\crcr\noalign{\kern+1pt\nointerlineskip} 
\hspace{.08in}$^{\alpha=0}$\hskip6pt\crcr\noalign{\kern3pt}}}}} 
\newcommand{\pan}{\mathop{\vtop{ialgin{##\crcr 
$\hfil\displaystyle{\oplus}\hfil$\crcr\noaligan{\kern+2pt\nointerlinkeskip} 
\hspace{.03in} $^{\alpha}$\hskip6pt\crcr\noalign{\kern3pit}}}}} 
\newcommand{\ka}{\mathop{\vtop{\ialign{##\crcr 
$\hfil\displaystyle{\longleftrightarrow}\hfil$\crcr\noalign{\kern-1pt\nointerlineskip} 
\hspace{.12in}$^K$\hskip6pt\crcr\noalign{\ker3pt}}}}} 
\newcommand{\bp}{\mathop{\vtop{ialign{##\crcr 
$\hfil\displaystyle{}\hfil$\crcr\noalign{\kern-13pt\nointerlineskip} 
\big{(}\hskip0pt\crcr\noalign{\kern3pt}}}}} 
\newcommand{\cbp}{\mathop{\vtop{ialign{##\crcr 
$\hfil\displaystyle{}\hfil$\crcr\noalign{\kern-13pt\nointerlineskip} 
\big{)}\hskip0pt\crcr\noalign{\kern3pt}}}}}

\newcommand{\s}{\sigma} 
\newcommand{\srange}{\sigma=0,...,N_c-1}

\newcommand{\ws}{\omega (h_\s)} 
\newcommand{\w}{\omega}

\newcommand{\sr}{\sqrt{\r}}

\newcommand{\hc}{$\hat{J}_{\gst}$} 
 
\newcommand{\tp}{{2\pi i}}

\newcommand{\sgb}{{\mbox{\scriptsize{\gb}}}}

\newcommand{\gfraks}{{\mbox{\scriptsize{\mbox{${\mathfrak g}$}}}}} 
\def\gb            {\mbox{$\hat{\mathfrak g}$}} 
 
\def\sm#1      {\mbox{\scriptsize $#1$}}

\def\d             {\mbox{$\mathbb D$}}

\def\srac#1#2{\smal{\frac{#1}{#2}}} 
\def\foot#1{\mbox{\footnotesize $#1$}} 
\def\scrs#1{\mbox{\scriptsize $#1$}} 
\def\tyny#1{\mbox{\tiny $#1$}} 
\def\smal#1{\mbox{\small $#1$}} 
\def\big#1{\mbox{\large $#1$}} 
\def\Big#1{\mbox{\Large $#1$}}

\def\hsp#1{\hspace{#1in}} 

\def\hjb{\hat{\bar{J}}}

\def\hE{{\hat{E}}} 
\def\he{{\hat{e}}}
 
\def\heb{{\hat{\bar{e}\hspace{.02in}}\hspace{-.02in}}} 
\def\hx{{\hat{x}}}
\def\hh{\hat{H}} 
\def\hbe{\hat{\beta}}
\def\hebi{{(\hat{\bar{e}\hspace{.02in}}^{-1})\hspace{-.02in}}}
\def\hei{(\hat{e}^{-1})}

\def\sdh{{\hat{\cal D}}}
\def\sdhb{{\hat{\bar{\cal D}\hspace{.02in}}}}
\def\hp{\hat{p}}
\def\hL{{\hat{L}}}

\def\hM{{\hat{M}}}
\def\hN{{\hat{N}}}

\def\hys{{{\hat{y}}_\ast}}

\def\gfrakh{\hat{\mathfrak g}} 
\def\hfrakh{\hat{\mathfrak h}}

\def\bfrak{\mathfrak b} 
\def\bfrakh{\mbox{$\hat{\mathfrak b}$}}

\def\dual{\underset{\s}{\longrightarrow}} 
\def\duals{\!{\underset{\s}{\rightarrow}}} 
 
\def\sg{\smal{\EuScript{G}}} 
\def\sj{{\cal J}} 
\def\sjb{{\bar{\cal J}}} 
\def\sjbh{{\hat{\sjb \hspace{.06in}}\hspace{-.06in}}}
\def\sjh{{\hat{\cal J}}}

\def\slh{\hat{{\cal L}}} 
\def\hc{^\dagger} 
\def\hcj{\dagger} 
\def\one{{\mathchoice {\rm 1\mskip-4mu l} {\rm 1\mskip-4mu} {\rm 1\mskip-4.5mu l} 
{\rm 1\mskip-5mu l}}} 
\def\d{\delta} 
\def\e{\eta}

\def\nnsrs{n\!+\!\srac{n(s)}{\r(\s)}} 
\def\nnsrsf{n\!+\!\frac{n(s)}{\r(\s)}} 
\def\nrm{{n(r)\m}} 
 
\def\nrmmode{{n(r)\m,m+\frac{n(r)}{\r(\s)}}} 
\def\nrn{{n(r)\n}} 
\def\mnrn{{-n(r),\n}} 
\def\nsn{{n(s)\n}} 
\def\ntd{{n(t)\d}} 
\def\ntdmode{{n(t)\d,p+\frac{n(t)}{\r(\s)}}} 
\def\nue{{n(u)\epsilon}} 
\def\nvz{{n(v)\zeta}} 
\def\nsnmode{{n(s)\n,n+\frac{n(s)}{\r(\s)}}} 
\def\mnnrnsrs{{m\!+\!n\!+\!\srac{n(r)+n(s)}{\r(\s)}}} 
\def\mnnrnsrsf{{m\!+\!n\!+\!\frac{n(r)+n(s)}{\r(\s)}}} 
\def\mnrrs{{m\!+\!\srac{n(r)}{\r(\s)}}} 
\def\mnrrsf{{m\!+\!\frac{n(r)}{\r(\s)}}} 
 
\def\scf{{\cal F}} 
\def\sG{{\cal G}}

\def\gfrak{\mbox{$\mathfrak g$}} 
\def\goto{\longrightarrow} 
\def\hj{\hat{J}} 
\def\nrn{{n(r)\n}} 
\def\nsn{{n(s)\n}} 
\def\schi{{\foot{\chi}}} 
\def\schisig{{\foot{\chi(\s)}}} 
\def\schizero{{\foot{\chi(0)}}} 
\def\ntd{{n(t)\delta}} 
\def\hc{^\dagger} 
\def\st{{\cal T}} 
\def\0b{\ } 
\def\pl{\partial} 
 
\def\Nrm{{N(r)\m}}

\def\Nsn{{N(s)\n}} 
\def\Ntd{{N(t)\d}}

\def\srange{\s=0,\ldots,N_c-1} 
\def\sm{{\cal M}} 
\def\sr{{\cal R}} 
\def\ho{{\hat{\Omega}}}
\def\sx{\smal{\EuScript{X}}}
\def\sxfoot{{\foot{\EuScript{X}}}}
\def\sxh{{\hat{\smal{\EuScript{X}}}}}
\def\sxtiny{\tyny{\EuScript{X}}}
\def\sxhs{{ \hat{\foot{\EuScript{X}}} }}
\def\sgtiny{\scrs{\EuScript{G}}}
\def\sb{{\cal B}}
\def\scp{{\smal{\cal P}}}

\def\jfj{{\frac{\hat{j}}{f_j(\s)}}}

\def\bigspc{{\quad \quad \quad \quad}} 
\def\bkspc{{\!\!\!\!\!}}

\def\Aj0{{| A (j;\st) \rangle_\s}}

\def\su{{ \mathfrak{su} }} 
\def\so{{ \mathfrak{so} }} 
 
\def\ugp{{ \mathfrak{u} }}
\def\nrrs{{ \srac{n(r)}{\r(\s)} }}

\def\hpl{{ \hat{\pl}}}
\def\tst{{ \tilde{\st}}}
\def\onefourpi{{ \frac{1}{4\pi}}}
\def\hs{{ \hat{S}}}
\def\slh{{ \hat{{\cL}} }}
\def\hG{{ \hat{G}}}
\def\ep{{ \epsilon}}
\def\duals{\!{\underset{\s}{\rightarrow}}} 
\def\eb{{ \bar{e}}}
\def\se{{\cal E}}
\def\seb{{\bar{\cal E}}}
\def\seh{{\hat{\cal E}}}
\def\bj{{ \bar{J}}}
\def\hphi{{ \hat{\phi}}}
\def\b{{ \beta}}
\def\bh{{ \hat{\beta}}}
\def\sgh{{ \hat{\sg}}}
\def\hpsi{{ \hat{\psi}}}
\def\hgamma{{ \hat{\gamma}}}
\def\half{{ \frac{1}{2}}}
\def\hlh{{ \hat{h}}}
\def\sh{{\cal H}}
\def\hsp{{\hat{{\cal P}}}}
\def\sa{{\cal A}}

\def\hY{{ \hat{Y}}}
\def\sY{{ \cal Y}}
\def\hnrm{{ \hat{n}(r)\m}}
\def\hnrhm{{ \hat{n} (r)\hat{\m}}}
\def\hnsn{{ \hat{n}(s) \n}}
\def\hnshn{{ \hat{n}(s) \hat{\n}}}
\def\hnvhz{{ \hat{n}(v) \hat{\zeta}}}
\def\hnwhl{{ \hat{n}(w) \hat{\lambda}}}

\makeatletter 
\renewcommand{\@makefnmark}{\mbox{$^{\ddagger\@thefnmark}$}} 
\renewcommand{\subsection}{\@startsection 
{subsection}{2}{0pt
}{-\baselineskip}{0.5\baselineskip} 
{\normalfont\normalsize\bf}} 
\renewcommand{\section}{\@startsection 
{section}{2}{0pt
}{-\baselineskip}{0.5\baselineskip} 
{\bf\large}} 

\makeatother 
\numberwithin{equation}{section} 
\numberwithin{table}{section}

\newcounter{myfigctr}
\def\myfig#1{\refstepcounter{myfigctr}%
 \label{#1}%
} 
 
\newcommand{\publititle}[8] 
{ 
  \vspace*{-3cm} 
  \begin{flushright} #1 \\ {\tt #2} \end{flushright} 
  \vfill 
  \begin{center}{\Large
    \bfseries #3}\end{center} 
  \vskip 8mm 
  \begin{center}{\large #4}\end{center} 
  \begin{center}{\normalsize #5}\end{center} 
  \vskip 8mm 
  \nopagebreak 
  \noindent #6 
  \vfill 
  \begin{flushleft} #7 
  \end{flushleft} 
  \hrule width 6.7cm \vskip.1mm 
  {\small #8} 
  \thispagestyle{empty} 
  \clearpage 
} 
 
\begin{document} 
 
\publititle{ ${}$ \\ UCB-PTH-02-61 \\ LBNL-51968 \\ hep-th/0212275}{}{Twisted Einstein Tensors
and Orbifold Geometry}{J. deBoer$^{1a}$ , M.B.Halpern$^{2b}$ and C. Helfgott$^{2c}$} 
{$^1${\em Institute for Theoretical Physics, University of Amsterdam,\\
Valckenierstraat 65, 1018 XE Amsterdam, The Netherlands}\\ ${}$
\\
$^2$ {\em Department of Physics, University of California and \\
Theoretical Physics Group,  Lawrence Berkeley National 
Laboratory \\ 
University of California, Berkeley, California 94720, USA}
\\[2mm]} {Following recent advances in the local theory of current-algebraic orbifolds, we study various geometric 
properties of the general WZW orbifold, the general coset orbifold and a large class of (non-linear) sigma model 
orbifolds. Phase-space geometry is emphasized for the WZW orbifolds -- while for the sigma model orbifolds we construct the 
corresponding {\it sigma model orbifold action}, which includes the previously-known general WZW orbifold action and 
general coset orbifold action as special cases. We focus throughout on the {\it twisted Einstein tensors} with diagonal
monodromy, including the twisted Einstein metric, the twisted $B$ field and the twisted torsion field of each orbifold 
sector. Finally, we present strong evidence for a conjectured set of {\it twisted Einstein equations} which should describe 
those sigma model orbifolds in this class which are also 1-loop conformal. 
} {$^a${\tt jdeboer@science.uva.nl} \\ $^b${\tt halpern@physics.berkeley.edu} \\ 
$^c${\tt helfgott@socrates.berkeley.edu} 
} 
 
\clearpage 
 
\renewcommand{\baselinestretch}{.4}\rm 
{\footnotesize 
\tableofcontents 
} 
\renewcommand{\baselinestretch}{1.2}\rm 
 
\section{Introduction} 

In the last few years there has been a quiet revolution in the local theory of {\it current-algebraic orbifolds}.
Building on the discovery of {\it orbifold affine algebras} \cite{Borisov:1997nc,Evslin:1999qb} in the cyclic
permutation orbifolds, Refs.~[3-5] gave the twisted currents and stress tensor in each twisted sector of any 
current-algebraic orbifold $A(H)/H$ - where $A(H)$ is any current-algebraic conformal field theory [6-10] with a discrete 
symmetry group $H$. The construction treats all current-algebraic orbifolds at the same time, using the method of {\it eigenfields} 
and the {\it principle of local isomorphisms} to map OPEs in the symmetric theory to OPEs in the orbifold. The orbifold results 
are expressed in terms of a set of twisted tensors or {\it duality transformations}, which are discrete Fourier transforms 
constructed from the eigendata of the $H$-{\it eigenvalue problem}.

More recently, the special case of the WZW orbifolds 
\begin{gather}
\frac{A_g(H)}{H} ,\quad H\subset Aut(g)
\end{gather}
was worked out in further detail [11-14], introducing the {\it extended $H$-eigenvalue problem} and the {\it linkage relation} to 
include the {\it twisted affine primary fields}, the twisted vertex operator equations and the {\it twisted KZ equations} of the 
WZW orbifolds. Ref.~\cite{so2n} includes a review of the general left- and right-mover twisted KZ systems. For detailed 
information on particular classes of WZW orbifolds, we direct the reader to the following references:

$\bullet$ the WZW permutation orbifolds [11-13]

$\bullet$ the inner-automorphic WZW orbifolds \cite{deBoer:2001nw, Perm}

$\bullet$ the (outer-automorphic) charge conjugation orbifold on $\su (n\geq 3)$ \cite{Halpern:2002ab}

$\bullet$ the outer-automorphic WZW orbifolds on $\so (2n)$, including the triality orbifolds \linebreak
\indent $\quad$ on $\so(8)$ \cite{so2n}.

\noindent Ref.~\cite{Halpern:2002ab} also solved the twisted vertex operator equations and the twisted KZ systems in an abelian 
limit\footnote{An abelian twisted KZ equation for the inversion orbifold $x \rightarrow -x$ was given earlier in Ref.~\cite{Froh}.}
to obtain the {\it twisted vertex operators} for each sector of a large class of orbifolds on abelian $g$. Moreover, Ref.~\cite{Perm} 
found the {\it general orbifold Virasoro algebra} (twisted Virasoro operators \cite{Borisov:1997nc,DV2}) of the WZW permutation 
orbifolds and used the general twisted KZ system to study {\it reducibility} of the general twisted affine primary field. Recent progress 
at the level of characters has been also reported in Refs.~\cite{KacTod,Borisov:1997nc,Ban,Birke}.

In addition to the operator formulation, Ref.~\cite{deBoer:2001nw} also gave the {\it general WZW orbifold action}, special cases
of which are further discussed in Refs.~\cite{Halpern:2002ab,so2n}. The general WZW orbifold action provides the classical description of each 
sector of every WZW orbifold $A_g (H)/H$ in terms of appropriate {\it group orbifold elements} with diagonal monodromy, which are
the classical limit of the twisted affine primary fields. Moreover, Ref.~\cite{Halpern:2002hw} gauged the general WZW orbifold action by general 
twisted gauge groups to obtain the {\it general coset orbifold action}, which describes each sector of the general coset orbifold $A_{g/h} (H)/H$.

The present paper is primarily an extension of the `action' series described in the previous paragraph: Our thrust here is to go beyond the group
orbifold elements of $A_g (H)/H$ and $A_{g/h} (H)/H$ to study the Einstein geometry and in particular the {\it twisted Einstein
tensors} of these orbifolds -- as well as their embedding in a much larger class of {\it sigma model orbifolds} $A_M (H)/H$.

In this development, we focus on the twisted Einstein tensors with diagonal monodromy on the cylinder $(\xi,t)$, for example the
{\it twisted Einstein metric} of sector $\s$:
\begin{equation}
\hG_{\nrm;\nsn} (\hx_\s (\xi +2\pi,t),\s) = e^{-\tp \srac{n(r)+n(s)}{\r(\s)}} \hG_{\nrm;\nsn} (\hx_\s (\xi,t),\s) \,.
\end{equation}
Here $\hx_\s (\xi,t)$ are the twisted Einstein coordinates of sector $\s$ and the indices $n(r),n(s),\m,\n$ and $\r(\s)$ are easily computed
for each orbifold. We are unaware of any prior study of twisted Einstein tensors in orbifold theory, although their existence in the twisted
sectors of each orbifold follows from the same principles as the more-familiar twisted currents.

For the WZW orbifolds $A_g(H)/H$ in particular we have chosen to approach the Einstein geometry via {\it phase space} dynamics, and, as a 
foundation for this discussion, we also provide the equal-time operator formulation of the general WZW orbifold. For the sigma model orbifolds 
$A_M(H)/H$, our central result is the construction of the corresponding {\it sigma model orbifold action} -- which includes as special cases 
the general WZW orbifold action and the general coset orbifold action, as well as the actions of a large number of other orbifolds such as 
orbifolds of the principal chiral models.

Finally, we present strong evidence for a conjectured set of {\it twisted Einstein equations} which should describe those sigma model orbifolds
in this class which are also 1-loop conformal. In analogy with the usual sigma-model Einstein equations [21-26], 
our twisted Einstein equations involve the twisted Ricci tensor, the twisted torsion field and the orbifold dilaton of sector $\s$.

A number of appendices are also included, and we mention in particular App.~E -- which notes in particular that the twisted Einstein metrics
are constant for all orbifolds on abelian $g$.

\section{Phase-Space Geometry of the General WZW Orbifold}

In this section we consider the phase-space dynamics and in particular the classical {\it phase-space geometry} of the general
WZW orbifold. The input here is the general left- and right-mover twisted current algebra and the corresponding left- and
right-mover twisted affine-Sugawara constructions \cite{Halpern:2000vj,deBoer:2001nw} of these orbifolds, as well as the classical group orbifold 
elements \cite{deBoer:2001nw} with diagonal monodromy. The twisted $B$ fields (see Subsecs.~$2.4, 2.5$) of the general WZW orbifold are needed 
for the canonical realization of the twisted currents, while the twisted Einstein metrics (see Subsec.~$2.8$) are first 
encountered in the canonical form of the general WZW orbifold Hamiltonian. As a final check on the results of this section, 
Subsec.~$2.9$ contains an alternate, phase-space derivation of the original coordinate-space form of the general WZW orbifold 
action given in Ref.~\cite{deBoer:2001nw}.

\subsection{The Twisted Current Modes and the Twisted Affine Primary Fields}

We begin our discussion with the {\it general twisted current algebra} \cite{deBoer:1999na,Halpern:2000vj,deBoer:2001nw} of the 
general WZW orbifold $A_g (H) /H$
\begin{subequations}
\label{tw-current-alg}
\begin{align}
[\hj_\nrm( \mnrrs), \hj_\nsn(\nnsrs) ] &= i\scf_{\nrm;\nsn}{}^{\!n(r)+n(s),\d}(\s) \hj_{n(r)+n(s),\d}(\mnnrnsrs ) \nn\\
 &+ (\mnrrs)\d_{m+n+\frac{n(r)+n(s)}{\r(\s)},\,0}\sG_{\nrm;-\nrn}(\s) \\
[\hjb_\nrm(\mnrrs), \hjb_\nsn(\nnsrs) ] &= i\scf_{\nrm;\nsn}{}^{\!n(r)+n(s),\d}(\s) \hjb_{n(r)+n(s),\d}(\mnnrnsrs ) \nn\\ 
& -(\mnrrs)\d_{m+n+\frac{n(r)+n(s)}{\r(\s)},\,0}\sG_{\nrm;-\nrn}(\s) \label{tw-r-CA}
\end{align}
\begin{equation}
[\hj_\nrm(\mnrrs), \hjb_\nsn(\nnsrs)] = 0, \quad \srange
\end{equation}
\begin{gather}
\hj_{n(r) \pm \r(\s),\m} (m +\srac{n(r) \pm \r(\s)}{\r(\s)}) =\hj_\nrm (m \pm 1+\nrrs) \\
\hjb_{n(r) \pm \r(\s),\m} (m +\srac{n(r) \pm \r(\s)}{\r(\s)}) =\hjb_\nrm (m\pm 1+\nrrs)
\end{gather}
\end{subequations}
where $g$ is any semisimple Lie algebra and $H \subset Aut (g)$ is any discrete symmetry group of $g$. In this result, $\r(\s)$ 
is the order of the element $h_\s \in H$, $N_c$ is the number of conjugacy classes of $H$ and $n(r) ,n(s)$ and $\m ,\n$ are 
respectively the spectral indices and degeneracy labels of the {\it $H$-eigenvalue problem} \cite{deBoer:1999na,Halpern:2000vj}
\begin{equation}
\label{H-eig}
\ws_a{}^b U^\hcj (\s)_b{}^\nrm \!=\! U^\hcj (\s)_a{}^\nrm E_{n(r)} (\s) ,\quad E_{n(r)}(\s) \!=\! e^{-\tp \nrrs} ,\quad \srange
\end{equation}
where $\ws$ is the action $J'=\ws J$ of $h_\s$ on the currents of untwisted affine $g$ \cite{Kac,Moody,Bardakci:1971nb,Halpern:1996js}.

The twisted tensors $\scf (\s)$ and $\sG (\s)$ in \eqref{tw-current-alg} are particular {\it duality transformations} \cite{deBoer:1999na}
called the {\it twisted structure constants} and the {\it twisted metric} of sector $\s$ respectively. (This is the twisted 
{\it tangent-space} metric, as opposed to the twisted Einstein-space metric which will be introduced later). The explicit 
forms of $\scf (\s)$ and $\sG (\s)$ were given in Refs.~\cite{deBoer:1999na,Halpern:2000vj}
\begin{subequations}
\label{sG-scf-Defn}
\begin{gather}
\sG_{\nrm;\nsn} (\s) \equiv \schisig_\nrm \schisig_\nsn U(\s)_\nrm{}^a U(\s)_\nsn{}^b G_{ab} \\
\scf_{\nrm;\nsn}{}^\ntd (\s) \equiv \schisig_\nrm \schisig_\nsn \schisig^{-1}_\ntd U(\s)_\nrm{}^a U(\s)_\nsn{}^b
   f_{ab}{}^c U\hc (\s)_c{}^\ntd
\end{gather}
\end{subequations}
where $f_{ab}{}^c$ are the structure constants of $g$, $G_{ab} = \oplus_I k_I \eta_{ab}^I$ is the generalized 
metric of affine $g$, $U\hc (\s)$ is the solution to the $H$-eigenvalue problem \eqref{H-eig} and $\schisig$ are normalization
constants with $\schizero =1$. In the untwisted sector $\s =0$, one has $\w (h_0) = U\hc (0) = \thickone$ and 
Eq.~\eqref{tw-current-alg} reduces to the original untwisted affine Lie algebra \cite{Kac,Moody,Bardakci:1971nb,Halpern:1996js} of $A_g(H)$.

As discussed in Ref.~\cite{deBoer:2001nw}, the {\it sign reversal} of the central term in the twisted right-mover current 
algebra \eqref{tw-r-CA} is quite correct, and is in agreement with earlier analysis at the level of orbifold characters \cite{Verlinde:1989oa}. 
See Refs.~[11-13] and in particular Ref.~\cite{so2n} for recent discussion of the closely related issue 
of {\it rectification} of the twisted right-mover current algebra into a copy of the twisted left-mover current algebra.

We will also need the commutators of the twisted current modes with the {\it twisted affine primary fields}
$\hg (\st,\xi,t,\s)$ of sector $\s$ on the cylinder $(\xi,t),\, 0\leq \xi \leq 2\pi$
\begin{subequations}
\label{Mink-spc}
\begin{gather}
[ \hj_\nrm (\mnrrs) ,\hg (\st,\xi,t,\s)] = \hg (\st,\xi,t,\s) \st_\nrm e^{i(\mnrrs ) (\xi +t)} \\
[\hjb_\nrm (\mnrrs) ,\hg (\st,\xi,t,\s)] = -e^{i (\mnrrs) (\xi -t)} \st_\nrm \hg (\st,\xi,t,\s) 
\end{gather}
\end{subequations}
where $\st_\nrm \equiv \st_\nrm (T,\s)$ is another set of duality transformations, called the {\it twisted
representation matrices} of sector $\s$
\begin{gather}
\st_\nrm (T,\s) \equiv \schisig_\nrm U(\s)_\nrm{}^a U(T,\s) T_a U\hc (T,\s) \,. \label{st-Defn}
\end{gather}
Here $U\hc (T,\s)$ is the unitary eigenvector matrix of the {\it extended $H$-eigenvalue problem} \cite{deBoer:2001nw}
\begin{subequations}
\label{extHeig}
\begin{gather}
W(h_\s ;T) U^\dagger (T,\s) = U^\dagger (T,\s) E(T,\s) ,\quad \srange \label{Ext-H-eig} \\
E(T,\s)_\Nrm {}^\Nsn \equiv \de_\Nrm{}^\Nsn e^{-\tp \frac{N(r)}{R(\s)}} ,\quad
\de_\Nrm{}^\Nsn \equiv \de_\m{}^\n \de_{N(r)-N(s),0\, \text{mod } R(\s)} 
\end{gather}
\end{subequations}
where $W(h_\s ;T)$ is the action of $h_\s \!\!\in \!\!H ,\, H \!\!\subset \!\!\text{Aut}(g)$ in matrix rep $T$ of $g$ and $R(\s)$ is the 
order of $W(h_\s ;T)$. The $H$-eigenvalue problem \eqref{H-eig}, which controls the form of the general twisted current algebra
\eqref{tw-current-alg}, is the special case of \eqref{extHeig} when $T$ is the adjoint representation. 

The twisted affine primary fields $\hg (\st,\s)$ and the twisted representation matrices $\st (T,\s)$ have the same matrix 
indices
\begin{equation}
\hg (\st, \xi,t,\s) \equiv \{ \hg (\st,\xi,t,\s)_\Nrm {}^\Nsn \} ,\quad \st (T,\s) \equiv \{ \st (T,\s)_\Nrm {}^\Nsn \} 
\end{equation}
where $N(r),N(s)$ and $\m ,\n$ are respectively the spectral indices and degeneracy labels of the extended $H$-eigenvalue 
problem. The commutators \eqref{Mink-spc} are the analogues on the cylinder of previous results \cite{deBoer:2001nw} on the sphere, using 
the correspondence
\begin{equation}
\label{hg-Rescale}
\hg(\st,\xi,t,\s) = \Big{\{} \bz^{\D_{\sgb (\s)} (\st)} \hg(\st,\bz,z,\s) z^{\D_{\sgb (\s)} (\st)} \Big{\}}_{z= e^{i(t+\xi)}
   ,\, \bz=e^{i(t-\xi)}}
\end{equation}
where $\D_{\sgb (\s)} (\st)$ is called the twisted conformal weight matrix \cite{deBoer:2001nw,Perm}. Reducibility of the twisted affine 
primary fields is discussed in Refs.~\cite{Halpern:2002ab,Perm}.

We mention here a number of properties of the duality transformations which are needed below, beginning with 
the {\it selection rules}:
\begin{subequations}
\label{Selection}
\begin{gather}
\sG_{\nrm;\nsn} (\s)= \de_{n(r)+n(s) ,0\, \text{mod } \r(\s)} \sG_{\nrm ;\mnrn} (\s) \label{sG-select}\\
\scf_{\nrm;\nsn}{}^\ntd (\s) = \de_{n(r)+n(s)-n(t) ,0\, \text{mod } \r(\s)} \scf_{\nrm;\nsn}
   {}^{n(r)+n(s) ,\de} (\s) 
\end{gather}
\begin{equation}
\label{T-select}
E_{n(r)}(\s)^\ast \st_\nrm (T,\s) =e^{\tp \frac{n(r)}{\r(\s)}} \st_\nrm (T,\s) = E(T,\s) \st_\nrm (T,\s) E(T,\s)^{\ast}\,. 
\end{equation}
\end{subequations}
More generally, the selection rules of duality transformations follow \cite{deBoer:1999na, Halpern:2000vj,deBoer:2001nw} directly from the 
$H$-invariance of the corresponding untwisted quantities. The indices of $\sG_{\bullet} (\s)$ and $\scf (\s)$ can be raised and 
lowered with $\sG_{\bullet} (\s)$ and its inverse $\sG^{\bullet} (\s)$, and moreover, $\scf_{\bullet} (\s)$ with 
all indices down is totally antisymmetric.

Other useful properties of the twisted representation matrices include \cite{deBoer:2001nw}
\begin{subequations}
\label{st-Props}
\begin{gather}
[ \st_\nrm (T,\s),\st_\nsn (T,\s)]= i\scf_{\nrm;\nsn}{}^{n(r)+n(s),\de} (\s) \st_{n(r)+n(s),\de} (T,\s) \label{gen-orb-Lie} \\
\widehat{Tr}(\sm(\st,\s) \st_\nrm(T,\s)  \st_\nsn(T,\s)) = \sG_{\nrm,\nsn}(\s)  \label{M-trace} \\
\widehat{Tr}(AB) \equiv \sum_{r,\m,s,\n} A_\Nrm{}^\Nsn B_\Nsn{}^\Nrm \\
[ \sm(\st,\s) ,\st_\nrm (T,\s)] =[ \sm (\st,\s),\hg (\st,\xi,t,\s)]=0 \label{M-g-Comm} \\
E(T,\s) \sm(\st(T,\s),\s) E(T,\s)^{\ast} =\sm(\st(T,\s),\s) ,\quad \srange \label{M-E-Comm}
\end{gather}
\end{subequations}
where the twisted structure constants $\scf (\s)$ of the {\it general orbifold Lie algebra} $\hg (\s)$ in 
\eqref{gen-orb-Lie} are the same as those in the general twisted current algebra \eqref{tw-current-alg}. The 
quantity $\sm (\st,\s)$ is another duality transformation called the {\it twisted data matrix} \cite{deBoer:2001nw,Halpern:2002ab, 
Halpern:2002hw,so2n} which stores the Dynkin indices of the untwisted representation matrices and the levels of the untwisted affine algebra.
The twisted data matrix is proportional to $\thickone$ when g has the form $g=\oplus_I \gfrak^I$, where $\gfrak^I$ with
$k_I =k$ is isomorphic to simple $\gfrak$ and $T^I \simeq T$.

\subsection{The General Twisted Equal-Time Operator Algebra}

Our next task is to translate the mode algebra of the previous subsection into corresponding {\it equal-time commutation
relations} on the cylinder $(\xi,t)$. For this purpose we define the twisted local currents on 
the cylinder:
\begin{subequations}
\label{Local-J-Defn}
\begin{gather}
\hj_\nrm (\xi,t) \equiv \sum_{m \in \Zint} \hj_\nrm (\mnrrs) e^{-i (\mnrrs)(\xi +t)} = 
   \hj_{n(r) \pm \r(\s) ,\m} (\xi,t) \quad \\
\hjb_\nrm (\xi,t) \equiv \sum_{m \in \Zint} \hjb_\nrm (\mnrrs) e^{-i (\mnrrs)(\xi -t)} = 
   \hjb_{n(r) \pm \r(\s) ,\m} (\xi,t) \\
\hj_\nrm (\xi +2\pi ,\s) =e^{-2\pi i\frac{n(r)}{\r(\s)}} \hj_\nrm (\xi ,\s) ,\quad 
   \hjb_\nrm (\xi +2\pi ,\s) =e^{-2\pi i\frac{n(r)}{\r(\s)}} \hjb_\nrm (\xi ,\s) \,. \label{JMonos}
\end{gather}
\end{subequations}
Note that the twisted left- and right- mover local currents have the same monodromy \cite{deBoer:2001nw} upon circumnavigation
of the cylinder.

Then we may use the mode algebra \eqref{tw-current-alg},\eqref{Mink-spc} to compute the general twisted equal-time current algebra
\begin{subequations}
\label{eq-time-current-alg}
\begin{align}
[ \hj_\nrm (\xi,t,\s) ,\hj_\nsn (\eta,t,\s) ] &= 2\pi i \Big{(} \scf_{\nrm;\nsn}{}^{n(r)+n(s),\de} (\s)
\hj_{n(r)+n(s) ,\de} (\eta,t,\s) \quad \nn \\ 
& +\de_{n(r)+n(s) ,0\, \text{mod } \r(\s)} \sG_{\nrm;-\nrn} (\s) \pl_\xi \Big{)} \de_{n(r)} (\xi -\eta) \label{eq-t-curr-alg-lm} \\
[ \hjb_\nrm (\xi,t,\s) ,\hjb_\nsn (\eta,t,\s) ] &= 2\pi i \Big{(} \scf_{\nrm;\nsn}{}^{n(r)+n(s),\de} (\s)
\hjb_{n(r)+n(s) ,\de} (\eta,t,\s) \quad \nn \\ 
& -\de_{n(r)+n(s) ,0\, \text{mod } \r(\s)} \sG_{\nrm;-\nrn} (\s) \pl_\xi \Big{)} \de_{n(r)} (\xi -\eta)
\end{align} \vsvs
\begin{equation}
[ \hj_\nrm (\xi,t,\s) ,\hjb_\nsn (\eta,t,\s)] =0 ,\quad \srange
\end{equation}
\end{subequations}
and the equal-time algebra of the twisted cylinder currents with the twisted affine primary fields $\hg$:
\vspace{-0.1in}
\begin{subequations}
\label{EqT-Chiral}
\begin{align}
[ \hj_\nrm (\xi,t,\s) ,\hg (\st,\eta,t,\s)] &= 2\pi \hg (\st,\eta,t,\s) \st_\nrm \de_{n(r)} (\xi-\eta) \\
[\hjb_\nrm (\xi,t,\s) ,\hg (\st,\eta,t,\s)] &=-2\pi \st_\nrm \hg (\st,\eta,t,\s) \de_{n(r)} (\xi-\eta) \,.
\end{align}
\end{subequations}
A central feature of the general twisted equal-time operator algebra \eqref{eq-time-current-alg}, 
\eqref{EqT-Chiral} is the appearance of the {\it monodromy factor} $\de_{n(r)} (\xi -\eta)$, which is a modified 
Dirac delta function with monodromy:
\begin{subequations}
\label{de_nr-Props}
\begin{gather}
\de_{n(r)} (\xi -\eta) \equiv  \frac{1}{2\pi} \sum_{m \in \Zint} e^{-i (\mnrrs )(\xi -\eta)} =
  e^{-i\frac{n(r)}{\r(\s)} (\xi -\eta)} \de (\xi -\eta) = \de_{-n(r)} (\eta-\xi) \\
\de_{n(r)} (\xi -\eta \pm 2\pi) = e^{\mp \tp \nrrs} \de_{n(r)} (\xi-\eta) \\
\de_{n(r) \pm \r(\s)} (\xi-\eta) = \de_{n(r)} (\xi-\eta) ,\quad \de_{0 \,\text{mod } \r(\s)} (\xi-\eta)
   =\de (\xi -\eta) \label{de-nr-period} 
\end{gather}
\begin{gather}
\de (\xi-\eta) \equiv \frac{1}{2\pi} \sum_{m\in \Zint} e^{-im(\xi-\eta)} ,\quad \de (\xi-\eta \pm 2\pi) =\de (\xi-\eta) \,.
\end{gather}
\end{subequations}
Here the ordinary $2\pi$-periodic Dirac delta function is denoted by $\de (\xi -\eta)$, and we note that the phase of
the monodromy factor is non-trivial when $\xi -\eta = \pm 2\pi$. Our notation $\de_{n(r)} (\xi-\eta)$ is a shorthand for the more
accurate form $\de_{n(r)/\r(\s)} (\xi-\eta)$.

In the computations above (and others to follow) we found it helpful to use identities of the following type
\begin{subequations}
\label{de_nr-Idents}
\begin{gather}
\hat{A}_\nrm (\xi,t) \de_{n(s)} (\xi \!-\!\eta) \!=\! \hat{A}_\nrm (\eta,t) \de_{n(r)+n(s)} (\xi \!-\!\eta) \bigspc \nn \\
   \bigspc \quad \text{ for any } \hat{A} \text{ s.t. } \hat{A}_\nrm (\xi +2\pi,t) = e^{-\tp \nrrs} \hat{A}_\nrm (\xi,t) \\
\hat{A}^\nrm (\xi,t) \de_{n(s)} (\xi \!-\!\eta) \!=\! \hat{A}^\nrm (\eta,t) \de_{n(s)-n(r)} (\xi \!-\!\eta) \bigspc \nn \\
   \bigspc \quad \text{ for any } \hat{A} \text{ s.t. } \hat{A}^\nrm (\xi +2\pi,t) = e^{\tp \nrrs} \hat{A}^\nrm (\xi,t) \\
\hat{A}_\nrm{}^{\nsn} (\xi,t) \de_{n(t)} (\xi -\eta) = \hat{A}_\nrm{}^{\nsn} (\eta,t)
   \de_{n(r)-n(s)+n(t)} (\xi-\eta) \bigspc \nn \\
   \bigspc \text{for any } \hat{A} \text{ s.t. } \hat{A}_\nrm{}^\nsn (\xi +2\pi,t) = e^{-\tp \frac{n(r)-n(s)}
   {\r(\s)}} \hat{A}_\nrm{}^\nsn (\xi,t) \label{A_^}
\end{gather}
\end{subequations}
which follow directly from the properties of the monodromy factor in \eqref{de_nr-Props}. Note that (by setting $n(r)$ or $n(s)$ to zero) 
the identities in ($2.15$a,b) are special cases of the identity in ($2.15$c). Moreover, Eq.~\eqref{de_nr-Idents} gives 
us integral identities such as
\begin{subequations}
\label{Int-Id2}
\begin{gather}
\hat{A}_\nrm{}^\nsn (\eta,t) \de_{n(s)-n(r)} (\eta-\xi) = \hat{A}_\nrm{}^\nsn (\xi,t) \de (\eta-\xi) \\
\int_0^{2\pi} \!\!d\eta \hat{A}_\nrm{}^\nsn (\eta,t) \de_{n(s)-n(r)} (\eta -\xi) = \hat{A}_\nrm{}^\nsn (\xi,t) \\
\int_0^{2\pi} \!\!d\eta \hat{A}_\nrm{}^\nsn (\eta,t) \pl_\eta \de_{n(s)-n(r)} (\eta-\xi) = -\pl_\xi \hat{A}_\nrm{}^\nsn (\xi,t) 
\end{gather}
\end{subequations}
which are useful in evaluating commutators or brackets of integrated quantities with local fields. Generalizations 
of these identities to objects with any number of indices are easily derived. For example, all these identities hold
for operators of the form $\hat{A}_{\nrm;\nsn}$ by replacing $n(s) \rightarrow -n(s)$ in the monodromy and the monodromy 
factors of \eqref{A_^} and \eqref{Int-Id2}.

Although the identities \eqref{de_nr-Idents} can be used to write the twisted equal-time algebra \eqref{eq-time-current-alg}, 
\eqref{EqT-Chiral} in alternate forms, e.g.
\begin{align}
[ \hj_\nrm (\xi,t,\s) ,\hj_\nsn (\eta,t,\s) ] &= 2\pi i \Big{(} \scf_{\nrm;\nsn}{}^{n(r)+n(s),\de} (\s)
\hj_{n(r)+n(s) ,\de} (\xi,t,\s) \quad \nn \\
& +\de_{n(r)+n(s) ,0\, \text{mod } \r(\s)} \sG_{\nrm;-\nrn} (\s) \pl_\xi \Big{)} \de_{-n(s)} (\xi -\eta) 
\end{align}
it is not possible to eliminate the phases of the monodromy factors $\de_{n(r)} (\xi-\eta)$ entirely. Moreover, it is not 
difficult to check {\it a posteriori} that the phases of the monodromy factors are necessary and sufficient to guarantee 
the consistency of the general twisted equal-time operator algebra with the monodromy \eqref{JMonos} of the twisted currents. 
For example, note that
\begin{equation}
\hj_{n(r)+n(s),\de} (\eta+2\pi,t,\s) \de_{n(r)} (\xi-\eta-2\pi) = e^{-\tp \srac{n(s)}{\r(\s)}} \hj_{n(r)+n(s),\de}
(\eta,t,\s) \de_{n(r)} (\xi-\eta)
\end{equation}
which matches the $\eta$ monodromy of the left side of \eqref{eq-t-curr-alg-lm}. We also remark that the $n(r) 
\rightarrow n(r) \!\pm \!\r(\s)$ periodicity of the monodromy factor in \eqref{de-nr-period} is consistent with the 
corresponding periodicity ($2.11$a,b) of the twisted currents.

\subsection{The Classical Limit and Twisted Equal-Time Brackets}

To describe the classical limit of WZW orbifold theory, we first replace each commutator $[ \,,\,]$ in 
Eqs.~\eqref{eq-time-current-alg}, \eqref{EqT-Chiral} by the rescaled Poisson bracket $[\,,\,] \rightarrow \{ \,,\, \}$ to obtain
the general twisted equal-time brackets:
\begin{subequations}
\label{EqTCA}
\begin{align}
\{ \hj_\nrm (\xi,t,\s) ,\hj_\nsn (\eta,t,\s) \} &= 2\pi i \Big{(} \scf_{\nrm;\nsn}{}^{n(r)+n(s),\de} (\s)
\hj_{n(r)+n(s) ,\de} (\eta,t,\s) \nn \\ \vsvsvs
& +\de_{n(r)+n(s) ,0\, \text{mod } \r(\s)} \sG_{\nrm;-\nrn} (\s) \pl_\xi \Big{)} \de_{n(r)} (\xi -\eta) \\
\{ \hjb_\nrm (\xi,t,\s) ,\hjb_\nsn (\eta,t,\s) \} &= 2\pi i \Big{(} \scf_{\nrm;\nsn}{}^{n(r)+n(s),\de} (\s)
\hjb_{n(r)+n(s) ,\de} (\eta,t,\s) \nn \\ \vsvsvs
& -\de_{n(r)+n(s) ,0\, \text{mod } \r(\s)} \sG_{\nrm;-\nrn} (\s) \pl_\xi \Big{)} \de_{n(r)} (\xi -\eta)
\end{align}\vsvsvs
\begin{equation}
\{ \hj_\nrm (\xi,t,\s) ,\hjb_\nsn (\eta,t,\s)\} =0 ,\quad \srange
\end{equation}
\end{subequations} \vsvs
\begin{subequations}
\label{EqTChiral}
\begin{align}
&\bigspc \{ \hj_\nrm (\xi,t,\s) ,\hg (\st,\eta,t,\s)\} = 2\pi \hg (\st,\eta,t,\s) \st_\nrm \de_{n(r)} (\xi-\eta) \\ \vsvsvs
&\bigspc \{\hjb_\nrm (\xi,t,\s) ,\hg (\st,\eta,t,\s)\} =-2\pi \st_\nrm \hg (\st,\eta,t,\s) \de_{n(r)} (\xi-\eta) \,.
\end{align}
\end{subequations}
Here $\{ \hg (\st,\xi,t,\s)\}$ are the {\it group orbifold elements} \cite{deBoer:2001nw,Halpern:2002ab,Halpern:2002hw,Perm,so2n}, which 
are the classical (high-level) limit of the twisted affine primary fields of sector $\s$.

The group orbifold elements are unitary matrices $\hg \hg^\dagger =\thickone$ which are locally group
elements but which exhibit the (two-sided) diagonal monodromy
\begin{subequations}
\label{gMono} 
\begin{gather}
\hg (\st (T,\s),\xi +2\p ,t,\s) = E(T,\s) \hg (\st ,\xi,t,\s) E(T,\s)^{\ast} \\
\hg (\st (T,\s),\xi +2\p ,t,\s)_\Nrm{}^\Nsn = e^{-\tp \frac{N(r)-N(s)}{R(\s)}} \hg(\st,\xi,t,\s)_\Nrm{}^\Nsn
\end{gather}
\end{subequations}
on circumnavigation of the cylinder. The eigenvalue matrices $E(T,\s)$ are defined by the extended $H$-eigenvalue
problem in Eq.~\eqref{extHeig}. We have checked that, thanks to the monodromy factors $\de_{n(r)} (\xi -\eta)$, 
the monodromies \eqref{JMonos} and \eqref{gMono} are consistent with the general twisted brackets \eqref{EqTCA} and 
\eqref{EqTChiral}. In particular, the selection rule \eqref{T-select} of the twisted representation matrices $\st$ is necessary 
to check this consistency for the results in  Eq.~\eqref{EqTChiral}. The group orbifold elements also inherit the vanishing 
commutator with the twisted data matrix
\vspace{-0.1in}
\begin{equation}
[\sm (T,\s) ,\hg (\st,\xi,t,\s)] =0
\end{equation}
from the same property \eqref{M-g-Comm} of the twisted affine primary fields.

\subsection{Twisted Einstein Coordinates on the Group Orbifold}

Using the group orbifold elements, we turn now to the definition and properties of various twisted geometric objects in 
sector $\s$ of the general WZW orbifold $A_g (H)/H$.

To begin, we remind \cite{deBoer:2001nw} the reader that the group orbifold elements can be expressed in terms of the twisted tangent-space 
coordinates $\hbe$ of sector $\s$
\begin{align}
&\!\!\!\hg (\st (T,\s),\xi ,t,\s) =\!e^{i \hbe^\nrm (\xi,t,\s) \st_\nrm (T,\s)} ,\,\,\,
    \hbe^\nrm (\xi +2\pi,t,\s) = \!\hbe^\nrm (\xi,t,\s) e^{\tp \nrrs} \label{Beta-mono} 
\end{align}
whose diagonal monodromy is equivalent (by the selection rule \eqref{T-select}) to the monodromy \eqref{gMono} of the group 
orbifold element. In what follows we often suppress the time label $t$.

Our next task is to choose {\it twisted Einstein coordinates} $\hx(\xi) \equiv \hx_\s(\xi)$ with diagonal 
\linebreak monodromy, a property which can hold only in a preferred class of coordinate systems. For 
simplicity, we choose the coordinate system $\hx(\xi)=\hbe(\xi)$ in which the twisted Einstein 
coordinates are equal to the twisted tangent-space coordinates
\begin{subequations}
\begin{gather}
\hx_\s^\nrm(\xi) \equiv \hbe^\nsn(\xi,\s) \he^{-1}(0,\s)_\nsn{}^{\nrm} = \hbe^\nrm(\xi,\s) 
\end{gather}
\begin{gather}
\he(0,\s)_\nrm{}^{\nsn} = \d_\nrm{}^{\nsn} \equiv \d_\m{}^\n \de_{n(r)-n(s) ,0 \,\text{mod }\r(\s)} \label{e(0)=1} \\
\hg(\st (T,\s),\xi,\s)= e^{i\hx_\s^\nrm(\xi) \st_\nrm(T,\s)} ,\quad \hx_\s^\nrm(\xi+2\p)= 
   \hx_\s^\nrm(\xi)e^{\tp \nrrs} \label{hx-mono}
\end{gather}
\end{subequations}
where $\he(0,\s)=\thickone$ is the twisted left-invariant vielbein (see below) evaluated at $\hx =0$ in sector $\s$ of the group
orbifold. 

In analogy with the standard construction on group manifolds, we may now use the group orbifold elements to 
define a {\it twisted left-invariant vielbein} $\he (\hx)$ and its inverse $\he^{-1} (\hx)$
\vspace{-0.15in}
\begin{subequations}
\label{l-i-viel}
\begin{gather}
\hpl_\nrm(\xi) \equiv \srac{\pl}{\pl \hx^\nrm(\xi)},\quad \hpl_\nrm(\xi+2\p)=  
   e^{-2\p i\frac{n(r)}{\r(\s)}} \hpl_\nrm(\xi)  \label{pl-Mono} \\
\he_\nrm(\st,\hx(\xi)) \equiv -i \hg^{-1}(\st,\xi,\s)\hpl_\nrm\hg(\st,\xi,\s) \equiv \he(\hx(\xi))_\nrm
   {}^\nsn \st_\nsn(T,\s) \label{Def-twVB} \\
\he(\hx(\xi))_\nrm{}^\nsn = \widehat{Tr}\Big{(}\, \sm(\st,\s) \he_\nrm(\st,\hx(\xi)) \st(T,\s)_\ntd 
   \,\Big{)}\sG^{\ntd;\nsn}(\s) \label{225c} \\
\he_\nrm(\st,\hx(\xi)) \equiv \he_\nrm(\st,\hx_\s(\xi),\s), \quad \he(\hx(\xi))_\nrm{}^{\nsn} 
\equiv \he(\hx_\s(\xi),\s)_\nrm{}^{\nsn} 
\end{gather} \vsvsvs
\begin{align}
& \he^{-1}(\hx)_\nrm{}^\ntd \,\he(\hx)_\ntd{}^\nsn \!=\! \d_\nrm{}^\nsn ,\,\,\,
   \he^{-1}(\hx)_\nrm{}^\nsn \,\he(\st,\hx)_\nsn \!=\! \st_\nrm(T,\s) 
\end{align}
\end{subequations}
for each twisted sector $\s$ of the group orbifold. The relation \eqref{225c} for the twisted component vielbeins follows 
from \eqref{M-trace} and \eqref{Def-twVB}. Similarly, for each sector $\s$ we have the {\it twisted right-invariant vielbein} 
$\heb (\hx)$ and its inverse $\heb^{-1}(\hx)$:
\begin{subequations}
\label{r-i-viel}
\begin{gather}
\heb_\nrm(\st,\hx(\xi)) \equiv -i \hg(\st,\xi,\s) \hpl_\nrm\hg^{-1}(\st,\xi,\s) \equiv 
   \heb(\hx(\xi))_\nrm{}^\nsn \st_\nsn(T,\s) \label{Def-twVB2} \\
\heb(\hx(\xi))_\nrm{}^\nsn = \widehat{Tr} \Big{(}\, \sm(\st,\s) \heb_\nrm(\st,\hx(\xi)) 
   \st(T,\s)_\ntd\,\Big{)}\sG^{\ntd;\nsn}(\s) 
\end{gather}
\begin{equation}
\heb_\nrm(\st,\hx(\xi)) \equiv \heb_\nrm(\st,\hx_\s(\xi),\s), \quad \heb(\hx(\xi))_\nrm{}^{\nsn} 
\equiv \heb(\hx_\s(\xi),\s)_\nrm{}^{\nsn} 
\end{equation}
\begin{align}
&\!\!\!\heb^{-1}(\hx)_\nrm{}^\ntd \,\heb(\hx)_\ntd{}^\nsn \!=\! \d_\nrm{}^\nsn ,\,\,\, \heb^{-1}(\hx)_\nrm{}^\nsn 
   \,\heb(\st,\hx)_\nsn \!=\! \st_\nrm(T,\s) \,.
\end{align}
\end{subequations}
The twisted vielbeins $\he (\hx) ,\heb (\hx)$ are called left- and right-invariant because they are invariant respectively 
under left and right translations $\hg \rightarrow \hg_0 \hg$ and $\hg \rightarrow \hg \hg_0$ by constant group orbifold
elements $\hg_0$ in each sector $\s$ of the group orbifold. 

The monodromies \eqref{gMono}, \eqref{pl-Mono} and the selection rule \eqref{T-select} then imply the diagonal monodromies
of the twisted vielbeins
\begin{subequations}
\label{heMonos}
\begin{gather}
\he_\nrm(\st,\hx(\xi+2\p))= e^{-\tp \nrrs} E(T,\s) \he_\nrm (\st, \hx(\xi)) E(T,\s)^\ast \\
\he(\hx(\xi+2\p))_\nrm{}^\nsn = e^{-2\p i\frac{n(r)-n(s)}{\r(\s)} } \he(\hx(\xi))_\nrm{}^\nsn \label{he-Mono1}
\end{gather}
\begin{gather}
\heb_\nrm(\st,\hx(\xi+2\p))= e^{-\tp \nrrs} E(T,\s) \heb_\nrm (\st,\xi(\hx)) E(T,\s)^\ast \\
\heb(\smal{\hx(\xi+2\p)})_\nrm{}^\nsn = e^{-2\p i \frac{n(r)-n(s)}{\r(\s)}} \heb(\smal{\hx
   (\xi)})_\nrm{}^\nsn \label{heb-Mono1}
\end{gather}
\end{subequations}
and the inverse twisted vielbeins have the same monodromies as the twisted vielbeins. 

We also define the {\it twisted adjoint action} $\ho (\hx) \equiv \ho (\hx_\s ,\s)$ of sector $\s$
\begin{subequations}
\label{ho-Defn}
\begin{gather}
\hg(\st,\xi,\s) \st_\nrm(T,\s) \hg^{-1}(\st,\xi,\s) = \ho(\hx_\s ,\s)_\nrm{}^\nsn \st_\nsn(T,\s) \label{ho-defn1} \\
\ho (\hx (\xi+2\pi))_\nrm{}^\nsn = e^{-\tp \frac{n(r)-n(s)}{\r(\s)}} \ho(\hx(\xi))_\nrm{}^\nsn \label{ho-Mono}
\end{gather}
\begin{gather}
\ho(\hx)_\nrm{}^\ntd \ho(\hx)_\nsn{}^\nue \sG_{\ntd;\nue}(\s) = \sG_{\nrm;\nsn}(\s) \label{hOmega-sG} \\
\!\!\ho(\hx)_\nrm{}^{\!\nue} \ho(\hx)_\nsn{}^{\!\nvz} \scf_{\nue ;\nvz}{}^{\!\ntd} (\s) \!=\! 
   \scf_{\nrm;\nsn}{}^{\!\nue} (\s) \ho(\hx)_\nue {}^{\!\ntd} \\
\heb(\hx)_\nrm{}^{\!\nsn} \!=\! -\he(\hx)_\nrm{}^{\!\ntd} \ho(\hx)_\ntd{}^{\!\nsn} \\ 
\heb^{-1}(\hx)_\nrm{}^{\!\nsn} \!=\!-\ho^{-1}(\hx)_\nrm{}^{\!\ntd} \he^{-1}(\hx)_\ntd{}^{\!\nsn} 
\end{gather}
\end{subequations}
which relates the twisted vielbeins $\he$, $\heb$ of sector $\s$. Other useful identities include
\begin{subequations}
\begin{equation}
\label{e-form-hint1}
\he^{-1} (\hx)_\nrm {}^\ntd \hpl_\ntd \ho (\hx)_\nsn {}^\nue = -\scf_{\nrm;\nsn}{}^\ntd (\s)
\ho(\hx)_\ntd {}^\nue
\end{equation}
\begin{equation}
\heb^{-1} (\hx)_\nrm {}^\ntd \hpl_\ntd \ho(\hx)_\nsn {}^\nue = \ho(\hx)_\nsn {}^\ntd
\scf_{\nrm;\ntd}{}^\nue (\s)
\end{equation}
\end{subequations}
and the twisted Cartan-Maurer and inverse twisted Cartan-Maurer relations of sector $\s$
\begin{subequations}
\begin{align}
& \!\! \hpl_\nrm\he(\hx)_\nsn{}^{\!\ntd} -\! \hpl_\nsn\he(\hx)_\nrm{}^{\!\ntd} \!=\! \he(\hx)_\nrm
{}^{\!\nue} \he(\hx)_\nsn{}^{\!\nvz} \scf_{\nue;\nvz}{}^{\!\ntd} (\s)
\end{align}
\begin{eqnarray}
\he^{-1}(\hx)_\nrm{}^\ntd\hpl_\ntd\he^{-1}(\hx)_\nsn{}^\nue - \he^{-1}(\hx)_\nsn{}^\ntd
\hpl_\ntd \he^{-1}(\hx)_\nrm{}^\nue \nn\\
   	= \scf_{\nsn;\nrm}{}^\ntd(\s)\he^{-1}(\hx)_\ntd{}^\nue
\end{eqnarray}
\end{subequations}
which also have the same form for $\he \rightarrow \heb$.

We turn next to our first {\it twisted Einstein tensors} in sector $\s$ of the group orbifold. In particular, the {\it twisted 
$B$ field} $\hb$ and the {\it twisted torsion field} $\hh$ of sector $\s$ are defined as follows:
\begin{subequations}
\label{hB-and-hH-defn} 
\begin{align}
&\!\!\!\!\hh_{\nrm;\nsn;\ntd}(\hx)\!\equiv \!\hpl_\nrm\hb_{\nsn;\ntd}(\hx) \!+\!\hpl_\nsn\hb_{\ntd;\nrm}(\hx) 
\!+\!\hpl_\ntd\hb_{\nrm;\nsn}(\hx) \\
&\quad \quad \equiv -i\,\widehat{Tr} \left(\sm(\st,\s) \he_\nrm(\st,\hx) [\he_\nsn(\st,\hx),\he_\ntd(\st,\hx)] \right) \\
&\quad \quad=\he(\hx)_\nrm{}^{\!n(r')\m'}\he(\hx)_\nsn{}^{\!n(s')\n'} \he(\hx)_\ntd
{}^{\!n(t')\d'}\scf_{n(r')\m';n(s')\n';n(t')\d'}(\s) \label{Fdown}
\end{align}
\end{subequations} \vsvsvs
\begin{subequations}
\label{mono-of-hB-and-hH}
\begin{align}
& \quad \quad \hh_{\nrm;\nsn;\ntd}(\hx(\xi+2\p)) =e^{-2\p i\frac{n(r)+n(s)+n(t)}{\r(\s)}} \hh_{\nrm;\nsn;\ntd}(\hx(\xi))  \\
&\!\! \hb_{\nrm;\nsn}(\hx(\xi+2\p)) = e^{-2\p i \frac{n(r)+n(s)}{\r(\s)}} \hb_{\nrm;\nsn}
   (\hx(\xi)) ,\,\,\, \hb_{\nrm;\nsn} = -\hb_{\nsn;\nrm} \label{hB-mono} \\
&\bigspc \bigspc \quad \quad \hh(\hx) \!\equiv \!\hh (\hx_\s (\xi),\s) ,\quad \hb (\hx) \! \equiv \!\hb(\hx_\s (\xi),\s) .
\end{align}
\end{subequations}
The diagonal monodromies here follow from the diagonal monodromy \eqref{heMonos} of the twisted vielbein.
We also note with Ref.~\cite{Halpern:2000vj} that the twisted structure constants in \eqref{Fdown} and hence the twisted torsion field
\begin{subequations}
\label{antisymm}
\begin{gather}
\scf_{\nrm ;\nsn ;\ntd} (\s) \equiv \scf_{\nrm;\nsn}{}^\nue (\s) \sG_{\nue ;\ntd} (\s) = 
   -\scf_{\nrm ;\ntd;\nsn} (\s) \\
\hh_{\nrm;\nsn;\ntd} (\hx(\xi)) = -\hh_{\nrm;\ntd;\nsn} (\hx(\xi))
\end{gather}
\end{subequations}
are totally antisymmetric.

Finally, we use the exponential form \eqref{hx-mono} of the group orbifold elements to obtain these twisted geometric 
objects in closed form, beginning with the explicit form of the twisted adjoint action
\begin{subequations}
\label{exp-form-of-ho}
\begin{gather}
\ho(\hx) \!\equiv \!\ho(\hx_\s,\s) \!\equiv \!e^{-i\hY(\hx)} ,\quad \hY(\hx(\xi)) \equiv \hY(\hx_\s(\xi),\s) 
   \equiv \hx_\s^\nrm(\xi) \tst_\nrm(T^{adj},\s) \label{ho=eihy} \\
\bkspc \!\!\tst_\nrm(T^{adj},\s)_\nsn{}^{\!\ntd} \!\equiv \! -i\scf_{\nrm;\nsn}{}^{\!\ntd}(\s)
   \!= \!\schisig_\nsn \schisig_\ntd^{-1} \st_\nrm(T^{adj},\s)_\nsn{}^{\!\ntd} \label{adj-resc-tw-rep} \\
[\tst_\nrm (T^{adj},\s) ,\tst_\nsn (T^{adj},\s) ]= i\scf_{\nrm;\nsn}{}^{n(r)+n(s),\de} (\s) \tst_{n(r)+n(s),\de} (T^{adj},\s) \\
\hY (\hx(\xi+2\pi))_\nrm{}^\nsn = e^{-\tp \srac{n(r)-n(s)}{\r(\s)} } \hY(\hx(\xi))_\nrm{}^\nsn  \label{hY-Mono}
\end{gather}
\end{subequations}
which follows from its definition in \eqref{ho-defn1}. Here $T^{adj}$ is the adjoint representation of untwisted Lie $g$, and the 
objects $\{ \tst_\nsn (T^{adj},\s) \}$ are examples of so-called rescaled twisted representation matrices (see App.~A). 
Using \eqref{Def-twVB}, we may also compute the twisted vielbein in closed form:
\begin{equation}
\he(\hx)_\nrm{}^\nsn = \Big{(} \smal{\frac{e^{i\hY (\hx)}-1}{i\hY (\hx)}} \Big{)}_\nrm{}^\nsn ,\quad 
   \he(0)_\nrm{}^\nsn=\d_\nrm{}^\nsn \,. \label{exp-form-of-ehat}
\end{equation}
Similarly (up to a twisted gauge transformation $\Delta \hb =\hpl \hat{\Lambda}$) the following closed form of the 
twisted $B$ field
\begin{align}
&\hb_{\nrm;\nsn}(\hx) \!=\! \smal{\left( \frac{e^{i\hY}-e^{-i\hY}-2i\hY} {(i\hY)^2} \right)} {}_\nrm
   {}^{\!\ntd} \sG_{\ntd;\nsn}(\s) \!=\! -\smal{\left( \frac{e^{i\hY}-e^{-i\hY}-2i\hY}{(i\hY)^2} \right)} 
   {}_\nsn{}^{\!\ntd} \sG_{\ntd;\nrm}(\s) \nn \\
&\bigspc \bigspc \bigspc =\! \smal{\left( \frac{e^{i\hY}-e^{-i\hY}-2i\hY}{(i\hY)^2}  \right)} {}_\nrm{}^{\! -n(s),\de}
   \sG_{-n(s),\de;\nsn}(\s)  \label{exp-form-of-hB} 
\end{align}
is obtained from \eqref{hB-and-hH-defn}. To see the antisymmetry of $\hb$ in \eqref{exp-form-of-hB}, we have used 
Eq.~\eqref{hOmega-sG} in the form 
\vspace{-0.1in}
\begin{equation}
\label{f-of-hOmega-id}
f(\ho )_\nrm {}^\ntd \sG_{\ntd;\nsn} (\s)= f(\ho^{-1} )_\nsn{}^\ntd \sG_{\ntd;\nrm} (\s) 
\end{equation}
which holds for any power series $f$ of $\ho$. The monodromy \eqref{hY-Mono} of $\hY$ is consistent with the monodromy 
\eqref{hB-mono} of $\hB$ because the twisted tangent-space metric $\sG_{\bullet}(\s)$ satisfies the selection rule 
\eqref{sG-select}.

The twisted Einstein metrics of the general WZW orbifold are discussed in Subsec.~$2.8$.

\subsection{Phase-Space Realization of the Twisted Currents}

We will also need to define {\it twisted Einstein momenta} $\hp^\s (\xi) \equiv \hp^\s (\xi,t)$, which are `canonically' 
conjugate to the twisted Einstein coordinates:
\begin{subequations}
\label{canon-brack}
\begin{gather}
\{ \hp_\nrm^\s (\xi,t) ,\hx^\nsn_\s (\eta,t) \} \equiv -i \de_\nrm{}^{\!\nsn} \de_{n(r)} (\xi-\eta) \label{tw-px-bracket} \\
\{ \hp^\s_\nrm (\xi,t),\hp^\s_\nsn (\eta,t) \} =\{ \hx_\s^\nrm (\xi,t),\hx_\s^\nsn (\eta,t) \} \equiv 0 \\
\hp_\nrm^\s (\xi +2\pi,t) = e^{-\tp \nrrs} \hp_\nrm^\s (\xi,t) ,\quad \hx_\s^\nrm (\xi +2\pi,t) =
   \hx_\s^\nrm (\xi,t) e^{\tp \nrrs}. \label{pxMonos-1}
\end{gather}
\end{subequations}
It is not difficult to check that, thanks to its monodromy factor $\de_{n(r)} (\xi-\eta)$, the twisted equal-time 
bracket \eqref{tw-px-bracket} is consistent with the monodromies in \eqref{pxMonos-1}. 

Using the twisted vielbein and twisted $B$ field of the previous subsection, we can now state the {\it canonical bracket 
realization} of the general twisted equal-time current algebra \eqref{EqTCA}
\begin{subequations}
\label{hb-form-of-twisted-currents}
\begin{gather}
\hj_\nrm(\xi,\s) = 2\pi \he^{-1}(\hx)_\nrm{}^\ntd \hp_\ntd(\hb) +  \half \pl_\xi \hx_\s^\ntd 
   \he(\hx)_\ntd{}^\nsn \sG_{\nsn;\nrm}(\s) 
\end{gather}
\begin{gather}
\hjb_\nrm(\xi,\s) = 2\pi \heb^{-1}(\hx)_\nrm{}^\ntd \hp_\ntd(\hb) -  \half \pl_\xi \hx_\s^\ntd 
   \heb(\hx)_\ntd{}^\nsn \sG_{\nsn;\nrm}(\s) \\
\hp_\nrm(\hb) \equiv \hp^\s_\nrm + \onefourpi \hb_{\nrm;\nsn}(\hx)\pl_\xi\hx_\s^\nsn ,\quad \srange
\end{gather}
\end{subequations}
in sector $\s$ of any current-algebraic orbifold. With the canonical brackets \eqref{canon-brack}, the identities 
\eqref{de_nr-Idents} and the relations \eqref{l-i-viel}-\eqref{antisymm}, we have checked at length that the 
realization \eqref{hb-form-of-twisted-currents} satisfies all the twisted equal-time brackets in \eqref{EqTCA} and 
\eqref{EqTChiral}. Two independent derivations of this result are discussed in Subsec.~$4.5$ and App.~B.

As in the untwisted case \cite{Club}, the classical realization \eqref{hb-form-of-twisted-currents} can in fact be promoted to 
an {\it operator realization} 
\vspace{-0.05in}
\begin{subequations}
\begin{gather}
[ \hp_\nrm^\s (\xi) ,\hx^\nsn_\s (\eta) ] = -i \de_\nrm{}^{\!\nsn} \de_{n(r)} (\xi-\eta) \\
[\hp^\s_\nrm (\xi) ,\hp^\s_\nsn (\eta)] = [\hx_\s^\nrm (\xi),\hx_\s^\nsn (\eta)] =0 
\end{gather}
\end{subequations}
using the same forms \eqref{exp-form-of-ehat}, \eqref{exp-form-of-hB} of $\he (\hx)$ and $\hb (\hx)$ and keeping the 
operator ordering precisely as shown in \eqref{hb-form-of-twisted-currents}. This derivation is outlined in App.~B using 
the method of {\it twisted affine Lie groups}, which generalizes the method of affine Lie groups \cite{Sochen,Club}.

The brackets of the twisted currents with the twisted Einstein coordinates can also be computed
\begin{subequations}
\begin{align}
\{ \hj_\nrm (\xi,t,\s) ,\hx_\s^\nsn (\eta,t,\s) \} &= -\tp \he^{-1} (\hx(\xi))_\nrm {}^\nsn \de_{n(s)} (\xi-\eta) \nn \\
&=-\tp \de_{n(r)} (\xi-\eta) \he^{-1} (\hx(\eta))_\nrm {}^\nsn \\
\{ \hjb_\nrm (\xi,t,\s) ,\hx_\s^\nsn (\eta,t,\s) \} &= -\tp \heb^{-1} (\hx(\xi))_\nrm {}^\nsn \de_{n(s)} (\xi-\eta) \nn \\
&=-\tp \de_{n(r)} (\xi-\eta) \heb^{-1} (\hx(\eta))_\nrm {}^\nsn
\end{align}
\end{subequations}
where Eq.~\eqref{A_^} is used to show the equivalence of the two forms on the right side of each result.

\subsection{The Operator WZW Orbifold Hamiltonian}

Our task in this and the following subsection is to obtain the classical Hamiltonian of sector $\s$ of the general WZW 
orbifold $A_g (H)/H$. As a foundation for this development, we begin again at the quantum level.

From the results of Refs.~\cite{deBoer:1999na,Halpern:2000vj,deBoer:2001nw}, we know that the operator stress tensors of the general WZW 
orbifold on the cylinder are the twisted affine-Sugawara constructions:
\begin{subequations}
\label{T-Defn}
\begin{gather}
\hat{T}_\s (\xi,t) \equiv \frac{1}{2\pi} {\cL}_{\sgb (\s)}^{\nrm;\mnrn} (\s) :\!\hj_\nrm (\xi,t) \hj_\mnrn (\xi,t) \!: ,\quad
   \hat{T}_\s (\xi +2\pi ,t) = \hat{T}_\s (\xi ,t) 
\end{gather}
\begin{gather}
\hat{\bar{T}}_\s (\xi,t) \equiv \frac{1}{2\pi} {\cL}_{\sgb (\s)}^{\nrm;\mnrn} (\s) :\!\hjb_\nrm (\xi,t) \hjb_\mnrn (\xi,t) 
   \!: ,\quad \hat{\bar{T}}_\s (\xi +2\pi ,t) = \hat{\bar{T}}_\s (\xi ,t) \\
{\cL}_{\sgb (\s)}^{\nrm ;\nsn} (\s) \!=\! \schisig^{-1}_\nrm \schisig^{-1}_\nsn U^\hcj (\s)_a {}^\nrm U^\hcj 
   (\s)_b {}^\nsn L^{ab}_g \nn \\
\quad \quad \quad =\de_{n(r)+n(s) ,0\,\text{mod } \r(\s)} {\cL}_{\sgb (\s)}^{\nrm;\mnrn} (\s) \,.\label{tw-Inv-Inert} 
\end{gather}
\end{subequations}
Here the twisted inverse inertia tensor ${\cL}_{\sgb (\s)} (\s)$ is another duality transformation, whose 
explicit form in \eqref{tw-Inv-Inert} will be needed below. The $\chi$'s in this expression are the same normalization 
constants which appeared in Eqs.~\eqref{sG-scf-Defn} and \eqref{st-Defn}, and $U^\hcj (\s)$ is the eigenvector matrix of 
the $H$-eigenvalue problem \eqref{H-eig}. We emphasize that ${\cL}_{\sgb (\s)} (\s)$ is the orbifold dual of the inverse 
inertia tensor $L_g^{ab}$ of the affine-Sugawara construction [6,7,32-34] on $g$
\begin{subequations}
\label{affSug}
\begin{gather}
T(\xi,t) = \frac{1}{2\pi} L_g^{ab} :J_a (\xi,t) J_b (\xi,t): 
\end{gather}
\begin{gather}
g = \oplus_I \gfrak^I ,\quad L_g^{ab} = \oplus_I \frac{\eta_I^{ab}}{2k_I + Q_I} ,\quad c_g = \sum_I \frac{2k_I \text{dim } 
   \gfrak^I} {2k_I + Q_I} \label{Labg-Defn} \\
L_g^{cd} \w (h_\s)_c{}^a \w (h_\s )_d{}^b = L_g^{ab} ,\quad \forall h_\s \in H \subset Aut(g)
\end{gather}
\end{subequations}
which describes the original $H$-symmetric theory $A_g (H)$.

The ordering $: \cdot :$ in Eqs.~\eqref{T-Defn} and \eqref{affSug} corresponds to operator product normal ordering 
\cite{Evslin:1999qb,deBoer:1999na,Halpern:2000vj,deBoer:2001nw} on the sphere. For the twisted left-mover current modes, the explicit form 
of operator product normal ordering is \cite{Halpern:2000vj}
\begin{subequations}
\label{OPENO}
\begin{align}
\!\!\!:\!\hj_{\nrm}(m\!+\!\nrrs) \hj_{\nsn}(n\!+\! \srac{n(s)}{\r(\s)})\!: \!&=\,:\!\hj_{\nrm}(\mnrrs) \hj_{\nsn}(\nnsrs)\!:_{M} \nn\\
  &\!\!-\!i \frac{\bar{n}(r)}{\r(\s)} \scf_{\nrm ;\nsn}{}^{\!\!\!\!n(r)+n(s) ,\de} (\s) \hj_{n(r)+n(s) ,\de} 
     (m\!+ \!n\!+ \!\srac{n(r) +n(s)}{\r(\s)}) \nn \\
   &\!\! +\frac{\bar{n} (r)}{2\r(\s)} (1-\frac{\bar{n} (r)}{\r(\s)})  \de_{\mnnrnsrsf ,0} \sG_{\nrm ;\mnrn} (\s)
\end{align}
\begin{align}
&\!\!\!:\!\hj_\nrm (\mnrrs )\hj_\nsn (\nnsrs ) \!:_M \equiv \theta (\mnrrs \geq 0) \hj_\nsn (\nnsrs) \hj_\nrm (\mnrrs) \nn \\
&\bigspc \bigspc+ \theta (\mnrrs <0) \hj_\nrm (\mnrrs )\hj_\nsn (\nnsrs ) \label{M-Defn}
\end{align}
\end{subequations}
and similarly for the twisted right-mover modes \cite{deBoer:2001nw}. The quantities $\bar{n} (r)$ are the pullbacks of $n(r)$ to 
the fundamental range $n(r) \in \{ 0,\ldots ,\r(\s)-1 \}$. 

The mode expansions of the orbifold stress tensors \eqref{T-Defn} are
\begin{subequations}
\label{Vir-Defn}
\begin{gather}
\hat{T}_\s (\xi,t) = \frac{1}{2\pi} \sum_m L_\s (m) e^{-im (\xi+t)} \\
\hat{\bar{T}}_\s (\xi,t) = \frac{1}{2\pi} \sum_m \bar{L}_\s (m) e^{+im (\xi-t)}
\end{gather}
\end{subequations}
where $L_\s (m) ,\,\bar{L}_\s (m)$ satisfy $\text{Vir} \oplus \text{Vir}$ with central charge $\hat{\bar{c}} =\hat{c}=c_g$.
Ordering identities such as \eqref{OPENO} give mode-ordered forms of the Virasoro generators, which lead directly to the conformal 
weight $\hat{\Delta}_0 (\s)$ of the scalar twist field [3-5,11-14]:
\begin{subequations}
\begin{gather}
(L_\s (m\geq 0) -\de_{m,0} \hat{\Delta}_0 (\s) )|0\rangle_\s = (\bar{L}_\s (m\geq 0) -\de_{m,0} \hat{\Delta}_0 (\s) )|0\rangle_\s =0 \\ 
\hat{\Delta}_0 (\s) = \sum_{r,\m,\n} {\cL}_{\sgb(\s)}^{\nrm;\mnrn} (\s) \frac{\bar{n}(r)}{2\r(\s)} (1-\frac{\bar{n}(r)}{\r(\s)})
   \sG_{\nrm;\mnrn} (\s) \,.
\end{gather}
\end{subequations}
We note in passing the simple forms of the twisted inverse inertia tensor ${\cL}_{\sgb(\s)} (\s)$ and these conformal weights 
\begin{subequations}
\begin{gather}
{\cL}_{\sgb (\s)}^{\nrm;\nsn} (\s)= \frac{x_\gfraks /2}{x_\gfraks +\tilde{h}_\gfraks} \sG^{\nrm;\nsn} (\s) ,\quad x_\gfraks =
   \frac{2k}{\psi^2_\gfraks} ,\quad \tilde{h}_\gfraks = \frac{Q_\gfraks}{\psi^2_\gfraks} \\
\hat{\Delta}_0 (\s) = \frac{x_\gfraks /2}{x_\gfraks +\tilde{h}_\gfraks} \sum_r \frac{\bar{n}(r)}{2\r(\s)} (1-\frac{\bar{n}(r)}{\r(\s)})
   \text{ dim} [\bar{n}(r)]
\end{gather}
\end{subequations} 
which hold in the special case when $\gfrak^I$ in \eqref{Labg-Defn} is isomorphic to simple $\gfrak$ and $k_I = k$ for all $I$.
These results apply to all WZW permutation orbifolds, even when the permutations are composed with inner or outer automorphisms 
of $\gfrak$, and further evaluation of these forms are given for specific cases in Refs.~[11-14].

Then we may use Eqs.~\eqref{T-Defn} and \eqref{Vir-Defn} to obtain the equal-time operator algebra of the orbifold stress tensors
\begin{subequations}
\begin{align}
&\!\![ \hat{T}_\s (\xi,t) ,\hat{T}_\s (\eta,t) ]=\!i \left( (\hat{T}_\s (\xi,t) \!+\!\hat{T}_\s (\eta,t))   
    -\frac{c_g}{24 \pi} (\pl^2_\xi +1)\right) \pl_\xi \de (\xi-\eta)  \\
&\!\![ \hat{\bar{T}}_\s (\xi,t) ,\hat{\bar{T}}_\s (\eta,t) ]= \!-i \left( (\hat{\bar{T}}_\s (\xi,t) \!+\!\hat{\bar{T}}_\s 
   (\eta,t)) -\frac{c_g}{24 \pi} (\pl^2_\xi +1) \right) \pl_\xi \de (\xi-\eta)
\end{align}
\begin{align}
&\!\![ \hat{T}_\s (\xi,t) ,\hj_\nrm (\eta,t,\s)] \!= \!i\hj_\nrm (\xi,t,\s) \pl_\xi \de_{n(r)} (\eta \!-\!\xi) \!
   = \!-i \pl_\eta \!\left( \hj_\nrm (\eta,t,\s) \de (\xi \!- \!\eta) \right)\\
&\!\! [ \hat{\bar{T}}_\s (\xi,t) ,\hjb_{\!\nrm} (\eta,t,\s)] \!= \!-i\hjb_{\!\nrm} (\xi,t,\s) \pl_\xi \de_{n(r)} (\eta \!-\!\xi) 
    \!= \!i \pl_\eta \!\left( \hjb_{\!\nrm} (\eta,t,\s) \de (\xi \!-\! \eta) \right) \\
&\quad \quad [\hat{T}_\s (\xi,t) ,\hat{\bar{T}}_\s (\eta,t)] =
   [\hat{T}_\s (\xi,t) ,\hjb_\nrm (\eta,t ,\s) ]= [\hat{\bar{T}}_\s (\xi,t) ,\hj_\nrm (\eta,t ,\s)] =0
\end{align}
\end{subequations}
as the Minkowski-space versions of known results \cite{deBoer:2001nw} on the sphere. The monodromy \eqref{JMonos} of the twisted 
currents is consistent with all these commutators.

Similarly, Eq.~\eqref{hg-Rescale} and the results of Ref.~\cite{deBoer:2001nw} can be used to obtain the equal-time commutators of the 
stress tensors with the twisted affine primary fields on the cylinder
\begin{subequations}
\label{Tg-Comm}
\begin{align}
[ \hat{T}_\s (\xi,t) ,\hg(\st,\eta,t,\s)] = & \Big{(} 2{\cL}_{\sgb (\s)}^{\nrm;\mnrn} (\s) :\!\hj_\nrm (\eta,t,\s)
   \hg(\st,\eta,t,\s) \st_\mnrn (T,\s)\!:  \nn \\
& \quad + i\hg(\st,\eta,t,\s) \D_{\sgb (\s)} (\st) \pl_\xi \Big{)} \de (\xi-\eta) \\
[ \hat{\bar{T}}_\s (\xi,t) ,\hg(\st,\eta,t,\s)] = & -\Big{(} 2{\cL}_{\sgb (\s)}^{\nrm;\mnrn} (\s) :\!\hjb_\nrm (\eta,t,\s)
   \st_\mnrn (T,\s) \hg(\st,\eta,t,\s)\!: \nn \\
& \quad +i\D_{\sgb (\s)} (\st) \hg(\st,\eta,t,\s) \pl_\xi \Big{)} \de (\xi-\eta) 
\end{align}
\end{subequations}
where $\D_{\sgb (\s)} (\st)$ is the twisted conformal weight matrix of Refs.~\cite{deBoer:2001nw,so2n}. The ordering $: \cdot :$ in 
Eq.~\eqref{Tg-Comm} is again equivalent to operator product normal ordering on the sphere, and takes the following form 
on the cylinder
\begin{subequations}
\begin{align}
:\!\hj_\nrm (\xi,t,\s) \hg (\st,\xi,t,\s)\!: =& :\!\hj_\nrm (\xi,t,\s) \hg (\st,\xi,t,\s)\!:_M \nn \\
   &\, - \srac{\bar{n}(r)}{\r(\s)} \hg (\st,\xi,t,\s) \st_\nrm (T,\s) \\
:\!\hjb_\nrm (\xi,t,\s) \hg (\st,\xi,t,\s)\!: =& :\!\hjb_\nrm (\xi,t,\s) \hg (\st,\xi,t,\s)\!:_{\bar{M}} \nn \\
   &\, +\srac{\overline{-n(r)}}{\r(\s)} \st_\nrm (T,\s) \hg (\st,\xi,t,\s) 
\end{align}
\end{subequations}
where $\overline{-n(r)}$ is the pullback of $-n(r)$ and these $M$ and $\bar{M}$ normal orderings are defined in Ref.~\cite{deBoer:2001nw}.  

For each sector $\s$, the {\it operator WZW orbifold Hamiltonian} $\hat{H}_\s$ and the operator WZW orbifold momentum 
$\hat{P}_\s$ are then defined as follows:
\begin{subequations}
\label{Q-orb-H&P}
\begin{align}
&\hat{H}_\s \!=\! \hat{P}_{0\s} \!=\!\int_0^{2\pi} \!\!\!\!d\xi \left( \hat{T}_\s (\xi) + \hat{\bar{T}}_\s (\xi) \right)
   \!=\! L_\s (0) + \bar{L}_\s (0) \nn \\
&\,\,\,\,=\! \frac{1}{2\pi} \int_0^{2\pi} \!\!\!\! d\xi \,{\cL}_{\sgb (\s)}^{\nrm ;\mnrn} (\s) 
   :\!\hj_\nrm (\xi,t) \hj_\mnrn (\xi,t) \!+\! \hjb_\nrm (\xi,t) \hjb_\mnrn (\xi,t)\!: \label{Q-orb-Hamil}
\end{align}
\begin{align}
&\hat{P}_\s \!=\! \hat{P}_{1\s} \!=\! \int_0^{2\pi} \!\!\!\!d\xi \left( \hat{T}_\s (\xi) - \hat{\bar{T}}_\s (\xi) \right)
   \!=\! L_\s (0) - \bar{L}_\s (0) \nn \\
&\,\,\,\,=\! \frac{1}{2\pi} \int_0^{2\pi}\!\!\!\! d\xi \,{\cL}_{\sgb (\s)}^{\nrm ;\mnrn} (\s) 
:\!\hj_\nrm (\xi,t) \hj_\mnrn (\xi,t) - \hjb_\nrm (\xi,t) \hjb_\mnrn (\xi,t)\!: \,.
\end{align}
\end{subequations}
Both operators are well-defined because the stress tensors have trivial monodromy, and the twisted operator
equations of motion are then obtained as usual:
\begin{subequations}
\begin{gather}
\pl_m \hat{A} (\xi,t) =i[ \hat{P}_{m\s} ,\hat{A} (\xi,t)] ,\quad \xi^m = (t,\xi) ,\quad \pl_m =(\pl_t ,\pl_\xi ) \\
\pl_- \hj_\nrm (\xi,t,\s) =\pl_+ \hjb_\nrm (\xi,t,\s) =0,\quad \pl_\pm \equiv \pl_t \pm \pl_\xi \\
\pl_- \hat{T}_\s (\xi,t) = \pl_+ \hat{\bar{T}}_\s (\xi,t) =0 \,. \label{Cons-Laws}
\end{gather}
\end{subequations}
Moreover, the conservation laws \eqref{Cons-Laws} tell us that the orbifold momentum operator $\hat{P}_\s$
is conserved.

This concludes our discussion of the equal-time operator formulation of WZW orbifolds. We remind the reader 
however, that Refs.~[11-14] give {\it twisted KZ equations} for correlators of the twisted 
affine primary fields in each sector $\s$ of all WZW orbifolds on the sphere. The equal-time results above (or 
Eq.~\eqref{hg-Rescale}) can similarly be used to obtain the corresponding twisted KZ equations of the WZW orbifolds 
on the cylinder.

\subsection{The Classical WZW Orbifold Hamiltonian}

From the operator formulation above, we know that the classical orbifold stress tensors of sector $\s$ are given by
\begin{subequations}
\label{cl-T_s}
\begin{gather}
\hat{T}_\s (\xi) =\frac{1}{4\pi} \sG^{\nrm ;\mnrn} (\s) \hj_\nrm (\xi ,\s) \hj_\mnrn (\xi,\s) ,\quad
   \hat{T}_\s (\xi +2\pi) =\hat{T}_\s (\xi) \\
\hat{\bar{T}}_\s (\xi) =\frac{1}{4\pi} \sG^{\nrm ;\mnrn} (\s) \hjb_\nrm (\xi ,\s) \hjb_\mnrn (\xi,\s) ,\quad
   \hat{\bar{T}}_\s (\xi +2\pi) =\hat{\bar{T}}_\s (\xi) 
\end{gather}
\begin{gather}
\sG^{\nrm;\nsn} (\s)= \schisig^{-1}_\nrm \schisig^{-1}_\nsn U^\hcj (\s)_a {}^\nrm
   U^\hcj (\s)_b {}^\nsn G^{ab} ,\quad G^{ab} = \oplus_I \frac{\eta_I^{ab}}{k_I} 
\end{gather}
\end{subequations}
where $G^{ab}$ is the inverse generalized metric of affine $g$, $U^\hcj (\s)$ is the eigenvector matrix of the 
$H$-eigenvalue problem \eqref{H-eig} and $\sG^\bullet (\s)$ is the inverse twisted metric of sector $\s$. 
To obtain these forms, we have followed the standard prescription: Drop the normal ordering in \eqref{T-Defn}, 
and take the high-level (classical) limit of the inverse inertia tensor $L_g^{ab}$ in \eqref{affSug}
\begin{subequations}
\begin{gather}
k_I \rightarrow \infty : \quad \quad L^{ab}_{g,\infty} = \frac{1}{2} G^{ab} \quad \label{Hi-Level1} \\
{\cL}_{\sgb (\s) ,\infty}^{\nrm ;\nsn} (\s) = \schisig^{-1}_\nrm \schisig^{-1}_\nsn U^\dagger (\s)_a{}^\nrm
   U^\dagger (\s)_b{}^\nsn L^{ab}_{g,\infty} = \frac{1}{2} \sG^{\nrm ;\nsn} (\s) \label{Hi-Level2}
\end{gather}
\end{subequations}
which gives the high-level limit of the twisted inverse inertia tensor shown in \eqref{Hi-Level2}.

Using \eqref{cl-T_s} and the twisted brackets \eqref{EqTCA}, \eqref{EqTChiral}, one finds that all the commutators 
of Subsec.~$2.6$ are replaced by the rescaled brackets:
\begin{subequations}
\label{TJg-Brack}
\begin{gather}
\{ \hat{T}_\s (\xi) ,\hat{T}_\s (\eta) \} = i \left( \hat{T}_\s (\xi) +\hat{T}_\s (\eta) \right) \pl_\xi \de
   (\xi-\eta) \\
\{ \hat{\bar{T}}_\s (\xi) ,\hat{\bar{T}}_\s (\eta) \} =-i \left( \hat{\bar{T}}_\s (\xi) +\hat{\bar{T}}_\s
   (\eta) \right) \pl_\xi \de (\xi -\eta) \\
   \{ \hat{T}_\s (\xi) ,\hat{\bar{T}}_\s (\eta) \} =0
\end{gather}
\begin{gather}
\{ \hat{T}_\s (\xi) ,\hj_\nrm (\eta,\s)\} = i\hj_\nrm (\xi) \pl_\xi \de_{n(r)} (\eta -\xi) 
   = -i \pl_\eta \!\left( \hj_\nrm (\eta,\s) \de (\xi- \eta) \right)\\
\{ \hat{\bar{T}}_\s (\xi) ,\hjb_\nrm (\eta,\s)\} = -i\hjb_\nrm (\xi) \pl_\xi \de_{n(r)} (\eta -\xi) 
    = i \pl_\eta \!\left( \hjb_\nrm (\eta,\s) \de (\xi- \eta) \right) \\
\{\hat{T}_\s (\xi) ,\hjb_\nrm (\eta ,\s) \}= \{\hat{\bar{T}}_\s (\xi) ,\hj_\nrm (\eta ,\s)\} =0
\end{gather}
\begin{gather}
\{ \hat{T}_\s (\xi) ,\hg (\st,\eta,\s) \} = \hg (\st,\eta,\s) \hj (\st,\eta,\s) \de (\xi-\eta) \label{Tg1} \\
\,\{ \hat{\bar{T}}_\s (\xi) ,\hg (\st,\eta,\s) \} = -\hjb (\st,\eta,\s) \hg (\st,\eta,\s) \de (\xi-\eta)  \label{Tg2}
\end{gather}
\end{subequations}
where $\hj (\st) ,\hjb (\st)$ are the twisted matrix currents:
\begin{subequations}
\label{Mat-Curr-Defn}
\begin{gather}
\hj (\st,\xi,t,\s) \equiv \hj_\nrm (\xi,t,\s) \sG^{\nrm ;\mnrn} (\s) \st_\mnrn (T,\s) \\
\hjb (\st,\xi,t,\s) \equiv \hjb_\nrm (\xi,t,\s) \sG^{\nrm ;\mnrn} (\s) \st_\mnrn (T,\s) \\
\hj_\nrm (\xi,t,\s) = \widehat{Tr} \left( \sm(\st,\s) \st_\nrm (T,\s) \hj (\st,\xi,t,\s) \right) \\
\hj_\nrm (\xi,t,\s) = \widehat{Tr} \left( \sm(\st,\s) \st_\nrm (T,\s) \hj (\st,\xi,t,\s) \right) \,.
\end{gather}
\end{subequations}
We remark in particular that the second terms in Eq.~($2.49$a,b) vanish in the classical limit ($2.55$g,h) because 
the twisted conformal weight matrix $\D_{\sgb (\s)} (\st)$ is $O(k^{-1})$.

We comment on the consistency of the monodromies which appear in the system \eqref{TJg-Brack}, \eqref{Mat-Curr-Defn}, 
beginning with the (two-sided) diagonal monodromy of the twisted matrix currents (see Eq.~\eqref{extHeig})
\begin{subequations}
\label{J-Mat-Mono}
\begin{gather}
\hj (\st,\xi +2\pi,\s) = E(T,\s) \hj (\st,\xi,\s) E(T,\s)^\ast \\
\hjb (\st,\xi+2\pi,\s) =E(T,\s) \hjb (\st,\xi,\s) E(T,\s)^\ast
\end{gather}
\end{subequations}
which follows from the monodromy \eqref{JMonos} and the selection rule \eqref{T-select}. Then it is straightforward to
check that the monodromy \eqref{J-Mat-Mono} of the matrix currents and the monodromy \eqref{gMono} of the group orbifold 
elements are consistent with the equal-time brackets \eqref{Tg1} and \eqref{Tg2}.

It follows from Eqs.~\eqref{Q-orb-H&P}, \eqref{cl-T_s} that the {\it classical WZW orbifold Hamiltonian} and the 
classical WZW orbifold momentum of sector $\s$ have the form
\begin{subequations}
\label{cl-tw-H&P}
\begin{align}
&\hat{H}_\s \!=\! \frac{1}{4\pi} \int_0^{2\pi} \!\!\! d\xi \,\sG^{\nrm ;\mnrn} (\s) \left( \hj_\nrm 
(\xi,t) \hj_\mnrn (\xi,t) \!+\! \hjb_\nrm (\xi,t) \hjb_\mnrn (\xi,t) \right) \label{eq:twistedHamil} \\
&\hat{P}_\s \!=\! \frac{1}{4\pi} \int_0^{2\pi} \!\!\! d\xi \,\sG^{\nrm ;\mnrn} (\s) \left( \hj_\nrm 
(\xi,t) \hj_\mnrn (\xi,t) \!-\! \hjb_\nrm (\xi,t) \hjb_\mnrn (\xi,t) \right)
\end{align}
\end{subequations}
where we have suppressed the $\s$-dependence of the twisted currents. This gives us the classical orbifold equations of motion
\begin{subequations}
\label{cl-Orb-EOM}
\begin{gather}
\pl_m \hat{A} (\xi,t) =i\{ \hat{P}_{m\s} ,\hat{A} (\xi,t)\}  \label{Orb-EOM1} \\
\pl_- \hj_\nrm (\xi,t) =\pl_+ \hjb_\nrm (\xi,t) =
   \pl_- \hat{T}_\s (\xi,t) =\pl_+ \hat{\bar{T}}_\s (\xi,t) =0
\end{gather}
\begin{gather}
\pl_t \hg (\st,\xi,t) =i( \hg(\st,\xi,t) \hj(\st,\xi,t) -\hjb(\st,\xi,t) \hg(\st,\xi,t)) \\
\pl_\xi \hg (\st,\xi,t) =i( \hg(\st,\xi,t) \hj(\st,\xi,t) +\hjb(\st,\xi,t) \hg(\st,\xi,t)) \\
\pl_+ \hg(\st,\xi,t) =2i\hg(\st,\xi,t) \hj(\st,\xi,t) ,\quad \pl_- \hg(\st,\xi,t) =-2i\hjb(\st,\xi,t)
   \hg(\st,\xi,t) \label{Cl-tw-VOE} \\
\hj (\st,\xi,t) =-\frac{i}{2} \hg^{-1} (\st,\xi,t) \pl_+ \hg(\st,\xi,t) ,\quad
   \hjb (\st,\xi,t) = -\frac{i}{2} \hg(\st,\xi,t) \pl_- \hg^{-1} (\st,\xi,t) \label{MatCDefn}
\end{gather}
\end{subequations}
which follow either by integration of the brackets in \eqref{TJg-Brack}, or directly from Eqs.~\eqref{EqTCA}, \eqref{EqTChiral}, 
and \eqref{cl-tw-H&P} -- using integral identities of the type \eqref{Int-Id2}. The twisted matrix currents $\hj (\st), \hjb (\st)$ 
are also conserved
\begin{subequations}
\label{Mat-Currs}
\begin{gather}
\pl_- \hj (\st,\xi,t,\s) = \pl_+ \hjb (\st,\xi,t,\s) =0 \label{MatC-Cons} \\
\hj_\nrm (\xi,t,\s) =-\frac{i}{2} \widehat{Tr} \left(\sm(\st,\s) \st_\nrm(T,\s) \hg^{-1} (\st,\xi,t,\s)
   \pl_+ \hg(\st,\xi,t,\s) \right) \\
\hjb_\nrm (\xi,t,\s) =-\frac{i}{2} \widehat{Tr} \left(\sm(\st,\s) \st_\nrm(T,\s) \hg(\st,\xi,t,\s)
   \pl_- \hg^{-1} (\st,\xi,t,\s) \right) 
\end{gather}
\end{subequations}
because $\hj_\nrm ,\hjb_\nrm$ are conserved. Note that the equations of motion in \eqref{Cl-tw-VOE} are the classical 
light-cone analogues on the cylinder of the twisted vertex operator equations of Ref.~\cite{deBoer:2001nw}.

\subsection{The Canonical Orbifold Hamiltonian and the Twisted Einstein Metric}

We may now give the {\it canonical realization} $\hat{P}_{m\s} =\hat{P}_{m\s} [\hx,\hp]$ of the classical WZW orbifold 
Hamiltonian and momentum
\begin{subequations} 
\label{Hamil-dens}
\begin{gather}
\hat{H}_\s = \int_0^{2\pi} \!\!\!\! d\xi \hat{\sh}_\s (\hx(\xi) ,\hp (\xi)) ,\quad \hat{P}_\s = \int_0^{2\pi} \!\!\!\!d\xi
   \hsp_\s (\hx(\xi),\hp(\xi)) ,\quad \srange \\
\hat{\sh}_\s (\hx(\xi),\hp(\xi)) \equiv 2\pi \hG^{\nrm;\nsn}(\hx) \hp_\nrm(\hb)\hp_\nsn(\hb) \quad \quad \quad \nn \\
   \bigspc + \frac{1}{8\p} \pl_\xi\hx_\s^\nrm \pl_\xi\hx_\s^\nsn \hG_{\nrm;\nsn}(\hx) 
\end{gather}
\begin{gather}
\hsp_\s (\hx(\xi),\hp(\xi)) \equiv \hp^\s_\nrm (\xi) \pl_\xi \hx_\s^\nrm (\xi) \\
\hsp_{m\s} (\hx (\xi+2\pi),\hp (\xi+2\pi)) = \hsp_{m\s} (\hx (\xi),\hp(\xi)) ,\quad \hsp_{m\s} = (\hat{\sh}_\s ,\hsp_\s )
\end{gather}
\end{subequations}
which are obtained by substitution of the canonical realization \eqref{hb-form-of-twisted-currents} of the twisted
currents into Eq.~\eqref{cl-tw-H&P}. 

In the canonical form \eqref{Hamil-dens} of the classical orbifold Hamiltonian, we have encountered for the first 
time the {\it twisted Einstein metric} $\hG_\bullet(\hx)$ of sector $\s$ and its inverse $\hG^\bullet (\hx)$
\begin{subequations}
\label{sGh}
\begin{align}
\hG_{\nrm;\nsn}(\hx(\xi)) &\equiv \he(\hx(\xi))_\nrm{}^\ntd \he(\hx(\xi))_\nsn{}^\nue \sG_{\ntd;\nue}(\s) \nn\\
&= \he(\hx(\xi))_\nrm{}^\ntd \he(\hx(\xi))_\nsn{}^{-n(t),\ep} \sG_{\ntd;-n(t),\ep}(\s) \nn \\
&= \hG_{\nsn;\nrm} (\hx (\xi)) \label{G-hat-down} \\
\hG^{\nrm;\nsn}(\hx(\xi)) &\equiv \sG^{\ntd;\nue}(\s) \he^{-1}(\hx(\xi))_\ntd{}^\nrm \he^{-1}
   (\hx(\xi))_\nue{}^\nsn \nn \\
&=\hG^{\nsn;\nrm} (\hx(\xi)) \label{G-hat-up}
\end{align}\vsvs
\begin{gather}
\hG_{\nrm;\ntd}(\hx(\xi)) \hG^{\ntd;\nsn}(\hx(\xi)) = \d_\nrm{}^{\nsn} \\
\hG_{\nrm;\nsn}(\hx) \equiv \hG_{\nrm;\nsn}(\hx_\s,\s), \quad  \hG^{\nrm;\nsn}(\hx) \equiv 
\hG^{\nrm;\nsn}(\hx_\s,\s) 
\end{gather}
\end{subequations}
where $\sG_\bullet(\s)$ and $\sG^\bullet (\s)$ are respectively the twisted tangent-space metric of sector $\s$ and 
its inverse (see Eqs.~\eqref{tw-current-alg} and \eqref{cl-T_s}). The diagonal monodromies of the twisted Einstein metric 
and its inverse
\begin{subequations}
\label{hG-Mono1}
\begin{gather}
\hG_{\nrm;\nsn}(\hx(\xi+2\p)) = e^{-2\p i \frac{n(r)+n(s)}{\r(\s)}}\hG_{\nrm;\nsn}(\hx(\xi)) \label{hG_Mono}\\
\hG^{\nrm;\nsn}(\hx(\xi+2\p)) = \hG^{\nrm;\nsn}(\hx(\xi)) e^{2\p i \frac{n(r)+n(s)}{\r(\s)}} \label{hG^Mono}
\end{gather}
\end{subequations}
follow from the monodromy \eqref{he-Mono1} of the twisted vielbein and its inverse. Using \eqref{hOmega-sG}, we note
that the definitions in \eqref{G-hat-down}, \eqref{G-hat-up} hold as well with $\he (\hx) \rightarrow \heb (\hx)$.

From Eqs.~\eqref{f-of-hOmega-id}, \eqref{G-hat-down} and the explicit form of the twisted vielbein in \eqref{exp-form-of-ehat}, 
we also obtain the closed form and symmetry of the twisted Einstein metric
\begin{subequations}
\label{exp-form-of-hG}
\begin{align}
& \!\! \hat{G}_{\nrm;\nsn} (\hx) =\! \smal{\left( \frac{e^{i\hY} +e^{-i\hY} -2}{(i\hY)^2} \right)}{}_\nrm 
   {}^{\!\ntd} \sG_{\ntd;\nsn} (\s) \\
& \quad =\! \smal{\left( \frac{e^{i\hY} +e^{-i\hY} -2}{(i\hY)^2} \right)}{}_\nsn {}^{\!\ntd} \sG_{\ntd;\nrm} (\s) 
  \!=\! \smal{\left( \frac{e^{i\hY} +e^{-i\hY} -2}{(i\hY)^2} \right)}{}_\nrm {}^{\! -n(s),\de} \sG_{-n(s),\de;\nsn} (\s) 
\end{align}
\end{subequations}
which should be considered together with the closed forms of $\he$ and $\hB$ in Eqs.~\eqref{exp-form-of-ehat} 
and \eqref{exp-form-of-hB}. The monodromy \eqref{hG-Mono1} of the twisted Einstein metric is again consistent with the 
monodromy \eqref{hY-Mono} of $\hY$ because the twisted tangent space metric $\sG_{\bullet}(\s)$ satisfies the selection rule 
\eqref{sG-select}.

The twisted Einstein metric allows us to define the {\it WZW orbifold interval}
\begin{equation}
\label{WZW-orb-ds2}
ds^2 (\s) \equiv \hat{G}_{\nrm ;\nsn} (\hx_\s) d\hx_\s^\nrm d\hx_\s^\nsn
\end{equation}
for each twisted sector $\s$ of the orbifold. It is clear from Eqs.~\eqref{hG_Mono} and \eqref{hx-mono}
that the WZW orbifold interval \eqref{WZW-orb-ds2} has trivial monodromy upon circumnavigation of the cylinder.
The interval can also be expressed in terms of the {\it induced} WZW orbifold metric $\hG_{mn}$ on 
the world sheet:
\begin{subequations}
\begin{gather}
ds^2 (\s) = \hat{G}_{mn} (\xi,t,\s) d\xi^m d\xi^n \\
\hat{G}_{mn} (\xi,t,\s) \equiv \hat{G}_{\nrm ;\nsn} (\hx_\s (\xi,t)) \pl_m \hx_\s^\nrm (\xi,t) \pl_n \hx_\s^\nsn
   (\xi,t) \\
\hat{G}_{mn} (\xi +2\pi,t,\s) = \hat{G}_{mn} (\xi,t,\s) ,\quad (\xi^0 ,\xi^1) =(t,\xi) ,\quad (\pl_0 ,\pl_1 )= 
   (\pl_t ,\pl_\xi) \,.
\end{gather}
\end{subequations}
In this direction lies the study of WZW orbifold geodesics and related topics, but we will not pursue this here.

Orbifold intervals and induced metrics are further discussed in Subsecs.~$3.3$ and $4.3$.

\subsection{Phase-Space Derivation of the General WZW Orbifold Action}

We turn finally to the classical {\it coordinate-space} formulation of the general WZW orbifold. In this development,
we use the orbifold Hamiltonian equation of motion \eqref{Orb-EOM1} for the twisted coordinates
\begin{equation}
\label{s-model-H-EOM}
\pl_t \hx_\s^\nrm = 4\p \hG^{\nrm;\nsn}(\hx)\hp_\nsn(\hb),\quad  \hp_\nrm(\hb) = \onefourpi 
   \hG_{\nrm;\nsn}(\hx) \pl_t \hx_\s^\nsn 
\end{equation}
to eliminate the twisted momenta $\hp$ in favor of $\pl_t \hx$. As a first example, one may use 
\eqref{s-model-H-EOM} and the canonical form \eqref{hb-form-of-twisted-currents} of the twisted currents
to obtain the following coordinate-space form of the twisted currents:
\begin{subequations}
\label{coord-form-hj}
\begin{gather} 
\hj_\nrm(\xi,\s) = \half \pl_+\hx_\s^\ntd(\xi) \he(\hx(\xi))_\ntd{}^{\nsn} \sG_{\nsn;\nrm}(\s) \\
\hjb_\nrm(\xi,\s) = \half \pl_-\hx_\s^\ntd(\xi) \heb(\hx(\xi))_\ntd{}^{\nsn} \sG_{\nsn;\nrm}(\s) \,.
\end{gather}
\end{subequations}
This result can also be obtained from Eqs.~\eqref{M-trace}, \eqref{l-i-viel}, \eqref{r-i-viel} and \eqref{Mat-Currs}.

Similarly, we may obtain the action formulation of sector $\s$ by the Legendre transformation
\begin{gather}
\label{orb-action}
\hat{{\cL}}_\s (\hx (\xi)) \equiv \hat{{\cL}}_\s (\hx (\xi),\pl_m \hx (\xi)) = \hp_\nrm^\s (\xi) \hx^\nrm_\s (\xi)
   -\hat{\sh}_\s (\hx (\xi),\hp(\xi)) 
\end{gather}
where $\hat{\sh}_\s$ is the canonical Hamiltonian density in Eq.~\eqref{Hamil-dens}. This gives the 
{\it twisted sigma model form} of the general WZW orbifold action
\begin{subequations}
\label{s-model-action}
\begin{gather} 
\hs_{\hat{g} (\s)} = \int  \! d^2\xi\, \slh_\s(\hx(\xi)) ,\quad d^2 \xi \equiv dt d\xi ,\quad \srange \\
\slh_\s (\hx(\xi)) = \frac{1}{8\p} \big{(}\hG_{\nrm;\nsn}(\hx)+\hB_{\nrm;\nsn}(\hx) \big{)} 
   \pl_+\hx_\s^\nrm \pl_-\hx_\s^\nsn \\
\slh_\s(\hx(\xi+2\p)) = \slh_\s(\hx(\xi)) \label{twLag-mono}
\end{gather}
\end{subequations}
for sector $\s$ of $A_g (H)/H$, where $\hG$ and $\hb$ are the twisted Einstein metric and the twisted $B$ field of that
sector. The trivial monodromy \eqref{twLag-mono} of the Lagrange density $\slh_\s$ follows from the diagonal monodromies of 
$\hx$, $\hG$ and $\hB$ in \eqref{mono-of-hB-and-hH}, \eqref{sGh} and \eqref{hb-form-of-twisted-currents}.

We may also reexpress the twisted sigma model form of the WZW orbifold action in terms of the group orbifold 
elements $\hg$. The central step here is the {\it orbifold Gauss' law}
\begin{subequations}
\label{s-model-WZW-term}
\begin{gather}
(\hat{g}^{-1} d\hat{g})^3 \equiv d^2\xi\,d\r\;\ep^{ABC}(\hat{g}^{-1} \pl_A\hat{g})(\hat{g}^{-1} 
   \pl_B\hat{g})(\hat{g}^{-1} \pl_C\hat{g}) \\
\smal{\widehat{Tr}(\sm\,(\hg^{-1}d\hg)^3) = d^2\xi d\r\,\pl_A j^A,\quad j^A \equiv 
   \frac{3}{2}\epsilon^{ABC} \pl_B\hx_\s^\nrm \pl_C\hx_\s^\nsn \hb_{\nrm;\nsn}} \\
\{ A,B,C \} = \{ t,\xi,\r \} ,\quad \epsilon^{t \xi \r} =1 \\
\int_\Gamma \widehat{Tr}(\sm\,(\hg^{-1}d\hg)^3) = -\frac{3}{2} \int \! d^2\xi\, \hb_{\nrm;\nsn} 
   \pl_+\hx_\s^\nrm \pl_-\hx_\s^\nsn 
\end{gather}
\end{subequations}
which follows from Eqs.~\eqref{M-trace}, \eqref{Def-twVB} and \eqref{hB-and-hH-defn}. Here $\Gamma$ is 
the usual solid cylinder $(t,\xi,\r)$ and $\epsilon^{ABC}$ is the Levi-Civita density. 
With Eqs.~\eqref{M-trace}, \eqref{Def-twVB}, \eqref{G-hat-down} and the orbifold Gauss' law
we finally obtain the original {\it group orbifold element form} of the general WZW orbifold action \cite{deBoer:2001nw}
\begin{subequations}
\label{gp-orbel-form}
\begin{align}
\!\!\hat{S}_{\hg (\s)} (\st) \!\equiv \!\hat{S}_{\hg (\s)} [\sm,\hg] =& -\!\frac{1}{8\pi}\!\int \!\!d^2\xi  
\widehat{Tr} \left( \sm(\st,\s) \hat{g}^{-1} (\st,\s) \pl_+\hat{g}(\st,\s) \hat{g}^{-1} (\st,\s)
\pl_-\hat{g}(\st,\s) \right) \nn\\
& -\frac{1}{12\pi}\int_{\Gamma} \!\widehat{Tr}\left( \sm(\st,\s) \Big{(} \hat{g}^{-1}(\st,\s) 
d\hat{g}(\st,\s) \Big{)}^3 \right)
\end{align}\vsvsvs
\begin{equation}
\hat{S}_{\hg (\s)} [\sm(\st,\s),\hg(\st,\xi+2\p,\s)] = \hat{S}_{\hg (\s)} [\sm(\st,\s),\hg(\st,\xi,\s)] , \,\,\,\, \srange
   \label{WZW-orb-S-mono}
\end{equation}
\end{subequations}
where $\sm (\st,\s)$ is the twisted data matrix of sector $\s$. The result \eqref{gp-orbel-form} was originally obtained in 
an entirely different manner -- namely by the {\it principle of local isomorphisms} reviewed in Appendices C,D and the 
following section. 

The trivial monodromy \eqref{WZW-orb-S-mono} of the WZW orbifold action follows from Eqs.~\eqref{M-E-Comm} and \eqref{gMono}.
Moreover, using Eq.~\eqref{M-g-Comm}, variation of the WZW orbifold action gives the conservation \eqref{MatC-Cons} of the twisted 
matrix currents as the orbifold equations of motion. The WZW orbifold action also reduces in the untwisted sector $\s=0$
to the ordinary WZW action \cite{Nov,W} for $A_g(H)$.

The geometry of general WZW orbifolds is further studied in Subsec.~$4.5$.

\section{Comments and Examples}

\subsection{A Limitation on the Principle of Local Isomorphisms}

The principle of local isomorphisms \cite{Borisov:1997nc,deBoer:1999na,Halpern:2000vj,deBoer:2001nw} was originally designed and verified for 
local orbifold operators and the singular terms of operator product expansions on the {\it sphere}. Our goal in this subsection 
is to clarify, by comparison with Sec.~2, the range of applicability of this principle when working on the 
{\it cylinder}.

It is clear in Ref.~\cite{deBoer:2001nw} that the principle of local isomorphisms can be used on the cylinder to obtain the form 
of all local orbifold operators and local products of classical fields. Indeed the general WZW orbifold action on 
the cylinder was originally derived from local isomorphisms
\vspace{-0.15in}
\begin{subequations}
\begin{gather}
\sg \dual \hg \\
\text{automorphic response } \dual \text{ monodromy} \bigspc
\end{gather}
\end{subequations}
in just this way, where $\sg$ is the eigengroup element and $\hg$ is the group orbifold element.
 
Let us consider another example, this time involving the currents, for which one needs the eigencurrents \cite{deBoer:1999na,Halpern:2000vj}:
\begin{align}
& \!\!\! \sj_\nrm (\xi,t,\s) \!=\! \schisig_\nrm U(\s)_\nrm {}^a J_a(\xi,t) ,\,\,\, \sjb_\nrm (\xi,t,\s) \!=\! \schisig_\nrm
   U(\s)_\nrm{}^a \bar{J}_a (\xi,t) \,. \label{sj-Defn}
\end{align}
Here $\{J ,\bar{J} \}$ are the untwisted left- and right-mover currents of $A_g (H)$, $U(\s)$ is the inverse of the eigenvector 
matrix of the $H$-eigenvalue problem \eqref{H-eig} and the $\schi$'s are the same normalization constants which appear in 
Eq.~\eqref{sG-scf-Defn}. Then we may obtain the form \eqref{cl-tw-H&P} of the classical WZW orbifold Hamiltonian $\hat{H}_\s$
and momentum $\hat{P}_\s$ from the untwisted classical Hamiltonian and momentum as follows:
\begin{subequations}
\begin{gather}
H =L(0) +\bar{L} (0) =\frac{1}{4\pi} \int_0^{2\pi} \!\!\!\!d\xi \,G^{ab} \left( J_a (\xi,t) J_b (\xi,t) + \bar{J}_a
(\xi,t) \bar{J}_b (\xi,t) \right) \\
P =L(0) -\bar{L} (0) =\frac{1}{4\pi} \int_0^{2\pi} \!\!\!\!d\xi \,G^{ab} \left( J_a (\xi,t) J_b (\xi,t) - \bar{J}_a
(\xi,t) \bar{J}_b (\xi,t) \right) 
\end{gather}
\begin{align}
G^{ab} J_a (\xi,t) J_b (\xi,t) &= \sG^{\nrm ;\mnrn} (\s) \sj_\nrm (\xi,t) \sj_\mnrn (\xi,t) \\
     & \dual \sG^{\nrm ;\mnrn} (\s) \hj_\nrm (\xi,t) \hj_\mnrn (\xi,t)
\end{align}
\begin{align}
G^{ab} \bar{J}_a (\xi,t) \bar{J}_b (\xi,t) &= \sG^{\nrm ;\mnrn} (\s) \bar{\sj}_\nrm (\xi,t) 
   \bar{\sj}_\mnrn (\xi,t) \\
  &\dual \sG^{\nrm ;\mnrn} (\s) \hjb_\nrm (\xi,t) \hjb_\mnrn (\xi,t) \,.
\end{align}
\end{subequations}
Here the local isomorphisms are
\begin{subequations}
\begin{gather}
\sj ,\bar{\sj} \dual \hj, \hjb \\
\text{automorphic response } E_{n(r)}(\s) \dual \text{ monodromy } E_{n(r)} (\s)
\end{gather}
\end{subequations}
where $E_{n(r)}(\s)$ is the eigenvalue matrix of the $H$-eigenvalue problem \eqref{H-eig}, the hatted objects are the twisted currents 
and $H,P \duals \hat{H}_\s, \hat{P}_\s$ are derived isomorphisms. Similarly, any local quantum or classical result in the orbifold can 
be obtained by local isomorphisms from the untwisted theory, for example the normal-ordered form of the orbifold stress tensors 
\cite{deBoer:1999na,Halpern:2000vj,deBoer:2001nw}, the bracket equations of motion \eqref{cl-Orb-EOM} and even the canonical realization 
\eqref{hb-form-of-twisted-currents} of the twisted currents (see Subsec.~$4.5$). 

For operators or fields with non-trivial monodromy however, the principle of local isomorphisms will {\it not} give
the correct twisted equal-time commutators or brackets -- because these objects are spatially non-local. As an example, consider 
the equal-time bracket form of the untwisted current algebra and its corresponding eigencurrent algebra:
\begin{subequations}
\begin{equation}
\{ J_a (\xi,t) ,J_b (\eta,t) \} = \tp \left( f_{ab}{}^c J_c (\eta,t) + G_{ab} \pl_\xi \right) \de (\xi-\eta)
\end{equation}
\begin{align}
\{ \sj_\nrm (\xi,t,\s) ,\sj_\nsn (\eta,t,\s) \} &= \tp \Big{(} \scf_{\nrm ;\nsn}{}^{n(r)+n(s),\de} (\s)
\sj_{n(r)+n(s),\de} (\eta,t,\s) \nn \\
&+ \de_{n(r)+n(s),0\, \text{mod } \r(\s)} \sG_{\nrm;\mnrn} (\s) \pl_\xi \Big{)} \de (\xi-\eta) .
\end{align}
\end{subequations}
It is then clear that a naive application of the local isomorphism $\sj \dual \hj$ would miss the phases $exp [-i \nrrs 
(\xi-\eta)]$ of the monodromy factors $\de_{n(r)} (\xi-\eta)$ in the twisted current algebra \eqref{EqTCA} -- and these 
phases are necessary for consistency with the monodromies of the twisted currents. More generally, the 
monodromy factors in the equal-time commutators and brackets of Subsecs.~$2.2$ and $2.3$ do not follow from local isomorphisms.

\subsection{Example: Einstein Geometry of the WZW Permutation Orbifolds}

As a large example of the development in Sec.~2, we work out here the explicit form of the twisted geometric objects $\he (\hx),
\hat{G} (\hx),\hb (\hx),$ and $\hh (\hx)$ for sector $\s$ of the general WZW permutation orbifold [11-13]. 
We begin with a brief review of what is known about these orbifolds at the tangent-space level.

For the WZW permutation orbifold $A_g (H)/H,$ one starts with the permutation-invariant system
\vspace{-0.1in}
\begin{equation}
g = \oplus_I \gfrak^I ,\quad \gfrak^I \simeq \gfrak ,\quad k_I =k ,\quad H(\text{permutation}) \subset Aut(g)
\end{equation}
where $H$(permutation) permutes the copies $\gfrak^I$ of simple $\gfrak$. We remind the reader that, in the cycle basis of 
Refs.~[5,11-13], each permutation is expressed as a product of disjoint cycles of size 
$f_j(\s)$, where $j$ indexes the cycles and $\hat{j}$ indexes the position within the $j$th cycle. In the 
notation of the orbifold program, this gives the labelling for sector $\s$ of the general WZW permutation orbifold 
\cite{deBoer:2001nw,Halpern:2002ab, Halpern:2002hw,Perm}
\begin{subequations}
\label{37}
\begin{gather}
\nrm \rightarrow \hat{j}aj ,\quad \Nrm \rightarrow \hat{j}\a j ,\quad \frac{n(r)}{\r(\s)} = \frac{N(r)}{R(\s)}
   = \jfj \label{nrrs=jfj} \\
\hat{j}=0,...,f_j(\s) -1 ,\quad a=1,...,\text{dim } \gfrak ,\quad \a =1,...,\text{dim } T ,\quad \srange
\end{gather}
\end{subequations} 
where $T$ is any matrix irrep of $\gfrak$. As a computational aid for the reader, we list some well-known examples of this labelling
\begin{subequations} 
\begin{align} 
& \!\!\Zint_\l : \,\, f_j(\s)\! = \!\rho(\s), \,\,\, \bar{\hat{j}} = 0,..., \rho(\s)\! -\!1, \,\,\, 
   j = 0,..., \frac{\lambda}{\rho(\s)}\! -\!1, \,\,\, \s = 0,..., \rho(\s)\!-\!1 \label{62a} \\ 
& \!\!\Zint_\l, \,\,\l = \text{prime}: \quad \r(\s) = \l , \quad \bar{\hat{j}}\! =\! 0,\dots, \l \!-\!1,  
\quad  j\!=\!0, \quad \s =1, \ldots, \l -1 \, \\  
& \!\! S_N : \quad \! f_j(\s) = \s_j, \quad \s_{j+1} \leq \s_j, \quad j = 0, \dots, n(\vec{\s})-1, 
\quad\sum_{j=0}^{n(\vec{\s})-1} \s_j = N \label{62c} 
\end{align} 
\end{subequations} 
so that e.g. the sectors of the $S_N$ permutation orbifolds are labelled by the ordered partitions of $N$. 

For sector $\s$ of all WZW permutation orbifolds, it is conventional to choose the normalization constants:
\vspace{-0.1in}
\begin{equation}
\label{chi=rt-fj}
\schisig_{\hat{j}aj} = \schisig_j= \sqrt{f_j(\s)} \,.
\end{equation}
Moreover, the duality transformations \eqref{sG-scf-Defn}, \eqref{st-Defn} and \eqref{tw-Inv-Inert} have been explicitly evaluated
[11-13] in this case as
\begin{subequations}
\label{Perm-Details}
\begin{gather}
\sG_{\hat{j}aj ;\hat{l}bl} (\s) = \de_{jl} kf_j(\s) \eta_{ab} \de_{\hat{j}+\hat{l} ,0\,\text{mod } f_j(\s)} ,\quad
   \scf_{\hat{j}aj;\hat{l}bl}{}^{\hat{m}cm} (\s) = \de_{jl} \de_j^m f_{ab}{}^c \de_{\hat{j}+\hat{l}-\hat{m},
   0\, \text{mod }f_j(\s)} \label{Perm-form-of-sG} \\
{\cL}_{\sgb (\s)}^{\hat{j}aj;\hat{l}bl} (\s) = \frac{\eta^{ab}}{2k +Q_{\gfraks}} \frac{1}{f_j(\s)} \de^{jl} \de_{\hat{j}
   +\hat{l} ,0\,\text{mod } f_j(\s)} ,\quad \st_{\hat{j}aj} (T,\s) = T_a \otimes t_{\hat{j}j} (\s) 
\end{gather}
\begin{gather}
[T_a ,T_b ]=if_{ab} {}^c T_c ,\quad t_{\hat{j}j} (\s)_{\hat{l}l} {}^{\hat{m}m} \equiv \de_{jl} \de_j^m
   \de_{\hat{j}+\hat{l} -\hat{m} ,0\, \text{mod } f_j\s } \\
\hat{j}=0,\ldots f_j(\s)-1 ,\quad a= 1\ldots \text{dim } \gfrak ,\quad \srange   
\end{gather}
\end{subequations}
where $f_{ab}{}^c$ and $\eta_{ab}$ are respectively the (untwisted) structure constants and Killing metric of 
$\gfrak$. The zero modes $\{ \hj_{0aj} (0) \}$ of the twisted current algebra and the corresponding subset $\{ \st_{0aj} 
(T,\s) \}$ of twisted representation matrices generate the residual symmetry algebra of sector $\s$. We also note that, 
because of the normalization \eqref{chi=rt-fj}, the rescaled twisted representation matrices $\tst (T^{adj},\s)$ in 
\eqref{exp-form-of-ho} reduce in this case to the ordinary twisted representation matrices $\st (T^{adj},\s)$. 

For the WZW permutation orbifolds, it is also known \cite{deBoer:2001nw} that the group  orbifold elements $\hg$ are reducible 
according to the disjoint cycles $j$ of $h_\s \in H$(permutation)
\begin{subequations}
\begin{gather}
\hg (\st,\xi,t,\s)_{\hat{j}\a j}{}^{\hat{l}\be l} = \de_j^l \hg_j (\st,\xi,t,\s)_{\hat{j}\a}{}^{\hat{l}\be} \\
\hg_j (\st,\xi +2\pi,t,\s)_{\hat{j}\a}{}^{\hat{l}\be} = e^{-\tp \frac{\hat{j}-\hat{l}}{f_j(\s)}} \hg_j (\st,\xi,t,\s)_{\hat{j}\a}{}^{\hat{l}\be} \\
\hat{S}_{\hg (\s)} [\sm ,\hg ]=\sum_j \hat{S}_{\s,j} [\hg_j] \label{gjAction}
\end{gather}
\end{subequations}
and the separable form of the WZW orbifold action in \eqref{gjAction} was given explicitly in Refs.~\cite{deBoer:2001nw,Halpern:2002hw}. 
As we will see below, all the twisted geometric objects of the WZW permutation orbifolds are similarly reducible 
and the orbifold Hamiltonian is separable, as expected from the reducibility of the operator theory \cite{Perm}.

Using Eqs.~\eqref{exp-form-of-ho} and \eqref{Perm-Details}, we find the closed form of the twisted adjoint action:
\begin{subequations}
\label{312}
\begin{gather}
\ho (\hx)_{\hat{j}aj}{}^{\hat{l}bl} = \hg^{-1} (\st (T^{adj},\s),\s)_{\hat{j}aj}{}^{\hat{l}bl} =\de_j^l
    \ho_j (\hx^j)_{\hat{j}a}{}^{\hat{l}b} ,\quad \ho_j (\hx^j) = e^{-i\hY_j (\hx^j)} \label{exp-form-of-perm-ho} \\
\hY (\hx)_{\hat{j}aj}{}^{\hat{l}bl} = \de_j^l \hY_j (\hx^j )_{\hat{j}a}{}^{\hat{l}b} ,\quad \hY_j (\hx^j )
    \equiv \sum_{a=1}^{\text{dim}\gfraks} \sum_{\hat{j}=0}^{\,f_j(\s)-1} \hx_\s^{\hat{j}aj} (\xi) T^{adj}_a \otimes
    \tau_{\hat{j}} (j,\s) \label{exp-form-of-perm-hY} \\
\tau_{\hat{j}} (j,\s)_{\hat{l}}{}^{\hat{m}} \equiv \de_{\hat{j}+\hat{l} -\hat{m},0\, \text{mod } f_j(\s)} \,.
\end{gather}
\end{subequations}
Then Eqs.~\eqref{exp-form-of-ehat}, \eqref{exp-form-of-hB}, \eqref{exp-form-of-hG}, and \eqref{Perm-form-of-sG} give the following
explicit forms of the twisted vielbein $\he (\hx)$ and the twisted Einstein tensors $\hat{G} (\hx), \hb(\hx)$ and $\hh(\hx)$
\begin{subequations}
\label{313}
\begin{gather}
\he (\hx)_{\hat{j}aj}{}^{\hat{l}bl} = \de_j^l \he_j (\hx^j)_{\hat{j}a}{}^{\hat{l}b} ,\quad \he_j (\hx^j) \equiv
   \Big{(} \smal{ \frac{e^{i\hY_j} -1}{i\hY_j}} \Big{)} ,\quad \he_j (0)_{\hat{j}a}{}^{\hat{l}b} = \de_a^b
   \de_{\hat{j}-\hat{l},0\, \text{mod } f_j(\s)} \\
\hat{G}_{\hat{j}aj ;\hat{l}bl} (\hx) = \de_{jl} \hat{G}_j (\hx^j)_{\hat{j}a;\hat{l}b} ,\quad \hb_{\hat{j}aj;\hat{l}bl} 
   (\hx) = \de_{jl} \hb_j (\hx^j)_{\hat{j}a;\hat{l}b} \\
\hat{G}_j (\hx^j)_{\hat{j}a;\hat{l}b} \equiv kf_j(\s) \left( \smal{ \frac{e^{i\hY_j} +e^{-i\hY_j} -2}{(i\hY_j)^2} } \right)
   {}_{\hat{j}a}{}^{-\hat{l},c} \eta_{cb} \label{exp-form-of-perm-hG} \\
\hb_j (\hx^j)_{\hat{j}a;\hat{l}b} \equiv kf_j(\s) \left( \smal{ \frac{e^{i\hY_j}-e^{-i\hY_j} -2i\hY_j}{(i\hY_j)^2}}
   \right){}_{\hat{j}a}{}^{-\hat{l},c} \eta_{cb} \label{exp-form-of-perm-hB} \\
\hh_{\hat{j}aj;\hat{l}bl;\hat{m}cm} (\hx^j) = \de_{jl} \de_{jm} \hh_j (\hx^j)_{\hat{j}a;\hat{l}b;\hat{m}c} \\
\hh_j (\hx^j)_{\hat{j}a;\hat{l}b;\hat{m}c} \equiv \hpl_{\hat{j}aj} \hb_j (\hx^j)_{\hat{l}b;\hat{m}c} + \hpl_{\hat{l}bj}
   \hb_j (\hx^j)_{\hat{m}c;\hat{j}a} + \hpl_{\hat{m}cj} \hb_j (\hx^j)_{\hat{j}a;\hat{l}b} \nn \\
\bigspc =kf_j (\s) \he_j (\hx^j )_{\hat{j}a}{}^{\hat{j}' a'} \he_j (\hx^j )_{\hat{l}b}{}^{\hat{l}' b'}
   \he_j (\hx^j )_{\hat{m}c}{}^{-\hat{j}' -\hat{l}' ,c'} f_{a' b' c'} \label{form-of-perm-hH} \\
\hat{j} ,\hat{l} ,\hat{m} = 0,\ldots f_j (\s) -1 ,\quad a,b,c =1,\ldots ,\text{dim } \gfrak ,\quad \srange 
\end{gather}
\end{subequations}
for each sector $\s$ of all WZW permutation orbifolds. Here $f_{abc} = f_{ab}{}^d \eta_{dc}$ are the totally 
antisymmetric structure constants of $\gfrak$.

Using the block-diagonal forms in \eqref{312},\eqref{313}, and the translation dictionary \eqref{37}, we find that the monodromies 
of all these objects
\begin{subequations}
\begin{gather}
\hx_\s^{\hat{j}aj} (\xi+2\pi) = \hx_\s^{\hat{j}aj} (\xi) e^{\tp \jfj} ,\quad
   \hY_j (\hx^j (\xi+2\pi))_{\hat{j}a}{}^{\hat{l}b} = e^{-\tp \frac{\hat{j}-\hat{l}}{f_j(\s)}} \hY_j (\hx^j (\xi))_{\hat{j}a}{}^{\hat{l}b} \\
\ho_j (\hx^j (\xi+2\pi))_{\hat{j}a}{}^{\hat{l}b} = e^{-\tp \frac{\hat{j}-\hat{l}}{f_j(\s)}} \ho_j (\hx^j (\xi))_{\hat{j}a}{}^{\hat{l}b} \\
\hG_j (\hx^j (\xi+2\pi))_{\hat{j}a;\hat{l}b} = e^{-\tp \frac{\hat{j}+\hat{l}}{f_j(\s)}} \hG_j (\hx^j (\xi))_{\hat{j}a; \hat{l}b} 
\end{gather}
\begin{gather}
\hb_j (\hx^j (\xi+2\pi))_{\hat{j}a;\hat{l}b} = e^{-\tp \frac{\hat{j}+\hat{l}}{f_j(\s)}} \hb_j (\hx^j (\xi))_{\hat{j}a; \hat{l}b} \\
\hh_j (\hx^j (\xi+2\pi))_{\hat{j}a;\hat{l}b; \hat{m}c} = e^{-\tp \frac{\hat{j} +\hat{l}+\hat{m}}{f_j(\s)}} \hh_j (\hx^j (\xi))_{\hat{j}a; 
   \hat{l}b;\hat{m}c} 
\end{gather}
\end{subequations}
follow from the general monodromies in Sec.~2.

Finally we find that, like the WZW orbifold action \cite{deBoer:2001nw,Halpern:2002hw}, the orbifold Hamiltonian density
\vspace{-0.1in}
\begin{subequations}
\begin{gather}
\hat{\sh}_\s (\hx,\hp) = \sum_j \hat{\sh}_j (\hx^j ,\hp_j ,\s) \\
\hat{\sh}_j (\hx^j ,\hp_j ,\s) \equiv 2\pi \hat{G}_j (\hx^j)^{\hat{j}a;\hat{l}b} \hp_{\hat{j}aj} (\hb_j)
   \hp_{\hat{l}bj} (\hb_j) +\frac{1}{8\pi} \pl_\xi \hx_\s^{\hat{j}aj} \pl_\xi \hx_\s^{\hat{l}bj} \hat{G}_j 
   (\hx^j)_{\hat{j}a;\hat{l}b} 
\end{gather}
\begin{gather}
\hp_{\hat{j}aj} (\hb_j) \equiv \hp_{\hat{j}aj}^\s +\frac{1}{4\pi} \hb_j (\hx^j )_{\hat{j}a;\hat{l}b} \pl_\xi
  \hx_\s^{\hat{l}bj} ,\quad \hp^\s_{\hat{j}aj} (\xi+2\pi) = e^{-\tp \jfj} \hp^\s_{\hat{j}aj} (\xi) \\
\{ \hp^\s_{\hat{j}aj} (\xi,t) , \hx_\s^{\hat{l}bl} (\eta,t) \} = -i \de_j^l \de_a^b \de_{\hat{j}-\hat{l} ,0\, \text{mod } f_j(\s)}
   \de_{\jfj} (\xi-\eta) \\
\de_{\jfj} (\xi -\eta) \equiv e^{-i\jfj (\xi-\eta)} \de (\xi-\eta) \\
\hat{G}_j (\hx^j )_{\hat{j}a;\hat{m}c} \,\hat{G}_j (\hx^j )^{\hat{m}c;\hat{l}b} =\de_a^b \de_{\hat{j}-\hat{l},0
  \,\text{mod }f_j(\s)}
\end{gather}
\end{subequations}
is separable into a sum of non-interacting terms $\hat{\sh}_j$.

It will be straightforward to work out the Einstein geometry of many other examples because (except for certain outer-automorphic WZW 
orbifolds of $\Zint_2$-type \cite{so2n}) the tangent-space data for all the basic WZW orbifolds is given in Refs.~[11-14].

\subsection{Subexample: The Permutation Orbifold $A_{\su (2) \oplus \su (2)} (\Zint_2 )/\Zint_2$}

As the simplest subexample in the WZW permutation orbifolds, we have worked out the WZW cyclic permutation orbifold
\vspace{-0.1in}
\begin{equation}
\frac{ A_{\su (2) \oplus \su (2)} (\Zint_2 )}{\Zint_2}
\end{equation}
in further detail. In this case, the unitary group elements $g$ and unitary group orbifold elements $\hg$ have the form
\begin{subequations}
\begin{gather}
g(T,\xi) = exp \left[ i \left( \begin{array}{cc} \beta^{0a}(\xi) T_a^{(0)} & 0 \\ 0 & \beta^{1a}(\xi) T_a^{(1)} \end{array} \right)
   \right] ,\quad T^{(0)} \simeq T^{(1)} \simeq T \\
\hg(\st(T),\xi) = exp [i (\hbe^{0a}(\xi) \one_2 + \hbe^{1a}(\xi) \tau_1) \otimes T_a ] ,\quad \tau_1 = \left( 
   \begin{array}{cc} 0 & \one \\ \one & 0 \end{array} \right) 
\end{gather}
\begin{gather}
\hbe^{\hat{j}a} (\xi+2\pi) = (-1)^{\hat{j}} \hbe^{\hat{j}a} (\xi)
\end{gather}
\end{subequations}
so that, in the single twisted sector $\s=1$, the residual symmetry of this orbifold is the diagonal $\su (2)$ generated by 
the zero modes of the twisted currents.

We may use Eq.~\eqref{exp-form-of-perm-hY} to find the explicit form of the matrix $\hY (\hx)$ in the twisted sector
\vspace{-0.1in}
\begin{subequations}
\label{su2Z2-data}
\begin{gather}
\hY (\hx (\xi)) = i \left( \hx^{0,a} (\xi) \one_2 + \hx^{1,a} (\xi) \tau_1 \right) \otimes I_a 
\end{gather}
\begin{gather}
\hx^{\hat{j}a} (\xi) \equiv \hx_{\s =1}^{\hat{j}a ,j=0} (\xi,t) ,\quad (T_a^{adj} )_{bc} =i (I_a )_{bc} 
   =-i \ep_{abc} ,\quad \ep_{123} =1 \\
\hx^{\hat{j}a} (\xi+2\pi) = (-1)^{\hat{j}} \hx^{\hat{j}a} (\xi) \label{hx-mono-su2Z2} \\
\r(1)=2 ,\quad j=0,\quad a,b,c \in \{ 1,2,3 \} ,\quad \hat{j} \in \{ 0,1 \}
\end{gather}
\end{subequations}
where $\hx(\xi)$ are the twisted Einstein coordinates with diagonal monodromy and we have chosen root length 
$\psi^2 =1$ for $\gfrak = \su (2)$. The matrices $I_a$ in \eqref{su2Z2-data} are the generators in the adjoint of 
the residual $\su(2)$ symmetry of the twisted sector.

Using \eqref{exp-form-of-perm-ho} and \eqref{su2Z2-data}, we find the explicit form of the twisted adjoint action:
\begin{subequations}
\label{form-of-su2Z2-ho}
\begin{gather}
\ho(\hx)_{0,a}{}^{0,b} = \ho(\hx)_{1,a}{}^{1,b} = \half \left(\ho ({\sxh}^+)_a{}^b + \ho({\sxh}^-)_a{}^b \right)\\
\ho(\hx)_{0,a}{}^{1,b} = \ho(\hx)_{1,a}{}^{0,b} = \half \left(\ho ({\sxh}^+)_a{}^b - \ho({\sxh}^-)_a{}^b \right)\\
\ho ({\sxh}^\pm ) \equiv \one +\frac{({\sxh}^\pm \cdot I)}{|{\sxh}^\pm |} \sin{|{\sxh}^\pm |}
  + \frac{({\sxh}^\pm \cdot I)^2}{|{\sxh}^\pm |^2} (\cos{|{\sxh}^\pm |} -1) \\
\sxh^{\pm,a} \equiv \sxh^{\pm,a} (\xi,t) \equiv \hx^{0,a}(\xi,t) \pm \hx^{1,a} (\xi,t) \,. \label{sxh-defn}
\end{gather}
\end{subequations}
Similarly, we obtain the explicit form of the twisted Einstein metric in this case
\begin{subequations}
\label{exp-form-of-su2Z2-hG}
\begin{gather}
\hat{G}_{0,a;0,b} (\hx(\xi)) = \hat{G}_{1,a;1,b} (\hx(\xi)) = k \left( G({\sxh}^+ (\xi)) +G({\sxh}^- (\xi)) \right)_{ab} 
\end{gather}
\begin{gather}
\hat{G}_{0,a;1,b} (\hx(\xi)) = \hat{G}_{1,a;0,b} (\hx(\xi)) = k \left( G({\sxh}^+ (\xi)) -G({\sxh}^- (\xi)) \right)_{ab} \\
G({\sxh}^\pm ) \equiv \one + \frac{({\sxh}^\pm \cdot I)^2}{| {\sxh}^\pm |^4} (2 \cos{| {\sxh}^\pm |}
   +|{\sxh}^\pm |^2 -2) 
\end{gather}
\end{subequations}
from Eq.~\eqref{exp-form-of-perm-hG}. It is not difficult to see that the twisted Einstein metric transforms as a second 
rank tensor under the residual $SU(2)$ symmetry group of the twisted sector
\begin{equation}
\hG_{\hat{j}a;\hat{l}b} (\hx) = \de_{ab} f_{\hat{j}\hat{l}}^{(0)} (\hx) + I_a I_b f_{\hat{j}\hat{l}}^{(2)} (\hx)
\end{equation}
but this is because we have chosen the preferred coordinate system $\hx = \hbe$.

We comment on monodromies in this example, starting from the diagonal monodromy \eqref{hx-mono-su2Z2} of the twisted Einstein 
coordinates:
\begin{subequations}
\label{su2Z2-Monos}
\begin{gather}
\{ \hx^{0,a} (\xi +2\pi) =\hx^{0,a} (\xi) ,\,\, \hx^{1,a} (\xi +2\pi) =-\hx^{1,a} (\xi) \}
   \,\,\,\Rightarrow \,\,\,\sxh^{\pm ,a} (\xi +2\pi) = \sxh^{\mp ,a} (\xi) \\
\Rightarrow \ho({\sxh}^\pm (\xi+2\pi)) = \ho({\sxh}^\mp (\xi)) ,\quad G({\sxh}^\pm (\xi+2\pi)) = 
   G({\sxh}^\mp (\xi)) \,.
\end{gather}
\end{subequations}
The coordinates $\sxh^{\pm,a}$ do not have definite monodromy and are in fact an example of the `coordinates with twisted 
boundary conditions' discussed more generally in App.~C. Then we may verify that the correct diagonal monodromies 
($3.14$b,c) of the twisted adjoint action and the twisted Einstein metric 
\begin{equation}
\ho (\hx (\xi+2\pi))_{\hat{j}a}{}^{\hat{l}b} = e^{-\pi i (\hat{j}-\hat{l})} \ho (\hx(\xi))_{\hat{j}a}{}^{\hat{l}b} ,\quad
\hat{G}_{\hat{j}a ;\hat{l}b} (\hx (\xi +2\pi)) = e^{-\pi i (\hat{j}+\hat{l})} \hat{G}_{\hat{j}a;\hat{l}b} (\hx (\xi))
\end{equation}
follow from Eqs.~\eqref{form-of-su2Z2-ho}, \eqref{exp-form-of-su2Z2-hG} and \eqref{su2Z2-Monos}.

The explicit form of the WZW orbifold interval \eqref{WZW-orb-ds2} in this case
\begin{eqnarray}
ds^2 (\s =1) &=& \hat{G}_{\hat{j}a ;\hat{l}b} (\hx) d\hx^{\hat{j}a} d\hx^{\hat{l}b} \nn \\
&=& k \left( G({\sxh}^+ )_{ab} d\sxh^{+,a} d\sxh^{+,b} + G({\sxh}^- )_{ab} d\sxh^{-,a} d\sxh^{-,b} \right) \label{su2Z2-ds2}
\end{eqnarray}
may be an appropriate starting point for further study of WZW orbifold geodesics and related topics. 

The corresponding explicit forms of the twisted $B$ field $\hB$ and the twisted torsion field $\hh$ of this orbifold are 
also easily worked out from Eqs.~\eqref{exp-form-of-perm-hB} and \eqref{form-of-perm-hH}, but we shall not do so here. 
As further examples, orbifolds on abelian $g$ are considered in App.~E.

\section{Sigma-Model Orbifolds}
\vspace{-.07in}

In this section we use eigenfields and local isomorphisms to find the action formulation of the orbifolds of a large class of non-linear 
sigma models. This sigma model orbifold action will include the general WZW orbifold action \cite{deBoer:2001nw} and the general coset 
orbifold action \cite{Halpern:2002hw} as special cases, as well as the action formulation of many other orbifolds such as orbifolds of the 
principal chiral models.

\subsection{Non-Linear Sigma Models with a Linear Symmetry}
\label{orbifolds-of-sigma-model}

We begin with a general nonlinear sigma model $A_M$ on a general target-space manifold $M$ 
\begin{subequations}
\label{s-model-defn}
\begin{gather}
S_{NLS} = \int \! d^2\xi\, {\cL}_{NLS}(x), \quad {\cL}_{NLS}(x) = \frac{1}{8\p}( G_{ij}(x) + B_{ij}(x) ) 
   \pl_+x^i \pl_-x^j \\
G_{ij} (x) = G_{ji} (x) ,\quad B_{ij} (x) = -B_{ji} (x) \\
H_{ijk} (x) = \pl_i B_{jk} (x) + \pl_j B_{ki} (x) + \pl_k B_{ij} (x) ,\quad \pl_i = \frac{\pl}{\pl x^i}
\end{gather} 
\end{subequations}
with a set of local\footnote{A complete treatment of the nonlinear sigma model should include discussion of global 
issues (in which the discussion of the text pertains to a coordinate patch), but we will not do so here.} 
Einstein coordinates $\{x^i ,\, i=1,\ldots,\text{dim } M \}$, as well as the usual Einstein metric $G_{ij} (x)$, 
$B$ field $B_{ij} (x)$ and torsion field $H_{ijk} (x)$. We will limit ourselves here to the class of sigma models 
$A_M (H)$ with a symmetry group $H$ which acts {\it linearly} on the coordinates and fields as follows:
\begin{subequations}
\label{lin-symm-defn}
\begin{gather}
{x^i}' = x^i (w\hc)_i{}^j  ,\quad \pl_i(w\hc)_j{}^{k} = 0 ,\quad {\pl_i}' =w_i{}^j \pl_j ,\quad w \equiv w(h_\s), 
   \;\forall h_\s \in H  \label{x'=xw} \\
G_{ij}(x)' \equiv G_{ij}(x') = w_i{}^l w_j{}^k G_{lk}(x),  \quad B_{ij}(x)' \equiv B_{ij}(x') = 
   w_i{}^l w_j{}^k B_{lk}(x) \label{GijBij} 
\end{gather}
\begin{gather}
H_{ijk} (x)' \equiv H_{ijk} (x') = w_i{}^l w_j{}^m w_k{}^n H_{lmn} (x) \\
{\cL}_{NLS}(x)' = {\cL}_{NLS}(x') = {\cL}_{NLS}(x) \,.
\end{gather}
\end{subequations}
Linearity of the symmetry transformations can be maintained only in certain preferred coordinate 
systems: In conventional terms, we are considering the special case of a symmetry group $H$ with 
one fixed point of $M$ in each sector $\s$, and the preferred coordinate systems are those in which 
the fixed points are at the origin for all $\s$. In a somewhat more general language, the symmetry group
$H$ must be a subgroup of the isometry group of $M$, and $H$ must also preserve the two-form $B$ on $M$.
The class of symmetric sigma models $A_M (H)$ includes as special cases the symmetric WZW models $A_g(H)$, 
the general $H$-invariant coset construction $A_{g/h} (H)$ and the $H$-invariant principal chiral models.

\subsection{The Sigma Model Orbifold Action}
 
Our primary task here is to use an Einstein-space generalization of the principle of local isomorphisms \cite{deBoer:1999na,
Halpern:2000vj,deBoer:2001nw} to obtain the sigma model orbifold action for the sigma model orbifolds $A_M (H)/H$
\begin{equation}
\frac{A_M (H)}{H} \supset \frac{A_g (H)}{H} , \frac{A_{g/h} (H)}{H} ,\ldots 
\end{equation}
in terms of their twisted Einstein fields.

As a first step, we define the Einstein-space $H$-eigenvalue problem
\vspace{-.04in}
\begin{subequations}
\label{Eins-eigen} 
\begin{gather}
w(h_\s)_i{}^{j}U\hc(\s)_j{}^{\nrm} \= U\hc(\s)_i{}^{\nrm}E_{n(r)}(\s) \\
U^\dagger (\s) U(\s) = \one ,\quad E_{n(r)}(\s) \= e^{-\tp \frac{n(r)}{\r(\s)}} \\
\w (h_0)_i{}^j =\de_i{}^j ,\quad U\hc (0) = \one ,\quad E_0 (0) =1 \\
i,j=1,\ldots ,\text{dim } M ,\quad \srange
\end{gather}
\end{subequations}
in analogy with the tangent-space $H$-eigenvalue problem in \eqref{H-eig}. Here $N_c$ is the number of conjugacy classes of $H$, 
$\r(\s)$ is the order of $h_\s \in H$, and $n(r),\m$ are respectively the spectral indices and degeneracy labels of the 
eigenvalue problem. In spite of appearances, there is no {\it a priori} relation between these Einstein-space indices and those 
defined by the $H$-eigenvalue problem in \eqref{H-eig} -- which applies only to the tangent-space description of group manifolds.

We now use the Einstein-space $H$-eigenvalue problem in the standard way \cite{deBoer:1999na,Halpern:2000vj,deBoer:2001nw}. 
The relevant eigenfields $\sx ,\sG, \sb ,\sh$ and their responses to the symmetry group are
\begin{subequations}
\label{Ein-Eigf-Defn}
\begin{gather}
\sx^\nrm(x) \equivs \sx_\s^\nrm(x) \equivs \schisig^{-1}_\nrm x^i U\hc(\s)_i{}^\nrm, \quad x^i(\sx) 
   \= \schisig_\nrm \sx_\s^\nrm(x) U(\s)_\nrm{}^i \label{sx-Defn} \\
\sG_{\nrm;\nsn} (\sx) \equiv \schisig_\nrm \schisig_\nsn U(\s)_\nrm{}^i U(\s)_\nsn{}^j G_{ij}(x(\sx)) \\
\sb_{\nrm;\nsn} (\sx) \equiv \schisig_\nrm \schisig_\nsn U(\s)_\nrm{}^i U(\s)_\nsn{}^j B_{ij}(x(\sx)) 
\end{gather} \vshalf
\begin{align}
& \sh_{\nrm;\nsn;\ntd} (\sx) \nn \\
& \quad \quad \equiv \! \schisig_\nrm \schisig_\nsn \schisig_\ntd U(\s)_\nrm {}^{\!i}
   U(\s)_\nsn {}^{\!j} U(\s)_\ntd {}^{\!k} H_{ijk} (x(\sx))
\end{align}
\end{subequations} \vsvsvs
\begin{subequations}
\label{Aut-Resp}
\begin{gather}
\sx^\nrm(x)' =  \sx^\nrm(x') = \schisig_\nrm^{-1} {x^i}' U\hc (\s)_i{}^\nrm = \sx^\nrm(x) E_{n(r)}(\s)^\ast \\
\sG_{\nrm;\nsn}(\sx)'\!=\,\sG_{\nrm;\nsn}(\sx(x'))\!=\!E_{n(r)}(\s)E_{n(s)}(\s)
  \sG_{\nrm;\nsn}(\sx)\\
\sb_{\nrm;\nsn}(\sx)'\!=\, \sb_{\nrm;\nsn}(\sx(x'))\!=\!E_{n(r)}(\s) 
  E_{n(s)}(\s)\sb_{\nrm;\nsn}(\sx) 
\end{gather}\vshalf
\begin{align}
& \sh_{\nrm;\nsn;\ntd} (\sx)' \nn \\
& \quad \quad = \! \sh_{\nrm;\nsn;\ntd} (\sx (x')) \!=\! E_{n(r)} (\s) E_{n(s)} (\s) E_{n(t)} (\s) \sh_{\nrm;\nsn;\ntd} (\sx)
\end{align}
\end{subequations}
where $\schisig$ are a set of normalization constants with $\schizero =1$. We may then rewrite the Lagrange density 
\eqref{s-model-defn} of the symmetric non-linear sigma model in terms of the coordinate eigenfields $\sx (x)$:
\begin{subequations}
\begin{gather}
\tilde{{\cL}}_{NLS}(\sx) \equiv {\cL}_{NLS}(x(\sx)) = \frac{1}{8\p} (\sG_{\nrm;\nsn}(\sx) + 
   \sb_{\nrm;\nsn}(\sx)) \pl_+\sx^\nrm\pl_-\sx^\nsn \label{tildeL} \\
\tilde{{\cL}}_{NLS}(\sx)' = \tilde{{\cL}}_{NLS}(\sx') = \tilde{{\cL}}_{NLS}(\sx) \,. \label{L'=L}
\end{gather}
\end{subequations}
In the form \eqref{tildeL}, the $H$-invariance \eqref{L'=L} of the Lagrange density follows directly from 
Eq.~\eqref{Aut-Resp}.

Next, we use the principle of local isomorphisms
\begin{subequations}
\label{Ein-PoLI}
\begin{gather}
\sx \dual \hx,\quad \sG(\sx) \dual \hG(\hx), \quad \sb(\sx) \dual \hB(\hx) \\
\sh (\sx) \dual \hat{H} (\hx) ,\quad \tilde{{\cL}}_{NLS}(\sx)
  \dual \hat{{\cL}}_\s^{NLS}(\hx)   \\
\text{ automorphic response }  E_{n(r)}(\s) \dual \text{ monodromy } E_{n(r)}(\s)
\end{gather}
\end{subequations}
to obtain the twisted Einstein fields $\hx, \hG, \hB, \hh$ with diagonal monodromy and the {\it sigma model orbifold action}
\vspace{-.03in}
\begin{subequations}
\label{sigma-model-orbifold-action}
\begin{gather}
\hat{S}_\s^{NLS} = \int \,d^2\xi\, \hat{{\cL}}_\s^{NLS}(\hx),\quad \srange \\
\hat{{\cL}}_\s^{NLS} (\hx) = \frac{1}{8\p} (\hG_{\nrm;\nsn}(\hx) + \hB_{\nrm;\nsn}(\hx)) 
  \pl_+\hx_\s^\nrm \pl_- \hx_\s^\nsn \\
\hh_{\nrm;\nsn;\ntd} (\hx) \!\equiv \! \left( \hpl_{\nrm} \hB_{\nsn;\ntd} (\hx) \!+ \text{cyclic in } \smal{(\nrm ,\nsn
  ,\ntd)} \right) \label{hh-Defn}
\end{gather}
\begin{gather}
\hx_\s^\nrm(\xi+2\p) =\hx_\s^\nrm (\xi) E_{n(r)} (\s)^\ast = \hx_\s^\nrm(\xi) e^{2\p i\frac{n(r)}{\r(\s)}} \label{x-Mono1} \\
\hG_{\nrm;\nsn}(\hx(\xi+2\p)) = e^{-2\p i\frac{n(r)+n(s)}{\r(\s)}} \hG_{\nrm;\nsn}(\hx(\xi)) \\
\!\!\hB_{\nrm;\nsn}(\hx(\xi+2\p)) \!=\! e^{-2\p i\frac{n(r)+n(s)}{\r(\s)}} \hB_{\nrm;\nsn}(\hx(\xi)) \\
\hh_{\nrm;\nsn;\ntd} (\hx (\xi+2\pi)) \!=\! e^{-\tp \frac{n(r)+n(s)+n(t)}{\r(\s)}} \hh_{\nrm;\nsn;\ntd} (\hx(\xi)) \\
\hat{{\cL}}_\s^{NLS}(\hx(\xi+2\p)) \!=\! \hat{{\cL}}_\s^{NLS}(\hx(\xi)) 
\end{gather}
\begin{gather}
\hx \equiv \hx_\s(\xi), \quad \hG(\hx) \equiv \hG(\hx_\s(\xi),\s), \quad \hB(\hx) \equiv \hB(\hx_\s(\xi),\s) ,\quad
  \hh(\hx) \equiv \hh (\hx_\s (\xi) ,\s) 
\end{gather}
\end{subequations}
which describes sector $\s$ of each sigma model orbifold $A_M (H)/H$. Here $\hx^\nrm$ are the twisted Einstein coordinates with 
$\hpl_\nrm \equiv \pl/ \pl \hx^\nrm$, $\hG (\hx)$ is the twisted Einstein metric, $\hB (\hx)$ is the twisted $B$ field and $\hh(\hx)$ 
is the twisted torsion field of sector $\s$. As expected, $\hG$ is symmetric in its indices, while $\hB$ and $\hh$ are
(totally) antisymmetric.

The eigenfields \eqref{Ein-Eigf-Defn} and the principle of local isomorphisms \eqref{Ein-PoLI} also give the 
{\it explicit functional forms} of the twisted Einstein tensors
\vspace{-.03in}
\begin{subequations}
\label{form-of-tw-G&B}
\begin{gather}
x^i = \schisig_\nrm \sx_\s^\nrm U(\s)_\nrm{}^i \dual \sxh^i(\hx) \equiv \sxh_\s^i(\hx_\s) \equiv 
  \schisig_\nrm \hx_\s^\nrm U(\s)_\nrm{}^i \label{x-to-sx}\\
\hG_{\nrm;\nsn}(\hx) = \schisig_\nrm \schisig_\nsn U(\s)_\nrm{}^i U(\s)_\nsn{}^j G_{ij} (x \,
  \duals\, \sxh(\hx)) \label{explicit-form-of-twisted-Einstein-tensor:G} \\
\hB_{\nrm;\nsn}(\hx) = \schisig_\nrm \schisig_\nsn U(\s)_\nrm{}^i U(\s)_\nsn{}^j B_{ij} 
  (x \,\duals \,\sxh(\hx)) \label{explicit-form-of-twisted-Einstein-tensor:B} 
\end{gather}\vsvsvs
\begin{align}
& \hh_{\nrm;\nsn;\ntd}(\hx) \\
& \bigspc =\! \schisig_\nrm \schisig_\nsn \schisig_\ntd U(\s)_\nrm{}^i U(\s)_\nsn{}^j U(\s)_\ntd{}^k 
   H_{ijk}(x \,\duals\, \sxh(\hx)) \nn
\end{align}
\end{subequations}
in terms of the untwisted Einstein tensors $G_{ij}(x)$, $B_{ij}(x)$ and $H_{ijk} (x)$. As an example, we may see this
for the twisted Einstein metric by following the steps shown in Fig.~1.

\begin{picture}(300,100)(0,0)
\put(50,85){$\sG_{\nrm;\nsn} (\sx_\s) = \schisig_\nrm \schisig_\nsn U(\s)_\nrm{}^i U(\s)_\nsn{}^j G_{ij} 
   (x(\sx))$}
\thicklines
\put(80,78){\vector(0,-1){40}}
\put(240,78){\vector(0,-1){40}}
\put(85,58){$\s$}
\put(245,58){$\s$}
\put(50,28){$\hG_{\nrm;\nsn} (\hx_\s) = \schisig_\nrm \schisig_\nsn U(\s)_\nrm{}^i U(\s)_\nsn{}^j G_{ij} 
  (x \,\duals\, \sxh_\s (\hx_\s)) $}
\put(128,4){Fig.\,\ref{fig:metric}:Metric and Twisted Metric}
\end{picture}
\myfig{fig:metric}

\noindent The arrow notation in ($4.10$b,c,d) means replacement in the functional form
\begin{equation}
\label{f(x-goto-hx)}
f ( x\, \duals \,\sxh(\hx)) \equiv f(x) \Big{|}_{x \, \duals \,\sxhs(\hx)} 
\end{equation}
and the objects $\sxh^i(\hx)$ (defined in \eqref{x-to-sx} as locally isomorphic to $x^i$) are identified in Appendix C as 
Einstein coordinates with twisted boundary conditions (matrix monodromy):
\begin{equation}
 \hat{\sx}_\s^i(\xi+2\p) = \hat{\sx}_\s^j(\xi)w\hc(h_\s)_j{}^i \,. \label{sxh-Mono1}
\end{equation}
Consequently, the twisted Einstein coordinates $\hx$ 
\begin{subequations}
\label{hx=sxhU}
\begin{gather}
\hx_\s^\nrm (\xi,t) = \schisig^{-1}_\nrm \sxh^i_\s (\xi,t) U\hc (\s)_i^\nrm =\schisig^{-1}_\nrm x^i (x \,\duals 
   \,\sxhs(\hx)) U\hc(\s)_i{}^\nrm \label{413a} \\
\hx_\s^\nrm (\xi+2\pi ,t) = \hx_\s^\nrm (\xi,t) e^{\tp \nrrs}
\end{gather}
\end{subequations}
are obtained as the monodromy decomposition of $\sxh$.

Note that we do not find any selection rules \cite{deBoer:1999na,Halpern:2000vj,deBoer:2001nw} for the twisted Einstein tensors $\hG$ 
or $\hB$. This is in contrast to the case of twisted tangent-space tensors (see e.g.~the remarks around \eqref{Selection}), and this 
absence can be traced back to the fact that the untwisted Einstein tensors are not invariant under the action 
\eqref{GijBij} of the automorphism group. In fact however, selection rules do exist for the moments of the twisted 
tensors -- as we shall see in Subsec.~$4.4$.

\subsection{The Sigma-Model Orbifold Interval}

The {\it sigma-model orbifold interval}
\begin{subequations}
\label{s-mod-ds2}
\begin{align}
ds^2 (\s) &\equiv \hG_{\nrm;\nsn} (\hx) d\hx_\s^\nrm d\hx_\s^\nsn \label{414a} \\
&= G_{ij} (x \,\duals\, \sxh_\s) d\sxh^i_\s d\sxh^j_\s \label{ds2-x-to-sxh} \\
&= ds^2 (\s =0) \Big{|}_{x \rightarrow \sxhs_\s} 
\end{align}
\end{subequations}
has the same basic form \eqref{414a} as that given in Eq.~\eqref{WZW-orb-ds2} for the WZW orbifolds. The alternate forms in 
($4.14$b,c) follow from Eqs.~\eqref{explicit-form-of-twisted-Einstein-tensor:G} and \eqref{413a}, the fields $\sxh^i$ are the 
Einstein coordinates \eqref{x-to-sx} with twisted boundary conditions and the untwisted interval is 
\vspace{-0.1in}
\begin{equation}
ds^2( \s=0) = G_{ij}(x) dx^i dx^j.
\end{equation} 
The relative simplicity of \eqref{ds2-x-to-sxh} (with no $\schisig$'s or $U(\s)$'s) is a consequence of the trivial monodromy 
of the orbifold interval.

All forms of the interval in \eqref{s-mod-ds2} apply as well to the WZW orbifolds, and we mention in particular
the specific WZW cyclic permutation orbifold
\begin{equation}
\frac{ A_{\su (2)\oplus \su (2)} (\Zint_2)}{\Zint_2} 
\end{equation}
which was discussed in Subsec.~$3.3$. In this case, the orbifold interval \eqref{s-mod-ds2} takes the explicit form
\vspace{-0.1in}
\begin{subequations}
\begin{gather}
x^{0,a} (\xi)' = x^{1,a} (\xi) ,\quad x^{1,a} (\xi)' = x^{0,a} (\xi) ,\quad a=1,2,3 \\
ds^2 (\s=0) = k (G(x^0)_{ab} dx^{0,a} dx^{0,b} + G(x^1)_{ab} dx^{1,a} dx^{1,b} ) \\
ds^2 (\s=1) \!=\! ds^2 (\s=0) \Big{|}_{x^{0,1} \rightarrow \sxhs^{\pm}} \!
   = \!k \left( G(\sxh^+ )_{ab} d\sxh^{+,a} d\sxh^{+,b} \!+\! G(\sxh^- )_{ab} d\sxh^{-,a} d\sxh^{-,b} \right)
\end{gather}
\end{subequations}
in agreement with Eq.~\eqref{su2Z2-ds2}, where $G(x)$ and $\sxh^\pm$ are defined in Eqs.~\eqref{form-of-su2Z2-ho} and
\eqref{exp-form-of-su2Z2-hG}. 

As illustrated in Eq.~\eqref{ds2-x-to-sxh}, orbifold quantities with trivial monodromy have simple forms when expressed in terms 
of the Einstein coordinates $\sxh$ with twisted boundary conditions. We note in particular the {\it induced} sigma-model orbifold 
metric $\hG_{mn}$ on the world sheet
\vspace{-0.1in}
\begin{subequations}
\begin{gather}
\hat{G}_{mn} (\xi,t,\s) \equiv \hat{G}_{\nrm;\nsn} (\hx_\s(\xi,t)) \pl_m \hx_\s^\nrm (\xi,t) \pl_n \hx_\s^\nsn (\xi,t) \\
\bigspc = G_{ij} (x \,\duals\, \sxh_\s ) \pl_m \sxh_\s^i (\xi,t) \pl_n \sxh_\s^j (\xi,t) \\
ds^2 (\s) = \hat{G}_{mn} (\xi,t,\s) d\xi^m d\xi^n ,\quad (\xi^0 ,\xi^1 )= (t,\xi) 
\end{gather}
\end{subequations}
and the closely-related alternate form
\begin{align}
&\hat{S}_\s^{NLS} \!= \! \frac{1}{8\pi} \int \!d^2 \!\xi \left( G_{ij} (x)+ B_{ij}(x) \right) \Big{|}_{x \,\duals\, \sxh_\s} 
   \pl_+ \sxh_\s^i (\xi,t) \pl_- \sxh_\s^j (\xi,t) ,\quad \srange
\end{align}
of the sigma model orbifold action \eqref{sigma-model-orbifold-action}. To our knowledge, fields with twisted boundary
conditions appeared first in Ref.~\cite{DHVW}.

\subsection{Moment Expansion of the Twisted Einstein Tensors}

In this subsection, we use moment expansions to further explore the results of the Subsec.~$4.2$.

We begin with the moment expansions of the untwisted Einstein tensors
\begin{subequations}
\label{untw-mom-exp}
\begin{gather}
G_{ij}(x) = \sum_{m=0}^\infty g_{ij,i_1...i_m} x^{i_1}\cdots x^{i_m} ,\quad B_{ij}(x) = 
   \sum_{m=0}^\infty b_{ij,i_1...i_m} x^{i_1}\cdots x^{i_m} \\
H_{ijk} (x) = \sum_{m=0}^\infty h_{ijk,i_1...i_m} x^{i_1}\cdots x^{i_m} \\
h_{ijk,i_1...i_m} \!=\! (m\!+\!1) \left( b_{ij,ki_1...i_m} + \text{ cyclic in } \foot{(ijk)} \right) \label{mom-b-h} 
\end{gather}
\end{subequations}
where the untwisted moments $g_{...} ,b_{...},$ and $h_{...}$ are totally symmetric in the indices $i_1 \ldots i_m$. 
Because $\sxh (\hx)$ in \eqref{x-to-sx} is linear in $\hx$, we know from the other relations in \eqref{form-of-tw-G&B} that 
the twisted Einstein tensors of sector $\s$ will have similar moment expansions
\begin{subequations}
\label{tw-mom-exp}
\begin{align}
&\hG_{\nrm;\nsn}(\hx_\s) = \sum_{m=0}^\infty \sG_{\nrm,\nsn,n(r_1)\m_1...n(r_m)\m_m} 
   \hx_\s^{n(r_1)\m_1}\cdots \hx_\s^{n(r_m)\m_m} \\
&\hB_{\nrm;\nsn}(\hx_\s) = \sum_{m=0}^\infty \sb_{\nrm,\nsn,n(r_1)\m_1...n(r_m)\m_m} 
    \hx_\s^{n(r_1)\m_1}\cdots \hx_\s^{n(r_m)\m_m} \\
&\hh_{\nrm;\nsn;\ntd}(\hx_\s) = \sum_{m=0}^\infty \sh_{\nrm,\nsn,\ntd,n(r_1)\m_1...n(r_m)\m_m} 
    \hx_\s^{n(r_1)\m_1}\cdots \hx_\s^{n(r_m)\m_m} 
\end{align}\vsvsvs
\begin{align}
\sh_{\nrm,\nsn,\ntd,n(r_1)\m_1...n(r_m)\m_m} =& (m\!+\!1) \Big{(} \sb_{\nrm,\nsn,\ntd,n(r_1)\m_1...n(r_m)\m_m} \nn \\
   & \bigspc + \text{cyclic in } \foot{(\nrm,\nsn,\ntd)} \Big{)} \label{mom-sb-sh}
\end{align}
\end{subequations}
in terms of the twisted Einstein coordinates $\hx_\s$. The twisted moments $\sG_{...} ,\sb_{...}$ and $\sh_{...}$ in 
Eq.~\eqref{tw-mom-exp} are also totally symmetric in the indices $n(r_1)\m_1 \ldots n(r_m)\m_m$. Substitution of the moment 
expansions \eqref{untw-mom-exp} and \eqref{tw-mom-exp} into \eqref{form-of-tw-G&B} gives the explicit forms of the twisted 
moments as an arbitrarily large set of {\it duality transformations} 
\vspace{-.03in}
\begin{subequations}
\label{sGh,sBh=...}
\begin{align}
&\!\!\sG_{\nrm,\nsn,n(r_1)\m_1...n(r_m)\m_m} \!\equiv\! \schi_\nrm \schi_\nsn U_\nrm{}^i U_\nsn{}^j
	 \!\left(\prod_{k=1}^m \schi_{n(r_k)\m_k} U_{n(r_k)\m_k}{}^{i_k}\right)\!g_{ij,i_1...i_m} \label{moment-sca} \\
&\!\!\sb_{\nrm,\nsn,n(r_1)\m_1...n(r_m)\m_m} \!\equiv\! \schi_\nrm \schi_\nsn U_\nrm{}^i U_\nsn{}^j
	 \!\left(\prod_{k=1}^m \schi_{n(r_k)\m_k} U_{n(r_k)\m_k}{}^{i_k}\right)\!b_{ij,i_1...i_m} \label{moment-scb} \\
& \!\!\sh_{\nrm,\nsn,\ntd,n(r_1)\m_1...n(r_m)\m_m} \nn \\
& \quad \quad \equiv \schi_\nrm \schi_\nsn \schi_\ntd U_\nrm{}^i U_\nsn{}^j U_\ntd{}^k
	 \!\left(\prod_{k=1}^m \schi_{n(r_k)\m_k} U_{n(r_k)\m_k}{}^{i_k}\right)\!h_{ijk,i_1...i_m} \label{moment-scc} 
\end{align}
\begin{gather}
\schi \equiv \schisig, \quad U \equiv U(\s)
\end{gather}
\end{subequations}
which express the twisted moments $\sG_{...} ,\sb_{...} ,\sh_{...}$ of $\hG ,\hB ,\hh$ in terms of the 
untwisted moments $g_{...} ,b_{...} ,h_{...}$ respectively. 

Next, we need the $H$-invariance of the untwisted moments 
\begin{subequations}
\label{transformation-of-untwisted-moments} 
\begin{gather}
(w\hc)_i{}^k (w\hc)_j{}^l (w\hc)_{i_1}{}^{j_1} \cdots (w\hc)_{i_m}{}^{j_m} g_{kl,j_1...j_m} = 
  g_{ij,i_1...i_m} \label{untw-g-xform} \\
(w\hc)_i{}^k (w\hc)_j{}^l (w\hc)_{i_1}{}^{j_1} \cdots (w\hc)_{i_m}{}^{j_m} b_{kl,j_1...j_m} = 
  b_{ij,i_1...i_m} \\
(w\hc)_i{}^l (w\hc)_j{}^m (w\hc)_k{}^n (w\hc)_{i_1}{}^{j_1} \cdots (w\hc)_{i_m}{}^{j_m} h_{lmn,j_1...j_m} = 
  h_{ijk,i_1...i_m} 
\end{gather}
\end{subequations}
which is a consequence of the symmetry transformations ($4.2$b,c) of the untwisted tensors. It follows from the 
duality transformations in \eqref{sGh,sBh=...} that the {\it selection rules} for the twisted moments
\begin{subequations}
\label{tw-mom-sel1}
\begin{gather}
\sG_{\nrm,\nsn,n(r_1)\m_1,...,n(r_m)\m_m} (1-E_{n(r)}^\ast E_{n(s)}^\ast E_{n(r_1)}^\ast 
   \cdots E_{n(r_m)}^\ast) = 0 \label{ex: sG-select}\\
\sb_{\nrm,\nsn,n(r_1)\m_1,...,n(r_m)\m_m} (1-E_{n(r)}^\ast E_{n(s)}^\ast E_{n(r_1)}^\ast \cdots E_{n(r_m)}^\ast) = 0 \\
\sh_{\nrm,\nsn,\ntd,n(r_1)\m_1...n(r_m)\m_m} (1-E_{n(r)}^\ast E_{n(s)}^\ast E_{n(t)}^\ast E_{n(r_1)}^\ast \cdots E_{n(r_m)}^\ast) = 0 
\end{gather}
\end{subequations}
and the solutions of these selection rules
\begin{subequations}
\label{tw-mom-sel2}
\begin{gather}
\sG_{\nrm,\nsn,n(r_1)\m_1,...,n(r_m)\m_m} \;\propto\; \d_{n(r)+n(s)+\sum_{j=1}^m n(r_j),0\,\text{mod}\,\r(\s) } \\
\sb_{\nrm,\nsn,n(r_1)\m_1,...,n(r_m)\m_m} \;\propto\; \d_{n(r)+n(s)+\sum_{j=1}^m n(r_j),0\,\text{mod}\,\r(\s) } \\
\sh_{\nrm,\nsn,\ntd,n(r_1)\m_1,...,n(r_m)\m_m} \;\propto\; \d_{n(r)+n(s)+n(t)+\sum_{j=1}^m n(r_j),0\,\text{mod}\,\r(\s) }
\end{gather}
\end{subequations}
are dual to Eq.~\eqref{transformation-of-untwisted-moments}. As an example, the selection rule \eqref{ex: sG-select} follows by 
substitution of \eqref{untw-g-xform} into the duality transformation \eqref{moment-sca} and subsequent use of the Einstein-space 
$H$-eigenvalue problem \eqref{Eins-eigen}.

Using the twisted moment expansions \eqref{tw-mom-exp} and the diagonal monodromy \eqref{x-Mono1} of the twisted Einstein 
coordinates $\hx$, it is easily checked that the selection rules \eqref{tw-mom-sel1} guarantee the correct diagonal monodromies 
of the twisted Einstein tensors $\hG ,\,\hB$ and $\hh$ in Eq.~\eqref{sigma-model-orbifold-action}.

One may also use eigenfields and the principle of local isomorphisms to discuss the phase-space formulation of the sigma model 
orbifolds. In this development, one finds the twisted coordinates and momenta
\begin{subequations}
\begin{gather}
p_i(\xi)' \equiv w_i{}^j p_j(\xi) ,\quad x^i(\xi)' = x^j(\xi) (w\hc)_j{}^{i} \\
\scp_\nrm(\xi) \equiv \schisig_\nrm U(\s)_\nrm{}^i p_i(\xi) ,\quad \scp_\nrm(\xi)' = e^{-\tp\nrrs} \scp_\nrm(\xi) 
   \label{scp-defn} 
\end{gather}
\begin{gather}
\scp \dual \hp ,\quad \sx \dual \hx \label{scp-to-hp}  \\
\hp^\s_\nrm (\xi +2\pi) = e^{-\tp\nrrs} \hp^\s_\nrm (\xi) ,\quad
   \hx_\s^\nrm (\xi +2\pi) = \hx_\s^\nrm (\xi) e^{\tp \nrrs}  \label{s-m-et-b-monos}  
\end{gather}
\end{subequations}
where $\scp_\nrm$ are the eigenmomenta and $\sx^\nrm$ are the eigencoordinates in \eqref{sx-Defn}. Similarly, the principle 
of local isomorphisms gives all local products of twisted classical fields in the sigma model orbifolds. For example, the sigma 
model orbifold Hamiltonian $\hat{H}_\s$ and momentum $\hat{P}_\s$ are those given in Eq.~\eqref{Hamil-dens}, now with the more 
general $\hG$ and $\hb$ of the sigma model orbifold. We must remember of course to include the monodromy factors 
\eqref{de_nr-Props} in the twisted equal-time canonical brackets
\begin{subequations}
\begin{gather}
\{  \hp^\s_\nrm(\xi),\; \hx_\s^\nsn(\e) \} = -i \d_\nrm{}^\nsn \d_{n(r)} (\xi-\e)  \label{hx hp =idel} \\
\{ \hp^\s_\nrm (\xi) ,\hp^\s_\nsn(\e) \} = \{ \hx_\s^\nrm (\xi) ,\hx_\s^\nsn (\e) \} =0
\end{gather}
\end{subequations}
in order to guarantee consistency of the brackets with the monodromies. One may also consider the twisted tangent-space 
formulation of the sigma model orbifolds, starting for example from the untwisted tangent-space formulation in 
Ref.~\cite{deB}.

The sigma model orbifolds described here are in general not conformal at the quantum level. To select those 
orbifolds which are 1-loop conformal, one will presumably need to consider the twisted Einstein equations 
which we present in Subsec.~$4.7$.

\vspace{-.07in}
\subsection{Example: More About WZW Orbifold Geometry}
\label{sigma-model-local-iso}

The WZW orbifolds discussed in Sec.~2 are special cases of the sigma model orbifolds above, but WZW orbifolds have
additional geometric structure which we will study now from the viewpoint of the preceding subsections. In particular,
the moment expansions and explicit functional forms of Subsec.~$4.3$ can be extended in this case to include the group orbifold 
elements and the twisted vielbeins, and all the local twisted phase-space geometry of Sec.~2 can be obtained by the principle of 
local isomorphisms.

We begin by recalling \cite{deBoer:2001nw} the automorphic responses of the untwisted unitary WZW group element $g(T)$ on the 
group manifold $M=G$ and the corresponding tangent-space coordinate $\beta$:
\begin{subequations}
\label{WZW-auto-resp}
\begin{gather}
g(T,\xi,t)' = W(h_\s ;T) g(T,\xi,t) W\hc (h_\s ;T) ,\quad \srange \label{WZW-g-resp} \\
g(T,\xi,t) = e^{i\be^a (\xi,t) T_a} ,\quad \be^a (\xi,t)' = \beta^b (\xi,t) w\hc (h_\s)_b{}^a \label{WZW-beta-resp} \\
W\hc (h_\s ;T) T_a W(h_\s ;T) = \w (h_\s)_a{}^b T_b ,\quad a,b =1,\ldots ,\text{dim } g \,. \label{Linkage}
\end{gather}
\end{subequations}
Here $\w(h_\s)$ is the action of $h_\s \in H \subset Aut(g)$ in the adjoint, $W(h_\s ;T)$ is the action of $h_\s$ in matrix rep $T$ 
of $g$ and the automorphic responses $g'$ and $\beta '$ in ($4.28$a,b) are consistent with each other due to the linkage relation 
\eqref{Linkage}. It is natural to choose the preferred Einstein coordinate system $x=\be$
\begin{subequations}
\label{x=beta}
\begin{gather}
x^i \equiv \be^a e(0)_a{}^{i}, \quad e(0)_i{}^{a} = \d_i{}^{a} ,\quad g(T,\xi,t) = e^{ix^i e(0)_i{}^a T_a} \label{e(0)=1,Sec3} \\
w(h_\s)_i{}^{j} \equiv e(0)_i{}^{a} w(h_\s)_a{}^{b} e(0)_b{}^{j} \\
U\hc(\s)_i{}^{\nrm} \equiv e(0)_i{}^{a} U\hc(\s)_a{}^{\nrm}  , \quad U(\s)_\nrm{}^{i} = U(\s)_\nrm{}^{a} 
  e(0)_a{}^{i} \label{Eins-U} \\
x^i{}'=x^j w\hc(h_\s)_j{}^{i} ,\quad i,j,a,b=1,...,\text{dim }g  \label{Lin-Resp}
\end{gather}
\end{subequations}
in which $e(0)=1$ is the left-invariant vielbein at the origin (see below) and the Einstein-space $H$-eigenvalue problem 
\eqref{Eins-eigen} is isomorphic (with the same $E_{n(r)} (\s)$) to the tangent-space $H$-eigenvalue problem \eqref{H-eig}. 
According to the linear response of the Einstein coordinates in Eq.~\eqref{Lin-Resp}, this is a special case of the preferred 
coordinate system in which our more general sigma model \eqref{s-model-defn}, \eqref{lin-symm-defn} was formulated. 

From these remarks alone, it follows that the WZW orbifolds are described by the special case \eqref{s-model-action} of 
the sigma model orbifold action \eqref{sigma-model-orbifold-action} with the specific forms of $\hG ,\hB$ and $\hh$
\begin{subequations}
\begin{align}
\hG_{\nrm;\nsn} (\hx) &= \he (\hx)_\nrm{}^\ntd \he(\hx)_\nsn{}^\nue \sG_{\ntd;\nue} (\s) \label{G-form-Geom} \\
&= \schisig_\nrm \schisig_\nsn U(\s)_\nrm{}^i U(\s)_\nsn{}^j G_{ij}^{WZW} (x \,\duals \,\sxh(\hx)) \label{G-form-smod} \\
&G_{ij}^{WZW} (x) = e(x)_i{}^a e(x)_j{}^b G_{ab} ,\quad G_{ab} = \oplus_I k_I \eta_{ab}^I
\end{align}
\begin{gather}
\hB_{\nrm;\nsn} (\hx) = \schisig_\nrm \schisig_\nsn U(\s)_\nrm{}^i U(\s)_\nsn{}^j B_{ij}^{WZW} (x \,\duals \,\sxh(\hx)) 
   \label{B-form-smod}
\end{gather}
\begin{gather}
\hh_{\nrm;\nsn;\ntd}(\hx) =\he (\hx)_{\nrm}{}^{\!n(r')\m'} \he (\hx)_\nsn{}^{\!n(s')\n'} \he (\hx)_\ntd{}^{\!n(t')\de'} 
   \scf_{n(r')\m' ;n(s')\n' ;n(t')\de'} (\s) \label{H-form-Geom} \\
\quad = \schisig_\nrm \schisig_\nsn \schisig_\ntd U(\s)_\nrm{}^i U(\s)_\nsn{}^j U(\s)_\ntd{}^k H_{ijk}^{WZW} 
   (x \,\duals\, \sxh(\hx)) \label{H-form-smod} \\
H_{ijk}^{WZW} (x) = e(x)_i{}^a e(x)_j{}^b e(x)_k{}^c f_{abc} ,\quad f_{abc} = f_{ab}{}^d G_{dc}
\end{gather}
\end{subequations}
where $G_{ij}^{WZW} ,B_{ij}^{WZW}$ and $H_{ijk}^{WZW}$ are the fields of the original untwisted WZW model $A_g (H)$.
The forms \eqref{G-form-Geom} and \eqref{H-form-Geom} were given in Sec.~2 and the forms \eqref{G-form-smod}, \eqref{B-form-smod} 
and \eqref{H-form-smod} are special cases of the more general results given for the sigma model orbifolds in 
Eq.~\eqref{form-of-tw-G&B}.

Moving beyond $\hG, \hB$ and $\hh$, we will now apply the method developed above for the sigma model orbifolds to study
the additional geometric structure of the WZW orbifolds -- namely the group orbifold elements $\hg$ and the twisted 
vielbeins $\he$.

The untwisted vielbeins $e,\bar{e}$ are defined as
\begin{equation}
e_i(x,T) = -i g^{-1}(T)\pl_i g(T) = e(x)_i{}^a T_a ,\quad \eb_i(x,T) = -i g(T)\pl_i g^{-1}(T) = 
  \eb(x)_i{}^a T_a 
\end{equation}
so the automorphic responses of the vielbeins
\begin{subequations}
\begin{gather}
e_i(x,T)' = e_i(x',T) = -i g^{-1}(T)' \pl_i{}' g(T)' = w_i{}^j W(h_\s;T) e_j(x,T) W\hc(h_\s;T) \\
\eb_i(x,T)' = \eb_i(x',T) = -i g(T)' \pl_i{}' g^{-1}(T)' = w_i{}^j W(h_\s;T) \eb_j(x,T) W\hc(h_\s;T) \\
(e(x)')_i{}^a{} =  e(x')_i{}^a{} = w_i{}^j e_j{}^b (w\hc)_b{}^a ,\quad (\eb(x)')_i{}^a =  
   \eb(x')_i{}^a = w_i{}^j \eb_j{}^b (w\hc)_b{}^a \label{e'=wewhc}
\end{gather}
\end{subequations}
follow from \eqref{x'=xw}, \eqref{Linkage} and the automorphic response of the group element in Eq.~\eqref{WZW-g-resp}. The corresponding 
eigengroup elements \cite{deBoer:2001nw} $\sg$ and eigenvielbeins $\se,\seb$ are then defined as
\begin{subequations}
\begin{gather}
\sg (\st(T,\s),\xi,t,\s) = U(T,\s) g(T,\xi,t) U\hc (T,\s) \label{sg-Defn} \\
\se_\nrm (\st,\sx) \equiv -i \sg^{-1}(\st,\xi) \pl_\nrm \sg(\st,\xi) \equiv \se(\sx)_\nrm{}^\nsn \st_\nsn , \quad \pl_\nrm \equiv 
   \frac{\pl}{\pl \sx^{\nrm}} \\
\seb_\nrm (\st,\sx) \equiv -i \sg(\st,\xi) \pl_\nrm \sg^{-1}(\st,\xi) \equiv \seb(\sx)_\nrm{}^\nsn \st_\nsn \\
\se(\sx)_\nrm{}^\nsn = \schi_\nrm \schi_\nsn^{-1} U_\nrm{}^i e(x(\sx))_i{}^a(U\hc)_a{}^\nsn 
   \label{eigenvielbein},\quad \se(\sx) = \schi U e(x(\sx)) U\hc \schi^{-1} \\
\seb(\sx)_\nrm{}^\nsn = \schi_\nrm \schi_\nsn^{-1} U_\nrm{}^i \eb(x(\sx))_i{}^a(U\hc)_a{}^\nsn 
   ,\quad \seb(\sx) = \schi U \eb(x(\sx)) U\hc \schi^{-1} 
\end{gather}
\end{subequations}
where $U\hc \equiv U\hc (\s)$ is the eigenvector matrix of the tangent-space $H$-eigenvalue problem \eqref{H-eig}, 
$U \equiv U(\s)$ is the eigenvector matrix \eqref{Eins-U} of the Einstein-space $H$-eigenvalue problem \eqref{Eins-eigen} 
and $U(T,\s)$ is the eigenvector matrix of the extended $H$-eigenvalue problem in \eqref{extHeig}. The twisted representation
matrices $\st \=\st(T,\s)$ are defined in Eq.~\eqref{st-Defn}. The automorphic responses of these eigenfields are
\begin{subequations}
\begin{gather}
\sg (\st(T,\s),\xi,t,\s)' = E(T,\s) \sg(\st(T) ,\xi,t,\s) E(T,\s)^\ast ,\quad \srange \label{sg-resp} \\
\se_\nrm (\sx)' \!=\! E_{n(r)}(\s) E(T,\s) \se_\nrm (\sx) E(T,\s)^\ast \\
\seb_\nrm (\sx)' \!=\! E_{n(r)}(\s) E(T,\s) \seb_\nrm (\sx) E(T,\s)^\ast \\
(\se(\sx)')_\nrm{}^\nsn{} = \se(\sx(x'))_\nrm{}^\nsn = E_{n(r)} (\s) \se(\sx)_\nrm{}^\nsn E_{n(s)} (\s)^\ast \\
(\seb(\sx)')_\nrm{}^\nsn{} = \seb(\sx(x'))_\nrm{}^\nsn = E_{n(r)} (\s) \seb(\sx)_\nrm{}^\nsn E_{n(s)} (\s)^\ast 
\end{gather}
\end{subequations}
where $E_{n(r)} (\s)$ and $E(T,\s)$ are respectively the eigenvalue matrices of the $H$-eigenvalue problem and the 
extended $H$-eigenvalue problem.

For these eigenfields, the principle of local isomorphisms reads
\begin{subequations}
\begin{gather}
 \sx \dual \hx, \quad \sg(\sx) \dual \hg(\hx) ,\quad \se(\sx) \dual \he(\hx), \quad \seb(\sx) \dual \heb(\hx) \\
 \text{ automorphic response }  E_{n(r)}(\s) ,\, E(T,\s) \dual  \text{ monodromy } E_{n(r)}(\s) ,\, E(T,\s)
\end{gather}
\end{subequations}
where $\hg ,\he,\heb$ are respectively the group orbifold elements and the left- and right-invariant twisted vielbeins. 
This gives the monodromy \eqref{gMono} of $\hg$ and the monodromies \eqref{heMonos} of $\he,\heb$, as well as the explicit 
functional forms of the group orbifold elements and the twisted vielbein
\begin{subequations}
\begin{gather}
\hg(\st(T),\xi,t,\s) = U(T,\s) g(T,x \,\duals \sxh(\hx)) U\hc (T,\s) = e^{i\hx_\s^\nrm (\xi,t) \st_\nrm (T,\s)} \label{hg-form} \\
\hx^\nrm_\s = \schisig_\nrm^{-1} \sxh^i U\hc(\s)_i{}^\nrm 
\end{gather}
\begin{gather}
\he(\hx_\s ,\s)_\nrm{}^\nsn \= \schisig_\nrm \schisig_\nsn^{-1} U (\s)_\nrm{}^i \,e(x \,\duals \sxh_\s (\hx_\s) )_i{}^a 
   U\hc (\s)_a{}^\nsn \label{he-form} \\
\he(\hx)\=\schi U e (x \,\duals \sxh(\hx))  U\hc \schi^{-1} 
\end{gather}
\end{subequations}
in terms of the functional forms of the untwisted group element $g(T,x)$ and the untwisted vielbein $e_i{}^{a}(x)$. The 
coordinates $\sxh(\hx)$ with twisted boundary conditions are defined in \eqref{x-to-sx}, the arrow notation is defined in 
\eqref{f(x-goto-hx)} and $\he ,\heb$ are related to $\hg$ as given in \eqref{Def-twVB} and \eqref{Def-twVB2}. The result 
\eqref{he-form} also holds for $\heb(\hx)$ with $e \goto \eb$. 

For the group orbifold elements $\hg$, the moment expansion in terms of the twisted Einstein coordinates $\hx$ is very simple
\begin{equation}
\hg(\st,\xi,t,\s) = \sum_{m=0}^\infty \frac{i^m}{m!} \st_{n(r_1)\m_1}(T,\s) ... \st_{n(r_m)\m_m}(T,\s) \hx_\s^{n(r_1)\m_1} (\xi)
\cdots \hx_\s^{n(r_m)\m_m} (\xi)
\end{equation}
and, as in the preceding subsection, the moments $\st_{n(r_1)\m_1} ...\st_{n(r_m)\m_m}$ of the group
orbifold elements satisfy the selection rules
\begin{equation}
E_{n(r_1)} (\s)^\ast ... E_{n(r_m)} (\s)^\ast \st_{n(r_1)\m_1} ... \st_{n(r_m)\m_m} = E(T,\s) \st_{n(r_1)\m_1} ...
   \st_{n(r_m)\m_m} E(T,\s)^\ast
\end{equation}
which follow from the selection rule \eqref{T-select} for $\st$. Moreover, as in Subsec.~$4.4$, these selection rules and the diagonal monodromy 
\eqref{x-Mono1} of $\hx$ guarantee the diagonal monodromy \eqref{gMono} of $\hg$. 

We may also introduce the moment expansions of the untwisted vielbein $e$ and the twisted vielbein $\he$
\vspace{-0.1in}
\begin{subequations}
\begin{gather}
e(x)_i{}^a = \sum_{m=0}^\infty e_{i,i_1...i_m}{}^{a} x^{i_1}\cdots x^{i_m} \\
(w\hc)_i{}^j (w\hc)_{i_1}{}^{j_1} \cdots (w\hc)_{i_m}{}^{j_m} e_{j,j_1...j_m}{}^{b} w_b{}^a = 
   e_{i,i_1...i_m}{}^{a} \label{wdw=d} \\
\he(\hx)_\nrm{}^\nsn =\sum_{m=0}^\infty \se_{\nrm,n(r_1)\m_1...n(r_m)\m_m}{}^{\nsn} \hx_\s^{n(r_1)\m_1}\cdots 
   \hx_\s^{n(r_m)\m_m} 
\end{gather}
\end{subequations}
where the $H$-invariance \eqref{wdw=d} of the untwisted vielbein moments follows from the automorphic responses \eqref{Lin-Resp} and 
\eqref{e'=wewhc} of $x$ and $e$. Then, we may use \eqref{x-to-sx} and \eqref{he-form} to obtain the twisted vielbein moments as an 
infinite set of duality transformations 
\begin{align}
& \se_{\nrm,n(r_1)\m_1...n(r_m)\m_m}{}^{\!\!\!\nsn} \!\=\! \schi_\nrm \schi_\nsn^{-1} U_\nrm{}^{\!i}
   \!\left(\,\,\prod_{j=1}^m \schi_{n(r_j)\m_j}U_{n(r_j)\m_j}{}^{i_j}\!\right)\! e_{i,i_1...i_m}{}^{\!a} 
   (U\hc)_a{}^{\!\nsn} \label{dual-of-d}
\end{align}
which express the twisted moments in terms of the untwisted moments. Similarly, we may use Eqs.~\eqref{wdw=d}, \eqref{dual-of-d} and both 
$H$-eigenvalue problems \eqref{H-eig} and \eqref{Eins-eigen} to obtain the selection rules for the twisted vielbein moments:
\begin{subequations}
\begin{gather}
\se_{\nrm,n(r_1)\m_1...n(r_m)\m_m}{}^{\nsn}(1-E_{n(r)}(\s)^\ast E_{n(r_1)}(\s)^\ast \cdots E_{n(r_m)}(\s)^\ast 
   E_{n(s)}(\s) ) = 0 \label{d(1-E)=0} \\
\se_{\nrm,n(r_1)\m_1...n(r_m)\m_m}{}^{\!\!\nsn} \propto \d_{n(r)-n(s)+\sum_{i=1}^m \!n(r_i),0\,\text{mod}
   \,\r(\s) } \,. \label{d=delta-d}
\end{gather}
\end{subequations}
As expected, these selection rules and the diagonal monodromy \eqref{hx-mono} of $\hx(\xi)$ guarantee the diagonal monodromy 
\eqref{he-Mono1} of the twisted vielbein, and similar moment expansions lead to duality transformations and 
selection rules which guarantee the monodromy of $\heb(\hx)$.

We remark in particular that the duality transformations \eqref{dual-of-d} and the untwisted boundary condition
\eqref{e(0)=1,Sec3} imply 
\begin{subequations}
\begin{gather}
e(0)_i{}^{a} = e_i{}^{a} = \d_i{}^{a} \Rightarrow 
\end{gather}
\begin{align}
& \!\!\!\!\he(0)_\nrm{}^{\!\!\!\nsn} \= \se_\nrm{}^{\!\!\!\nsn} \= \se(0)_\nrm{}^{\!\!\!\nsn} \= \schi_\nrm \schi^{-1}_\nsn 
   U_\nrm{}^{i} e(0)_i{}^{\!a} (U\hc)_a{}^{\!\!\nsn} \= \d_\nrm{}^{\!\!\!\nsn}
\end{align}
\end{subequations}
so that the preferred twisted coordinate system $\hx=\bh ,\, \he(0)=1$ chosen in Subsec.~$2.4$ is the orbifold dual 
of the preferred untwisted coordinate system $x=\b ,\, e(0)=1$ chosen in Eq.~\eqref{x=beta} for the symmetric WZW model. 

Using \eqref{e(0)=1,Sec3}, \eqref{eigenvielbein} and \eqref{he-form}, the vielbein, the eigenvielbein and the twisted vielbein 
can also be evaluated in closed form
\begin{subequations}
\begin{gather}
e(x)_i{}^a = e(0)_i{}^b \left( \smal{ \frac{e^{iY(x)}-1}{iY(x)} } \right)_{\!b}^{\,\,a} \!,\quad Y(x) \equiv x^i e(0)_i{}^a T_a^{adj} 
   ,\quad e(0)=1 \\
 \se(\scrs{\sx}) =  \smal{ \frac{e^{i\sY({\sxtiny})}-1}{i\sY(\sxtiny)} },\quad \sY(\sx) \equiv \schi U Y(x(\sx)) U\hc \schi^{-1} = \sx^\nrm \tst_\nrm(T^{adj}) ,\quad \se(0)=1
\end{gather}\vspace{-.35in}
\begin{gather}
 \he(\hx) = \smal{ \frac{e^{i\hY(\hx)}-1}{i\hY(\hx)} },\quad \hY(\hx(\xi)) = \hx_\s^\nrm (\xi) \tst_\nrm(T^{adj},\s),\quad \he(0)=1
\end{gather}
\end{subequations}
in agreement with Eqs.~\eqref{exp-form-of-ho} and \eqref{exp-form-of-ehat}. 

Finally, the canonical form \eqref{hb-form-of-twisted-currents} of the twisted currents follows by local isomorphisms 
from Bowcock's untwisted canonical form \cite{Bow}
\vspace{-.04in}
\begin{subequations}
\label{Bowcock}
\begin{gather}
J_a(\xi)=2\p e(x(\xi))_a{}^i p_i(B,\xi)+\half \pl_\xi x^i(\xi) e(x(\xi))_i{}^b G_{ba}\\
\bj_a(\xi)=2\p \eb(x(\xi))_a{}^i p_i(B,\xi)-\half  \pl_\xi x^i(\xi) \eb(x(\xi))_i{}^b G_{ba} \\
p_i(B,\xi)= p_i(\xi)+\frac{1}{4\p}B_{ij}(x(\xi)) \pl_\xi x^j(\xi) \\
\pl_i B_{jk}(x) + \pl_j B_{ki}(x) + \pl_k B_{ij}(x) = -i\, Tr \big{(}\,M(k,T)  e_i(x,T) [e_j(x,T), 
   e_k(x,T)] \,\big{)} \,.
\end{gather}
\end{subequations}
To see this, one checks first that the eigencurrents $\sj, \sjb$ have the same form 
\begin{subequations}
\begin{eqnarray}
\sj_\nrm (\xi,t,\s) &=& 2\pi\se^{-1} (\sx(\xi))_\nrm{}^\nsn \scp_\nsn (\sb,\xi) \nn \\
   & &+ \frac{1}{2} \pl_\xi \sx^\nsn \se(\sx(\xi))_\nsn{}^\ntd \sG_{\ntd;\nrm} (\s) \\
\sjb_\nrm (\xi,t,\s) &=& 2\pi\seb^{-1} (\sx(\xi))_\nrm{}^\nsn \scp_\nsn (\sb,\xi) \nn \\
   & & - \frac{1}{2} \pl_\xi \sx^\nsn \seb(\sx(\xi))_\nsn{}^\ntd \sG_{\ntd;\nrm} (\s) \\
\scp_\nrm (\sb,\xi) &=& \scp_\nrm^\s (\xi) + \frac{1}{4\pi} \sb_{\nrm;\nsn} (\sx(\xi)) \pl_\xi \sx^\nsn
\end{eqnarray}
\end{subequations}
when rewritten in terms of the eigenfields $\sx, \scp, \se,\seb$ and $\sb$, and then the principle of local isomorphisms
gives the canonical form of the twisted currents in Eq.~\eqref{hb-form-of-twisted-currents}. Similarly, the familiar 
coordinate-space form of the untwisted WZW currents
\begin{gather}
J_a(\xi) = \half  \pl_+x^i(\xi) e(x(\xi))_i{}^b  G_{ba}  ,\quad \bj_a(\xi) = \half \pl_-x^i(\xi) 
   \eb(x(\xi))_i{}^b G_{ba}
\end{gather}
leads directly to the coordinate-space form of the twisted currents in Eq.~\eqref{coord-form-hj}.

\vspace{-.07in}
\subsection{Example: The General Coset Orbifold}

The general coset orbifold action \cite{Halpern:2002hw} was obtained by gauging the general WZW orbifold action by general twisted gauge groups.
In this section, we will use the general coset orbifold action to find the specific form of the sigma model orbifold action 
\eqref{sigma-model-orbifold-action} which describes the general coset orbifold.

The general coset orbifold $A_{g/h}(H)/H$ has been discussed at the operator level in Refs.~\cite{Evslin:1999ve,Halpern:2000vj,Halpern:2002hw}. 
In these orbifolds $A_{g/h}(H)$ is any coset construction \cite{Bardakci:1971nb,Halpern:1971ay,GKO,Halpern:1996js} with a discrete symmetry 
group $H$, and the untwisted coset construction is mapped by local isomorphisms to the twisted coset constructions of the orbifold
\begin{equation}
\frac{g}{h} = \frac{g}{h} (H) = \frac{g(H)}{h(H)} \dual \frac{\gfrakh (\s)}{\hfrakh (\s)}  ,\quad \srange
\end{equation}
where $\hfrakh (\s) \!\subset \!\gfrakh (\s)$ are the twisted affine algebras of sector $\s$.

At the classical level, the corresponding orbifold Lie algebras $\hat{h}(\s) \!\subset \!\hg(\s)$ of the coset 
orbifold
\begin{equation}
\frac{\gfrakh (\s)}{\hfrakh (\s)} \leftrightarrow \frac{\hg (\s)}{\hat{h} (\s)}
\end{equation}
are used to label the {\it general coset orbifold action} \cite{Halpern:2002hw}
\begin{align} 
&\widehat{S}_{\hg(\s)/\hat{h}(\s)}[\sm, \hg, \hat{A}_{\pm}] =  \label{empty} \\
& \quad = \widehat{S}_{\hg(\s)}[\sm, \hg] + \frac{1}{4\pi}\!\int \!\!d^2 \xi \0b \widehat{Tr} \Big(\sm \big(
   \hat{g}^{-1} \pl_+ \hg (i \hat{A}_-) -\!i \hat{A}_+ \pl_-\hg \hat{g}^{-1} -\!\hat{g}^{-1} \hat{A}_+ 
   \hg \hat{A}_- +\!\hat{A}_+ \hat{A}_-\big)\Big)\, \nn
\end{align}
which describes sector $\s$ of the general coset orbifold $A_{g/h} (H)/H.$ Here $\hg,\sm$ and $\widehat{S}_{\hg(\s)}[\M, \hg]$ are
respectively the group orbifold element, the twisted data matrix and the WZW orbifold action \eqref{gp-orbel-form} for sector 
$\s$ of the WZW orbifold $A_g (H)/H$, while $\{ \ha_\pm \}$ with $[\ha_\pm ,\sm ]=0$ are the {\it twisted matrix gauge fields} 
valued on the orbifold Lie subalgebra $\hat{h} (\s)\! \subset \!\hg(\s)$. The general coset orbifold action reduces in the untwisted sector 
$\s=0$ to the ordinary gauged WZW action [41-45] for $A_{g/h}(H)$. Many other properties of this action are discussed in 
Ref.~\cite{Halpern:2002hw}, but we note here only that the general coset action has trivial monodromy under the field monodromies
\begin{subequations}
\begin{gather}
\hg (\st,\xi +2\pi,t,\s) = E(T,\s) \hg (\st,\xi,t,\s) E(T,\s)^\ast \\
\ha_\pm (\st,\xi +2\pi,t,\s) = E(T,\s) \ha_\pm (\st,\xi,t,\s) E(T,\s)^\ast
\end{gather}
\end{subequations}
as well as a general twisted gauge invariance. Our task here is to integrate out the twisted matrix gauge fields $\{\hat{A}_\pm \}$
to find the equivalent twisted sigma model form of the general coset orbifold action. 

We begin by reminding the reader that the twisted $\hfrakh (\s)$ currents $\hj_{\hnrhm}$ and the 
twisted representation matrices $\st_\hnrhm$ of the orbifold Lie subalgebra $\hat{h} (\s)$ have the form
\begin{subequations}
\begin{gather}
\hj_{\hnrhm} (\xi,t,\s) \equiv R_r(\s)_{\hat{\m}}{}^\m \hj_\hnrm (\xi,t,\s) ,\quad \forall \hat{n}(r) ,\hat{\m} \in \hat{h}(\s) \subset \hg(\s) \\
\st_\hnrhm (T,\s) \equiv R_r(\s)_{\hat{\m}}{}^\m \st_\hnrm (T,\s) \\
\ha_\pm (T,\xi,t,\s) = \ha_\pm^\hnrhm (\xi,t,\s) \st_\hnrhm (T,\s) \in \hat{h}(\s) \\
\{ \hat{n} (r) \} \subset \{ n(r) \} ,\quad \text{dim} \{\hat{\m} \} \leq \text{dim} \{\m \}
\end{gather}
\end{subequations}
where $\hj_\nrm$ are the $\gfrakh (\s)$ currents, $\st_\nrm$ are the twisted representation matrices of the orbifold Lie 
algebra $\hg(\s)$, and $R(\s)$ is the embedding matrix \cite{Evslin:1999ve,Halpern:2000vj,Halpern:2002hw} of the twisted affine subalgebra 
$\hfrakh (\s) \!\subset \!\gfrakh(\s)$. With Ref.~\cite{Halpern:2002hw} we emphasize that the twisted affine algebras and their corresponding 
orbifold Lie algebras
\begin{subequations}
\begin{gather}
[\hj_{\sgb (\s)} (\cdot) , \hj_{\sgb (\s)} (\cdot) ] = i\scf_{\sgb (\s)} \hj_{\sgb (\s)} (\cdot) +\sG_{\sgb (\s)} \longrightarrow
[\st_{\hg (\s)} ,\st_{\hg (\s)} ]=i\scf_{\sgb (\s)} \st_{\hg (\s)} \\
[\hj_{\hfrakh (\s)} (\cdot) ,\hj_{\hfrakh (\s)} (\cdot) ]=i\scf_{\hfrakh (\s)} \hj_{\hfrakh (\s)} (\cdot) +
   \sG_{\hfrakh (\s)} \longrightarrow
[\st_{\hat{h} (\s)} ,\st_{\hat{h} (\s)}] = i\scf_{\hfrakh (\s)} \st_{\hat{h} (\s)}
\end{gather}
\end{subequations}
share the same twisted structure constants $\scf_{\gfrakh (\s)}$ and $\scf_{\hfrakh (\s)}$.

We will also need the definitions
\begin{subequations}
\begin{align}
\sG_{\nrm;\hnshn} (\s) &\equiv \widehat{Tr} \left(\sm (T,\s) \st_\nrm (T,\s) \st_\hnshn (T,\s) \right) \nn \\
& = R_s(\s)_{\hat{\n}}{}^\n \sG_{\nrm ;\hnsn} (\s) = R_s(\s)_{\hat{\n}}{}^\n \sG_{\hnsn;\nrm}(\s) \equiv \sG_{\hnshn ;\nrm} (\s) \\
\sG_{\hnrhm ;\hnshn} (\s) & \equiv \widehat{Tr} \left(\sm (T,\s) \st_\hnrhm (T,\s) \st_\hnshn (T,\s) \right) \nn \\
&= R_r(\s)_{\hat{\m}}{}^\m R_s(\s)_{\hat{\n}}{}^\n \sG_{\hnrm ;\hnsn} (\s) = \sG_{\hnshn ;\hnrhm} (\s)
\end{align}
\begin{gather}
\sG_{\hnrhm ;\nsn} (\s) = \de_{\hat{n}(r) +n(s),0\, \text{mod }\r(\s)} \sG_{\hnrhm ;-\hat{n}(r),\n}(\s) \\
\sG_{\hnrhm ;\hnshn} (\s) =\de_{\hat{n}(r) +\hat{n}(s),0\, \text{mod }\r(\s)} \sG_{\hnrhm ;-\hat{n}(r) ,\hat{\n}} (\s) 
\end{gather}
\begin{align}
& \ho (\hx)_\hnrhm{}^{\!\nsn} \!\equiv \!R_r(\s)_{\hat{\m}}{}^\m \ho (\hx)_\hnrm{}^{\!\nsn} ,\quad \ho (\hx)_{\hnrhm ;\hnshn} 
   \!\equiv \!\ho(\hx)_\hnrhm{}^{\!\ntd} \sG_{\ntd ;\hnshn} (\s) 
\end{align}
\end{subequations}
where $\sm (T,\s)$ is the twisted data matrix in \eqref{st-Props}, $\sG_{\hfrakh (\s)} \simeq \{\sG_{\hnrhm ;\hnshn} 
(\s)\}$ is the induced metric on the twisted affine subalgebra $\hfrakh (\s)$, and $\ho$ is the twisted adjoint action
in \eqref{ho-Defn}. The induced metric $\sG_{\hfrakh (\s)}$ is invertible in the $\hat{h}(\s)$ subspace \cite{Halpern:2000vj}
\begin{equation}
\sG_{\hnrhm ;\hat{n}(t)\hat{\de}} (\s) \sG^{\hat{n}(t)\hat{\de} ;\hnshn} (\s) = \de_{\hat{\m}}{}^{\hat{\n}} 
   \de_{\hat{n}(r) -\hat{n}(s),0\, \text{mod }\r(\s)}
\end{equation}
when the original $H$-symmetric coset construction $\frac{g}{h}(H)$ was a reductive coset space.

Following the conventional procedure, we may integrate out the twisted matrix gauge fields by solving their equations
of motion:
\begin{subequations}
\begin{gather}
\widehat{Tr} \left\{ \sm \left( -i\pl_- \hg \hg^{-1} -\hg \ha_- \hg^{-1} +\ha_- \right) \st_\hnrhm \right\} =0 \quad \quad \quad \\
\widehat{Tr} \left\{ \sm \left( -i\hg^{-1} \pl_+ \hg -\hg^{-1} \ha_+ \hg +\ha_+ \right) \st_\hnrhm \right\} =0  ,\quad 
\forall \hat{n}(r) ,\hat{\m} \in \hat{h}(\s) \,.
\end{gather}
\end{subequations}
After some algebra, we find that sector $\s$ of the general coset orbifold $A_{g/h} (H)/H$ is described by the twisted 
sigma model action \eqref{sigma-model-orbifold-action} with twisted Einstein metric $\hG$ and twisted $B$ field $\hB$:
\begin{subequations}
\begin{align} 
\!\!\!\!\left( \hG (\hx) +\hb(\hx) \right)&\!{}_{\nrm;\nsn} = \left( \hG(\hx) +\hb(\hx) \right)^{WZW}_{\nrm;\nsn} \nn \\
  \quad \quad -2 \Big{(} \he (\hx) & \!{}_\nrm {}^\ntd \sG_{\ntd ;\hnvhz} (\s) \Big{)} \Big{(} \heb (\hx)_\nsn{}^\nue 
  \sG_{\nue ;\hnwhl} (\s) \Big{)} X^{-1} (\hx)^{\hnwhl ;\hnvhz} 
\end{align}
\begin{align}
& X_{\hnrhm ;\hnshn} (\hx) \!\equiv \!X_{\hnrhm;\hnshn} (\hx_\s,\s) \!\equiv \!(\sG (\s) \!- \!\ho (\hx_\s,\s) )_{\hnrhm ;\hnshn} 
   ,\,\,\,\s =0,..., N_c -1 . \label{X-Defn}
\end{align}
\end{subequations}
The matrix $X(\hx)$ in \eqref{X-Defn} is invertible in the $\hat{h} (\s)$ subspace because $\sG_{\hnrhm ;\hnshn} (\s)$ is 
invertible. This set of twisted sigma models should be considered in a fixed gauge such as
\begin{align}
0=\widehat{Tr} \left( \sm \hx_\s^\nsn (\xi,t) \st_\nsn \st_\hnrhm \right) & = \hx_\s^\nsn (\xi,t) \sG_{\nsn ;\hnrhm} (\s) \nn \\
   &= \hx_\s^{\hat{n}(r) \n} (\xi,t) \sG_{\hat{n}(r)\n ;-\hat{n}(r),\hat{\m}} (\s) ,\quad \forall \hat{n}(r) ,\hat{\m} \in \hat{h}(\s)
\end{align}
where we have chosen $\hx = \hbe$, as above.

The general forms of other twisted tensors can be obtained in these twisted coset sigma models, including e.g. the separate 
forms of $\hG$ and $\hb$ and the twisted Riemann tensor of $\hG$, but we refrain for brevity from doing so here. Using the coset 
orbifold examples discussed in Ref.~\cite{Halpern:2002hw}, it is also straightforward to work out the twisted Einstein tensors explicitly 
for a large number of special cases. The contribution of the orbifold dilaton in the coset orbifolds is included in the 
discussion of the following subsection.

\vspace{-.07in}
\subsection{The Twisted Einstein Equations}
\label{OEE}

It is well-known [21-26] that the general nonlinear sigma model $A_M$ in \eqref{s-model-defn} is 1-loop conformal
when the sigma-model Einstein equations
\begin{gather}
 R_{ij} + \frac{1}{4}H_{ki}{}^{l} H^{k}{}_{lj} - 2\nabla_i \nabla_j \phi = 0 ,\quad (\nabla^k -2\nabla^k \phi ) H_{kij} = 0 \label{untwisted-Einstein}
\end{gather}
are satisfied, where $\phi$ is the dilaton field.

Using eigenfields and local isomorphisms (see below), we have been able to find a set of 
{\it twisted Einstein equations} in sector $\s$ of each sigma model orbifold $A_M (H)/H$
\begin{subequations}
\label{orbifold-EE}
\begin{gather}
\hat{R}_{\nrm;\nsn}(\hx) + \frac{1}{4} \hh_{\ntd;\nrm}{}^\nue (\hx) \; \hh^\ntd{}_{\nue;\nsn} (\hx) 
   -2 \hat{\nabla}_\nrm  \hat{\nabla}_\nsn \hphi(\hx) = 0 \\
\big{(}\,\hat{\nabla}^\ntd -2 \hat{\nabla}^\ntd \hphi(\hx)\, \big{)}\; \hh_{\ntd;\nrm;\nsn}(\hx) = 0, 
   \quad \srange \\
\hx \equiv \hx_\s, \quad \hat{R}(\hx) \equiv \hat{R}(\hx_\s,\s), \quad \hh(\hx) \equiv \hh(\hx_\s,\s), 
   \quad \hphi(\hx) \equiv \hphi(\hx_\s,\s)
\end{gather}
\end{subequations}
which are satisfied when the Einstein equations \eqref{untwisted-Einstein} hold in the symmetric sigma model $A_M(H)$. 
Here $\hx$ are the twisted Einstein coordinates, $\hh(\hx)$ is the twisted torsion field \eqref{hh-Defn}, and $\hat{\phi}(\hx)$ is 
the orbifold dilaton. Moreover, it turns out that all the twisted tensors in \eqref{orbifold-EE} are constructed in 
analogy with the usual untwisted Einstein tensors, following the {\it duality algorithm} \cite{deBoer:1999na,Halpern:2000vj}
\vspace{-.12in}
\begin{subequations}
\label{dual-algorithm} 
\begin{gather}
i \dual \nrm,\quad x^i \dual \hx_\s^\nrm ,\quad \pl_i \dual \hat{\pl}_\nrm, \quad  \nabla_i \dual 
   \hat{\nabla}_\nrm \\
G_{ij}(x) \dual \hG_{\nrm;\nsn}(\hx),\quad   G^{ij}(x) \dual \hG^{\nrm;\nsn}(\hx) \\
\Gamma_{ij}{}^k(x) \dual \hat{\Gamma}_{\nrm;\nsn}{}^\ntd(\hx) ,\quad R_{ijkl}(x) \dual \hat{R}_{\nrm;\nsn;
   \ntd;\nue}(\hx) 
\end{gather}
\end{subequations}
where, as above, the indices $n(r),\m$ are the spectral indices and degeneracy labels of the Einstein-space $H$-eigenvalue 
problem \eqref{Eins-eigen}. For example, one finds that
\begin{subequations}
\begin{gather}
\hh_{\nrm;\nsn}{}^\ntd (\hx) \equiv \hh_{\nrm;\nsn;\nue} (\hx) \hG^{\nue;\ntd} (\hx) \\
\hat{\Gamma}_{\nrm;\nsn}{}^\ntd(\hx) \equiv \half \big{(}\,\hat{\pl}_{(\nrm} \hG_{\nsn);\nue}(\hx) - 
   \hat{\pl}_\nue \hG_{\nrm;\nsn}(\hx)\,\big{)}\,\hG^{\nue;\ntd}(\hx) 
\end{gather}
\end{subequations}
where $\hat{\Gamma}$ is the twisted Christoffel connection, and all indices in \eqref{orbifold-EE} are raised and 
lowered with the twisted Einstein metric $\hG_\bullet$ and its inverse $\hG^\bullet$.
Other examples include the twisted Ricci tensor and the covariant derivatives of the orbifold dilaton
\begin{subequations}
\begin{gather}
\hat{R}_{\nrm;\nsn}(\hx) \equiv \hat{R}_{\nrm;\nsn;\ntd;\nue}(\hx)\hG^{\ntd;\nue}(\hx) \label{Ricci-Defn} \\
\hat{\phi}_\nrm(\hx) \equiv \hat{\nabla}_\nrm\hat{\phi}(\hx) = \hat{\pl}_\nrm \hat{\phi}(\hx) \\
\hat{\nabla}_\nrm \hat{\phi}_\nsn(\hx) = \hpl_\nrm \hat{\phi}_\nsn(\hx) - \hat{\Gamma}_{\nrm;\nsn}
   {}^{\ntd}(\hx)\hat{\phi}_\ntd(\hx) 
\end{gather}
\end{subequations}   
where the four-index symbol in \eqref{Ricci-Defn} is the twisted Riemann tensor -- constructed in the usual way from $\hat{\Gamma}$.
   
In this development, we also obtain the diagonal monodromies of all the twisted fields, including the monodromies of 
$\hG_{\bullet}, \,\hB$ and $\hh$ in Eq.~\eqref{sigma-model-orbifold-action}, and for example:
\begin{subequations}
\begin{gather}
 \hx^\nrm(\xi + 2\p) = \hx^\nrm (\xi) e^{2\p i\frac{n(r)}{\r(\s)}} \\
\hat{\phi}(\hx(\xi+2\p)) = \hat{\phi}(\hx(\xi)) ,\quad
   \hat{\phi}_\nrm(\hx(\xi+2\p)) = e^{-2\p i\frac{n(r)}{\r(\s)}}\hat{\phi}_\nrm(\hx(\xi)) \label{phi-monos} \\
\hat{\nabla}_\nrm \hat{\phi}_\nsn(\hx(\xi+2\p)) = e^{-2\p i\frac{n(r)+n(s)}{\r(\s)}} \hat{\nabla}_\nrm 
   \hat{\phi}_\nsn(\hx(\xi)) \label{phi-monos2} \\
\hG^{\nrm;\nsn}(\hx(\xi+2\p)) =\hG^{\nrm;\nsn}(\hx(\xi))  e^{2\p i\frac{n(r)+n(s)}{\r(\s)}} 
\end{gather}
\begin{gather}
\hat{\Gamma}_{\nrm;\nsn}{}^\ntd (\hx(\xi+2\p)) = e^{-2\p i\frac{n(r)+n(s)-n(t)}{\r(\s)}} 
   \hat{\Gamma}_{\nrm;\nsn}{}^\ntd (\hx(\xi)) \\
\hat{R}_{\nrm;\nsn;\ntd;\nue}(\hx(\xi+2\p)) = e^{-2\p i\frac{n(r)+n(s)+n(t)+n(u)}{\r(\s)}} 
   \hat{R}_{\nrm;\nsn;\ntd;\nue}(\hx(\xi)) \\
\hat{R}_{\nrm;\nsn}(\hx(\xi+2\p)) = e^{-2\p i\frac{n(r)+n(s)}{\r(\s)}} \hat{R}_{\nrm;\nsn}(\hx(\xi)) \,.
\end{gather}
\end{subequations} 
Moreover, as given for $\hG_{\bullet} ,\hB$ and $\hh$ in Eq.~\eqref{form-of-tw-G&B}, we obtain the explicit forms 
of all the twisted Einstein tensors, including:
\begin{subequations}
\label{explicit-forms-of-EE-tensors}
\begin{gather}
\hat{\phi}(\hx) = \phi(x \,\duals\, \sxh(\hx)) \\
\hG^{\nrm;\nsn}(\hx) = \schi_\nrm^{-1} \schi_\nsn^{-1} G^{ij} (x \,\duals\, \sxh(\hx)) (U\hc)_i{}^\nrm (U\hc)_j{}^\nsn \\
\hat{\Gamma}_{\nrm;\nsn}{}^\ntd(\hx) = \schi_\nrm \schi_\nsn \schi_\ntd^{-1} U_\nrm{}^i U_\nsn{}^j
   \Gamma_{ij}{}^k (x \,\duals\, \sxh(\hx)) (U\hc)_k{}^\ntd 
\end{gather}
\begin{gather}
\hat{R}_{\nrm;\nsn;\ntd;\nue} (\hx) \bigspc \bigspc \bigspc \bigspc \bigspc \\
\bigspc \!=\! \schi_\nrm \schi_\nsn \schi_\ntd \schi_\nue U_\nrm{}^i U_\nsn{}^j U_\ntd{}^k
   U_\nue{}^l R_{ijkl} (x \,\duals\, \sxh(\hx)) \nn \\
\hat{R}_{\nrm;\nsn} (\hx)= \schi_\nrm \schi_\nsn U_\nrm{}^i U_\nsn{}^j (R_{ijkl}G^{kl}) (x \,\duals\, \sxh(\hx)) \\
U_\nrm{}^{l} \equiv U(\s)_\nrm{}^{l}, \quad (U\hc)_l{}^\nrm \equiv U\hc(\s)_l{}^\nrm ,\quad 
   \schi_\nrm \equiv \schisig_\nrm \, . 
\end{gather}
\end{subequations}
In these results, the quantities $\sxh(\hx)$ are the coordinates with twisted boundary conditions, whose explicit form is given in 
Eq.~\eqref{x-to-sx} -- so that once again each twisted tensor is completely determined as a function of the twisted coordinates 
$\hx$, given the coordinate dependence of the untwisted Einstein tensors. As in Subsec.~$4.4$, selection rules 
which guarantee the indicated monodromies can be obtained for the moments of each twisted Einstein tensor.

The derivation of the results above proceeds as follows: In Eq.~\eqref{lin-symm-defn} we specified the automorphic 
response of the untwisted objects $x,G_{\bullet},B$ and $H$. From these transformations we easily deduce the 
automorphic response of all the other tensors in the symmetric sigma model. The conclusion is that all objects in 
the symmetric theory transform under the symmetry group $H$ with a factor $w$ on the left for a down index 
and a factor of $w\hc$ on the right for an up index. For example:
\vspace{-.05in}
\begin{subequations}
\begin{gather}
G^{ij}(x)' = G^{ij}(x') = G^{kl}(x)(w\hc)_k{}^i (w\hc)_l{}^j ,\quad \nabla_i{}' =  w_i{}^j \nabla_j \\
\Gamma_{ij}{}^l(x)' = \Gamma_{ij}{}^l(x') = w_i{}^p w_j{}^q \Gamma_{pq}{}^s(x) (w\hc)_s{}^l ,\quad 
   \phi(x)' = \phi(x') = \phi(x) \\
R_{ijkl} (x)' =R_{ijkl} (x') = w_i{}^{i'} w_j{}^{j'} w_k{}^{k'} w_l{}^{l'} R_{i'j'k'l'} (x) \,.
\end{gather}
\end{subequations}
This tells us that the corresponding eigenfields should be defined with a factor $\schi U$ or $\schi^{-1} 
U\hc$ for each up or down index respectively. It is then straightforward to verify that the Einstein equations 
have the same form in terms of the eigenfields as they do in terms of the original fields, although the indices 
$i \rightarrow \nrm$ of the eigenfields are now different. Then the twisted Einstein equations \eqref{orbifold-EE} and 
the explicit functional forms \eqref{explicit-forms-of-EE-tensors} of the twisted tensors follow by local isomorphisms. 
These explicit forms can also be used to check directly that the twisted Einstein equations hold when the untwisted 
Einstein equations are satisfied. Local isomorphisms also tell us that each twisted down index carries a monodromy 
$e^{-2\p i\frac{n(r)}{\r(\s)}}$ and each twisted up index carries a monodromy $e^{+2\p i\frac{n(r)}{\r(\s)}}$, 
so that objects which are scalars, such as the orbifold curvature scalar $\hat{R}$ and the orbifold dilaton 
$\hat{\phi}$, have trivial monodromy.

The twisted Einstein equations \eqref{orbifold-EE} are almost certainly the condition that twisted sector
$\s$ of a sigma model orbifold is conformal, but a direct check (a one-loop computation using the orbifold 
sigma model action $\hat{S}_\s^{NLS}$ in \eqref{sigma-model-orbifold-action}) is necessary to be absolutely sure. 
This situation is analogous to that of the general {\it orbifold Virasoro master equation} \cite{Evslin:1999qb,
deBoer:1999na,Halpern:2000vj} which follows from local isomorphisms, but which also needed a direct check of
the Virasoro algebra using orbifold OPE's. It is expected that this check will go through for the sigma model 
orbifolds because (as seen for the general orbifold Virasoro master equation) the short-distance singularities 
in the untwisted and twisted sectors differ only by the duality algorithm \eqref{dual-algorithm}.

Finally, it would be interesting to study possible transitions among the $N_c$ orbifold sectors (space-times) induced
by the existence of twist fields.

\vspace{-.02in}
\bigskip

\noindent
{\bf Acknowledgements} 

For helpful discussions, we thank O. Ganor, N. Obers, C. Park, E. Rabinovici, F. Wagner and J. Wang.  

This work was supported in part by the Director, Office of Energy Research, Office of High Energy and Nuclear
Physics, Division of High Energy Physics of the U.S. Department of Energy under Contract DE-AC03-76SF00098 
and in part by the National Science Foundation under grant PHY00-98840.

\appendix
\vspace{-.05in}

\section{Rescaled Group Orbifold Elements}

To begin this discussion we recall the forms of the twisted structure constants and the twisted representation
matrices \cite{deBoer:2001nw}
\begin{subequations}
\begin{gather}
\scf_{\nrm;\nsn}{}^\ntd(\s) \equiv \schisig_\nrm \schisig_\nsn \schisig^{-1}_\ntd U(\s)_\nrm{}^a U(\s)_\nsn{}^b 
   f_{ab}{}^c U^\hcj (\s)_c^\ntd \\
\st_\nrm (T,\s) \equiv \schisig_\nrm U(\s)_\nrm{}^a U(T,\s) T_a U^\hcj (T,\s)
\end{gather}
\end{subequations}
where $U\hc(\s)$ and $U\hc(T,\s)$ are the eigenvector matrices of the $H$-eigenvalue problem \eqref{H-eig} and the 
extended $H$-eigenvalue problem \eqref{extHeig} respectively. The $\schi$'s are the standard normalization constants 
in Eqs.~\eqref{sG-scf-Defn}, \eqref{st-Defn} and \eqref{T-Defn} with $\foot{\chi (\s =0)} =1$.

In terms of these objects we may define the {\it rescaled} group orbifold elements $\tilde{g}$
\begin{subequations}
\begin{gather}
\tilde{g}(\tst,\xi,\s) \equiv \schi(T,\s) \hg(\st,\xi,\s) \schi(T,\s)^{-1} = e^{i\hx_\s^\nrm(\xi) 
   \tst_\nrm(T,\s)} \quad (\bh=\hx) \\
\tst_\nrm (T,\s)_\Nsn {}^\Ntd \equiv \schi (T,\s)_\Nsn \schi (T,\s)_\Ntd^{-1} \st_\nrm (T,\s)_\Nsn{}^\Ntd \label{tst=xstx} \\
[\tst_\nrm (T,\s) ,\tst_\nsn (T,\s) ]= i\scf_{\nrm;\nsn}{}^\ntd (\s) \tst_\ntd (T,\s) \\
\schi(T,\s)_\Nrm{}^{\Nsn} \equiv \d_\Nrm{}^{\Nsn} \schi (T,\s)_\Nrm 
\end{gather}
\begin{gather}
\tilde{g}(\tst,\xi+2\p,\s))_\Nrm{}^{\Nsn} = e^{-\frac{2\p i}{R(\s)}(N(r)-N(s))} \tilde{g}
   (\tst,\xi,\s)_\Nrm{}^{\Nsn} \\
\hg(\st,\xi,\s) = \schi(T,\s)^{-1} \tilde{g}(\tst,\xi,\s) \schi(T,\s) \label{tildeg}
\end{gather}
\end{subequations}
where $\hg$ is the group orbifold element, $\tst_\nrm (T,\s)$ are the {\it rescaled} twisted 
representation matrices and the quantities $\schi (T,\s)$ are another set of normalization constants which satisfy:
\begin{equation}
\schi (T,0) = \one ,\quad \schi (T^{adj} ,\s)_\nrm = \schisig_\nrm \,.
\end{equation} 
The special case $\{ \tst_\nrm (T^{adj} ,\s) \}$ appears in 
Eq.~\eqref{exp-form-of-ho}, where we have already encountered an example of the rescaled group orbifold 
elements, namely the twisted adjoint action $\ho$:
\begin{subequations}
\label{ho=tginv}
\begin{gather}
\ho(\hx(\xi)) = e^{-i\hY (\hx(\xi))} = e^{-i\hx_\s^\nrm(\xi) \tst_\nrm (T^{adj},\s)} = \tilde{g}^{-1}
   (\tst(T^{adj}),\xi,\s)\\
i \left( \hpl_\nrm \ho(\hx) \right) \ho^{-1} (\hx) = \he(\hx)_\nrm {}^\nsn \tst_\nsn (T^{adj},\s) \,. \label{he-ho-eqn}
\end{gather}
\end{subequations}
Although the rescaled quantities $\tilde{g}$ and $\tst$ are quite natural from the point of view of 
Eq.~\eqref{ho=tginv}, we have refrained from using them in the text because the unitarity of the rescaled group 
orbifold elements depends on the choice of the normalization constants $\schi(T,\s)$. In particular, using the 
unitarity of the group orbifold element $\hg$, one finds that
\begin{equation}
\tilde{g}\hc(\tst,\xi,\s) = (\schi(T,\s)\schi(T,\s)^\ast )^{-1} \tilde{g}^{-1}(\tst,\xi,\s) (\schi(T,\s)
   \schi(T,\s)^\ast )
\end{equation}
so that the rescaled group orbifold elements $\tilde{g}$ are unitary only when $\schi \schi^\ast \propto 
\thickone $ .

Nevertheless, all the classical results of WZW orbifold theory can be rewritten in terms of the rescaled quantities, for example,
\begin{subequations}
\begin{align}
\hat{S}_\s(\tst) \equiv \hat{S}_\s(\st(\tst))=& -\!\frac{1}{8\pi}\!\int \!d^2\xi \widehat{Tr}\big{(}\,
    \tilde{\sm}(\tst,\s) \tilde{g}^{-1} (\tst,\s) \pl_+\tilde{g}(\tst,\s)\tilde{g}^{-1}
    (\tst,\s)\pl_-\tilde{g}(\tst,\s)\,\big{)} \nn\\
 &-\frac{1}{12\pi}\int_{\Gamma} \widehat{Tr}\big{(}\,\tilde{\sm}(\tst,\s)\0b(\;\tilde{g}^{-1}(\tst,\s) 
    d\tilde{g}(\tst,\s)\;)^3\,\big{)}
\end{align}\vshalf
\begin{gather}
\tilde{\sm}(\tst,\s) \equiv \schi(T,\s) \sm(\st,\s) \schi(T,\s)^{-1} \\ 
\{ \hj_\nrm(\xi,\s), \tilde{g}(\tst,\e,\s)  \} = 2\p  \d_{n(r)}(\xi-\e) \tilde{g}(\tst,\e,\s) \tst_\nrm(T,\s) \\
\{ \hjb_\nrm(\xi,\s), \tilde{g}(\tst,\e,\s)  \} = -2\p  \d_{n(r)}(\xi-\e) \tst_\nrm(T,\s) \tilde{g}(\tst,\e,\s) \\ 
\hj_\nrm(\xi,\s)\sG^{\nrm,\mnrn}(\s) \tst_{\mnrn}(T,\s) = -\frac{i}{2} \tilde{g}^{-1}(\tst,\xi,\s) \pl_+ 
   \tilde{g}(\tst,\xi,\s) \\
\hjb_\nrm(\xi,\s)\sG^{\nrm,\mnrn}(\s) \tst_{\mnrn}(T,\s) = -\frac{i}{2} \tilde{g}(\tst,\xi,\s) 
   \pl_- \tilde{g}^{-1}(\tst,\xi,\s)
\end{gather}
\end{subequations}
where we have used Eq.~\eqref{tst=xstx} and the inverse relation \eqref{tildeg}.

\section{Twisted Affine Lie Groups}

In Refs.~\cite{Sochen,Club}, the {\it method of affine Lie groups} was used to obtain both {\it quantum} and {\it 
classical} canonical realizations of the left- and right-mover currents of any untwisted affine Lie algebra.
In this construction, the modes of the currents are identified as the reduced {\it affine Lie derivatives} on the
affine Lie group, and this quantum canonical realization generalizes Bowcock's \cite{Bow} classical canonical realization.

In what follows we will work out the extension of this method to the case of {\it twisted affine Lie groups}, 
which gives the quantum and classical realizations of the general left- and right-mover twisted
current algebra discussed in the text. The results of this appendix were obtained in collaboration with 
C.~Park.

We begin with a set of abstract modes $\hE$ which satisfy the general twisted current algebra $\gfrakh(\s)$ 
in Ref.~\cite{Halpern:2000vj}
\begin{align}
[ \hE_\nrm(m+\srac{n(r)}{\r(\s)}), \hE_\nsn(n+\srac{n(s)}{\r(\s)}) ] =& i\scf_{\nrm;\nsn}
   {}^{n(r)+n(s),\d}(\s) \hE_{n(r)+n(s),\d}(m+n+\srac{n(r)+n(s)}{\r(\s)}) \nn\\
&+ (\mnrrs)\d_{\mnnrnsrsf,0} \sG_{\nrm;-n(r),\n}(\s) \label{Abs-CA}
\end{align}
where $\scf(\s)$ and $\sG(\s)$, given explicitly in Eq.~\eqref{sG-scf-Defn}, are respectively the twisted structure constants 
and twisted metric of sector $\s$. 

By including the central term among the generators, the algebra \eqref{Abs-CA} can be recast as an infinite-dimensional 
Lie algebra 
\begin{subequations}
\begin{gather}
[\seh_\hL,\seh_\hM ] = i\tilde{\scf}_{\hL\hM}{}^\hN\seh_\hN \\
\seh_\hL=(\hE_\nrm(m+\srac{n(r)}{\r(\s)}), 1),\quad \hL= ((\nrm,\mnrrs ),\hys),\quad \seh_\hys=1 \\
\tilde{\scf}_{\nrm,\mnrrsf;\nsn,\nnsrsf}{}^{\ntd,p+\frac{n(t)}{\r(\s)}} = \d_{m+n-p+\frac{n(r)+n(s)-n(t)}{\r(\s)},0}
   \,\scf_{\nrm;\nsn}{}^\ntd(\s)  \label{non-zero-structure-constant-1}
\end{gather}
\begin{gather}
\tilde{\scf}_{\nrm,\mnrrsf;\nsn,\nnsrsf}{}^\hys = -i (\mnrrs) \d_{\mnnrnsrsf,0} \, \sG_{\nrm;\nsn}(\s) 
   \label{non-zero-structure-constant-2}
\end{gather}
\end{subequations}
whose non-zero structure constants are given in \eqref{non-zero-structure-constant-1} and 
\eqref{non-zero-structure-constant-2}. The elements $\hat{\gamma}$ of the corresponding twisted affine 
Lie group are defined as
\vspace{-.05in}
\begin{subequations} 
\begin{gather} 
\hat{\gamma} = e^{iy}\hpsi(\hx(t)) ,\quad \hpsi(\hx(t)) \equiv exp \big{(} i\hx^{\nrm;\mnrrsf}(t) 
   \hE_\nrm(\mnrrs)\big{)} \\
\hx^{n(r) \pm \r(\s),\m; m \mp 1 + \frac{n(r) \pm \r(\s)}{\r(\s)} }(t) = \hx^{\nrm; \mnrrs }(t) \\
\quad \he(0)_\hL{}^{\hM} = - \heb(0)_\hL{}^{\hM} = \d_\hL{}^{\hM} \label{preferred-coord}
\end{gather}
\end{subequations}
where $\hx(t)$ are the twisted mode coordinates at time $t$, the coordinate $y$ corresponds to the 
central term and $\hpsi(\hx(t))$ is the reduced twisted affine Lie group element. Here we have chosen the 
preferred coordinate system \eqref{preferred-coord} on the affine Lie group, where $\he(0)=1$ is the twisted 
mode vielbein at the origin (see below). This choice ultimately guarantees that the local operators of the 
construction have definite monodromy. More generally, the twisted mode vielbeins are defined as
\begin{subequations}
\begin{gather} 
\he_{\hL} \equiv -i \hgamma^{-1} \pl_{\hL} \hgamma = \he_{\hL}{}^{\hM} \seh_{\hM} ,\quad \heb_{\hL} 
   \equiv -i \hgamma \pl_{\hL} \hgamma^{-1} = \heb_{\hL}{}^{\hM} \seh_{\hM} ,\quad \pl_{\hL} \equiv 
   \frac{\pl}{\pl \hx^{\hL}} \\
\he \equiv  \he(t),\quad \heb \equiv \heb(t), \quad \pl_{\hL} \equiv \pl_{\hL}(t) \label{suppression-of-t}
\end{gather}
\end{subequations}
in terms of the elements $\hat{\gamma}$ of the twisted affine Lie group. As seen in \eqref{suppression-of-t}, 
we often suppress the fixed time label $t$.

The method of Refs.~\cite{Sochen,Club} then gives the general twisted current modes as the reduced {\it twisted affine 
Lie derivatives}
\vspace{-.05in}
\begin{subequations}
\label{reduced-twisted-affine-Lie-derivatives}
\begin{align}
\hj_\nrm(\mnrrs,t) &= -i \hei_{\nrm,\mnrrsf}{}^{\nsn,\nnsrsf}\sdh_{\nsn,\nnsrsf} \\
\hjb_\nrm(\mnrrs,t) &= -i \hebi_{\nrm,\mnrrsf}{}^{\nsn,\nnsrsf}\sdhb\!\!_{\nsn,\nnsrsf}
\end{align}\vshalf
\begin{align}
\sdh_{\nrm,\mnrrsf} &\equiv \pl_{\nrm,\mnrrsf} - i\; \he_{\nrm,\mnrrsf}{}^\hys \\
\sdhb\!\!_{\nrm,\mnrrsf} &\equiv \pl_{\nrm,\mnrrsf} + i\; \heb_{\nrm,\mnrrsf}{}^\hys
\end{align}\vshalf
\begin{align}
 \hj_\nrm(\mnrrs,t)\hpsi(\hx(t))  &= \hpsi(\hx(t)) \hE_\nrm(\mnrrs) \\
 \hjb_\nrm(\mnrrs,t)\hpsi(\hx(t)) &= -\hE_\nrm(\mnrrs)\hpsi(\hx(t))
\end{align}
\end{subequations}
which provide a {\it differential realization} (at fixed $t$) of the general left- and right-mover twisted current algebra 
in Eq.~\eqref{tw-current-alg}. The {\it sign reversal} of the right-mover central term in \eqref{tw-current-alg} is in fact 
completely natural from the viewpoint of affine Lie groups (see Eqs.~($2.12$) and ($2.13$) of Ref.~\cite{Club}) and twisted 
affine Lie groups. Indeed, this derivation of the general twisted current algebra provides us with a third 
independent derivation (see Ref.~\cite{deBoer:2001nw}) of this sign reversal.

Following Ref.~\cite{Club}, we also find the equivalent $\hb$ form of the twisted current modes
\vspace{-.05in}
\begin{subequations}
\label{twisted-bforms-of-jhat}
\begin{align}
\!\!& \!\!\hj_\nrm(m\!+\! \srac{n(r)}{\r(\s)} ,t) \!= \!-i{\hei}_{\nrm,\mnrrsf}{}^{\!\!\!\!\!\nsn,\nnsrsf} \sdh_{\!\!\nsn,\nnsrsf}(\hb) 
   \!+\!\srac{1}{2} \ho_{\nrm,\mnrrsf}{}^{\!\hys} \\
\!\!& \!\!\hjb_{\!\nrm} (m\!+\! \srac{n(r)}{\r(\s)} ,t) \!= \!-i{\hebi}_{\nrm,\mnrrsf}{}^{\!\!\!\!\!\nsn,\nnsrsf} 
   \sdh_{\!\!\nsn,\nnsrsf}(\hb) \!-\!\srac{1}{2} (\ho^{-1})_{\nrm,\mnrrsf}{}^{\!\hys} 
\end{align}\vsvsvs
\begin{align}
&\sdh_{\nrm,\mnrrsf}(\hb) \equiv  \pl_{\nrm,\mnrrsf} \\
&\; +\!\srac{i}{2} \hb_{\nrm,\mnrrsf;\nsn,\nnsrsf} \hei_{\ntd,p+\frac{n(t)}{\r(\s)}} {}^{\!\!\!\!\!\nsn,\nnsrsf}
   \sG^{\ntd,p+\frac{n(t)}{\r(\s)};\nue,q+\frac{n(u)}{\r(\s)}}(\s) \ho_{\nue,q+\frac{n(u)}{\r(\s)}}
   {}^{\!\hys} \nn
\end{align}
\begin{gather}
\ho_\hL{}^\hM \equiv (e^{-i\hlh})_\hL{}^\hM, \quad \hlh \equiv \sum_{r,\m,m}\hx^{\nrm,\mnrrsf}
   \tst^{adj}_{\nrm,\mnrrsf} ,\quad (\tst^{adj}_\hL)_\hM{}^\hN \equiv -i \tilde{\scf}_{\hL;\hM}{}^\hN
\end{gather}\vsvsvs
\begin{align}
\!\!& \!\!\hB_{\nrmmode;\nsnmode} \!\equiv \!\smal{\left( \frac{e^{i\hlh}-e^{-i\hlh}-2i\hlh}{(i\hlh)^2} \right)}
   {}_\nrmmode {}^{\!\!\!\!\ntdmode} \sG_{\ntdmode;\nsnmode}(\s) \label{hB-mode}
\end{align}
\begin{gather}
\sG_{\nrmmode;\nsnmode}(\s) \equiv \d_{m+n+\frac{n(r)+n(s)}{\r(\s)},\,0}  \sG_{\nrm;\nsn}(\s) 
\end{gather}
\end{subequations}
where $\hB$ in \eqref{hB-mode} is a mode form of the relation \eqref{exp-form-of-hB} for the local field $\hB$ of the text.
Evaluating this result in the untwisted sector\footnote{For $\s = 0$ and simple $g$, our results can be 
compared to those of Ref.~\cite{Club} (with $E \leftrightarrow J$) in the special coordinate system 
$e_{i\m}{}^{am}(0) = \d_{i\m}{}^{am}$ where $am \simeq i\m$. We find that our $B$ field is identified 
with $-kB$ in that reference, and moreover our quantities $y$, $e_{i\m}{}^{y^{\ast}}$, $(e^{-1})_{i\m}
{}^{y^{\ast}}$, $\Omega_{am}{}^{y^{\ast}}$ and $(\Omega^{-1})_{am}{}^{y^{\ast}}$ also contain an extra 
factor of $k$.}, we find some typos in Ref.~\cite{Club}: The second (barred) equations in (2.30a), 
(2.33a), (3.6), (4.6a) and (4.31a) of that reference should all appear with $\ho^{-1}$ instead of $\ho$. Also, 
the right hand sides of Eqs.~(4.22a,b) should have an extra factor -1.

As discussed in Ref.~\cite{Club}, we may also define twisted local fields on the cylinder as the sums of the twisted modes above:
\begin{subequations}
\label{tw-aff-mode-exp}
\begin{align}
\hj_\nrm(\xi,t) &\equiv \sum_m e^{-i(\mnrrsf)\xi} \hj_\nrm(\mnrrs,t) \\
\hjb_\nrm(\xi,t) &\equiv \sum_m e^{-i(\mnrrsf)\xi} \hjb_\nrm(\mnrrs,t)
\end{align}\vsvsvs
\begin{gather}
\hx^\nrm(\xi,t) \equiv \sum_m e^{i(\mnrrsf)\xi} \hx^{\nrm,\mnrrsf}(t)  \\
\hat{p}_\nrm(\xi,t) \equiv \frac{1}{i} \frac{\de}{\de \hx^\nrm (\xi,t)} \equiv 
   -\frac{i}{2\p} \sum_m e^{-i(\mnrrsf)\xi}  \pl_{\nrm,\mnrrsf}(t) \label{tw-aff-mode-exp:hatp} \\
[\hat{p}_\nrm(\xi,t), \hx^\nsn(\e,t)] = -i \d_\nrm{}^{\nsn} \d_{n(r)} (\xi-\e) \label{AppC-px-brk}
\end{gather}
\begin{gather}
\he(\hx(\xi,t))_\nrm{}^{\nsn} \equiv \sum_m e^{i(n-m+\frac{n(s)-n(r)}{\r(\s)}) \xi} \he(t)_{\nrm,\mnrrsf}
   {}^{\nsn,\nnsrsf}  ,\quad \forall n \in {\mathbb Z} \label{mode-exp-vielbein} \\
\hb(\hx(\xi,t))_{\nrm;\nsn} \equiv \sum_m e^{-i(n+m+\frac{n(s)+n(r)}{\r(\s)}) \xi} \hb(t)_{\nrm,\mnrrsf;\nsn,\nnsrsf} 
   ,\quad \forall n \in {\mathbb Z} \,. \label{mode-exp-bfield}
\end{gather}
\end{subequations}
Here $\de_{n(r)} (\xi-\eta)$ in \eqref{AppC-px-brk} is the monodromy factor \eqref{de_nr-Props} of the text, and, as noted in
Ref.~\cite{Club}, the expansions (B$.7$f,g) are independent of the integer $n$. The time dependence of the twisted current modes
\vspace{-.05in}
\begin{subequations}
\label{time-dependence-of-twisted-current-modes}
\begin{align}
\hj_\nrm(\mnrrs,t) &= e^{-i(\mnrrsf)t}\hj_\nrm(\mnrrs) \\
\hjb_\nrm(\mnrrs,t) &= e^{i(\mnrrsf)t}\hjb_\nrm(\mnrrs)
\end{align}
\end{subequations}
follows from the operator WZW orbifold Hamiltonian $\hh_\s= L_\s(0) + \bar{L}_\s(0)$, where  $L_\s(0)$ and 
$\bar{L}_\s(0)$ are the zero modes of the left- and right-mover twisted affine-Sugawara constructions 
in Eq.~\eqref{T-Defn}. The time-independent modes in \eqref{time-dependence-of-twisted-current-modes} are the 
twisted current modes of the text. 

Using \eqref{twisted-bforms-of-jhat} and the mode expansions \eqref{tw-aff-mode-exp}, one finds the {\it 
operator analogue} of the local canonical realization \eqref{hb-form-of-twisted-currents} with the same forms 
\eqref{exp-form-of-ehat}, \eqref{exp-form-of-hB} for $\he,\hB$ and the same operator ordering shown there
\begin{subequations}
\begin{gather}
\hj_\nrm(\xi,\s) = 2\pi \he^{-1}(\hx)_\nrm{}^\ntd \hp_\ntd(\hb) +  \half \pl_\xi \hx_\s^\ntd 
   \he(\hx)_\ntd{}^\nsn \sG_{\nsn;\nrm}(\s) \\
\hjb_\nrm(\xi,\s) = 2\pi \heb^{-1}(\hx)_\nrm{}^\ntd \hp_\ntd(\hb) -  \half \pl_\xi \hx_\s^\ntd 
   \heb(\hx)_\ntd{}^\nsn \sG_{\nsn;\nrm}(\s) 
\end{gather}
\begin{gather}
\hp_\nrm(\hb) = \hp^\s_\nrm + \onefourpi \hb_{\nrm;\nsn}(\hx)\pl_\xi\hx^\nsn ,\quad \srange \\
[\hat{p}^\s_\nrm(\xi,t), \hx_\s^\nsn(\e,t)] = -i \d_\nrm{}^{\nsn} \d_{n(r)} (\xi-\e) \\
[\hp^\s_\nrm (\xi,t) ,\hp^\s_\nsn (\e,t)]= [\hx_\s^\nrm (\xi,t), \hx_\s^\nsn (\e,t)] = 0
\end{gather}
\end{subequations}
but here the twisted momentum operator $\hp_\nrm^\s (\xi,t)$ is the functional derivative in \eqref{tw-aff-mode-exp:hatp}. 
This operator construction satisfies the equal-time (operator) twisted current algebra \eqref{eq-time-current-alg}, and 
the classical realization \eqref{hb-form-of-twisted-currents} of the twisted currents is obtained by replacing operator 
$\hp$ above by classical $\hp$.

\section{The Commuting Diagrams of Orbifold Theory}

In this appendix, we point out that the {\it commuting diagram} given for the left-mover currents $J$ in Ref.~\cite{deBoer:1999na}
can be extended to all the fields of orbifold theory.

Including the right-mover currents $\bj$, the commuting diagram of this reference takes the form:

\begin{picture}(328,178)(0,0)
\put(123,165){$(J,\bj)$}
\put(152,158){\line(1,0){5}}
\put(173,158){\line(1,0){5}}
\put(193,158){\line(1,0){5}}
\put(213,158){\line(1,0){5}}
\put(233,158){\line(1,0){5}}
\put(162,158){\line(1,0){5}}
\put(182,158){\line(1,0){5}}
\put(203,158){\line(1,0){5}}
\put(223,158){\line(1,0){5}}
\put(243,158){\line(1,0){5}}
\put(254,150){\oval(15,15)[tr]}
\put(261,147){\line(0,-1){5}}
\put(261,139){\line(0,-1){5}}
\put(261,131){\line(0,-1){5}}
\put(261,123){\line(0,-1){5}}
\put(261,114){\line(0,-1){5}}
\put(261,106){\vector(0,-1){10}}
\put(166,165){$\foot{\schi U (J,\bj) = (\sj,\sjb)}$}
\thicklines
\put(146,158){\vector(0,-1){60}}
\put(146,98){\vector(0,1){60}}
\put(123,86){$(\sjh,\sjbh)$}
\put(261,165){$(\sj,\sjb)$}
\put(166,86){$\foot{\schi U (\sjh,\sjbh) =(\hj,\hjb)} $}
\put(265,158){\vector(0,-1){60}}
\put(265,98){\vector(0,1){60}}
\put(261,86){$(\hj,\hjb)$}
\put(100,70) {{\footnotesize Each vertical double arrow is a local isomorphism }}
\put(75,59) {$\foot{J,\bj}$}
\put(100,59) {{\footnotesize = currents: trivial monodromy, mixed under automorphisms}}
\put(75,47) {$\foot{\sj,\sjb}$}
\put(100,47) {{\footnotesize = eigencurrents: trivial monodromy, diagonal under automorphisms}}
\put(75,33) {$\foot{\hj,\hjb}$}
\put(100,33) {{\footnotesize = twisted currents with diagonal monodromy}}
\put(75,21) {$\foot{\sjh,\sjbh}$}
\put(100,21) {{\footnotesize = currents with twisted boundary conditions}}
\put(118,4) {Fig.\,\ref{fig:currents}: Currents and orbifold currents}
\end{picture}
\myfig{fig:currents}
\vspace{.2in}

\noindent In particular, the fields connected by vertical double-arrows are {\it locally isomorphic}, that is,
they have the same OPE's -- but the automorphic response of the unhatted fields becomes the monodromy of the hatted fields. 
The method of eigenfields and local isomorphisms \cite{Borisov:1997nc,deBoer:1999na,Halpern:2000vj,deBoer:2001nw} follows the dashed 
line in this figure to obtain the twisted currents $\hj ,\hjb$ with diagonal monodromy. For example:
\begin{subequations}
\begin{gather}
J_a (\xi)' = \w(h_\s)_a{}^b J_b (\xi) \\
\sj_\nrm (\xi,\s) = \schisig_\nrm U(\s)_\nrm{}^a J_a (\xi) ,\quad \sj_\nrm (\xi,\s)' = e^{-\tp \nrrs} \sj_\nrm (\xi,\s) \\
\hj_\nrm (\xi+2\pi,\s) = e^{-\tp \nrrs} \hj_\nrm (\xi,\s) \,.
\end{gather}
\end{subequations}
It is also possible to proceed in the other direction $J \rightarrow \sjh \rightarrow \hj$ around the commuting diagram, 
in which case we encounter the left- and right-mover currents $\sjh$, $\sjbh$ with {\it twisted boundary conditions}
\begin{subequations}
\begin{equation}
\sjh_a (\xi+2\p,\s) = w(h_\s)_a{}^{b} \sjh_b (\xi,\s), \quad \sjbh_{\!\!a} (\xi+2\p,\s) = w(h_\s)_a{}^{b} 
   \sjbh_{\!\!b} (\xi,\s)
\end{equation}
\begin{align}
\!\!& \!\!\sjh_a (\xi,\s) \= U\hc(\s)_a{}^{\!\nrm}\schisig^{-1}_\nrm \hj_\nrm(\xi,\s), \,\,\, \sjbh_{\!\!a} (\xi,\s) 
   \= U\hc(\s)_a{}^{\!\nrm}\schisig^{-1}_\nrm \hjb_{\!\nrm} (\xi,\s) \\
&\hj_\nrm(\xi,\s) \= \schisig_\nrm U(\s)_\nrm{}^{a} \sjh_a(\xi,\s) ,\,\,\, \hjb_\nrm(\xi,\s) \= 
   \schisig_\nrm U(\s)_\nrm{}^{a} \sjbh_{\!\!a} (\xi,\s) \label{jhat=schiUsjhat}
\end{align}
\end{subequations}
where $U\hc (\s)$ is the eigenvector matrix of the $H$-eigenvalue problem \eqref{H-eig} of sector $\s$.
We emphasize that $\sjh ,\sjbh$ have a matrix or non-diagonal monodromy and, according to \eqref{jhat=schiUsjhat}, 
the {\it monodromy decompositions} of $\sjh$, $\sjbh$ give the twisted currents $\hj$, $\hjb$ with diagonal monodromy. 

Similarly, the commuting diagram for the unitary group elements and group orbifold elements is

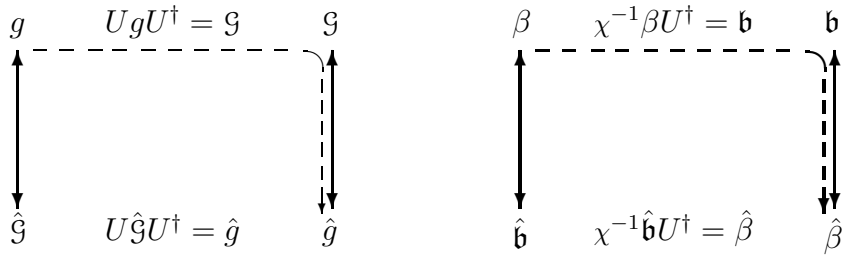
\begin{picture}(370,178)(0,0)
\put(53,165){$g$}
\put(62,158){\line(1,0){5}}
\put(83,158){\line(1,0){5}}
\put(103,158){\line(1,0){5}}
\put(123,158){\line(1,0){5}}
\put(143,158){\line(1,0){5}}
\put(72,158){\line(1,0){5}}
\put(92,158){\line(1,0){5}}
\put(113,158){\line(1,0){5}}
\put(133,158){\line(1,0){5}}
\put(153,158){\line(1,0){5}}
\put(164,150){\oval(15,15)[tr]}
\put(171,147){\line(0,-1){5}}
\put(171,139){\line(0,-1){5}}
\put(171,131){\line(0,-1){5}}
\put(171,123){\line(0,-1){5}}
\put(171,114){\line(0,-1){5}}
\put(171,106){\vector(0,-1){10}}
\put(89,165){$UgU\hc = \sg$}
\thicklines
\put(56,158){\vector(0,-1){60}}
\put(56,98){\vector(0,1){60}}
\put(53,86){$\hat{\sg}$}
\put(171,165){$\sg$}
\put(89,86){$ U\hat{\sg}U\hc=\hg$}
\put(175,158){\vector(0,-1){60}}
\put(175,98){\vector(0,1){60}}
\put(171,86){$\hg$}

\put(243,165){$\b$}
\put(252,158){\line(1,0){5}}
\put(273,158){\line(1,0){5}}
\put(293,158){\line(1,0){5}}
\put(313,158){\line(1,0){5}}
\put(333,158){\line(1,0){5}}
\put(262,158){\line(1,0){5}}
\put(282,158){\line(1,0){5}}
\put(303,158){\line(1,0){5}}
\put(323,158){\line(1,0){5}}
\put(343,158){\line(1,0){5}}
\put(354,150){\oval(15,15)[tr]}
\put(361,147){\line(0,-1){5}}
\put(361,139){\line(0,-1){5}}
\put(361,131){\line(0,-1){5}}
\put(361,123){\line(0,-1){5}}
\put(361,114){\line(0,-1){5}}
\put(361,106){\vector(0,-1){10}}
\put(274,165){$\schi^{-1}\b U\hc = \bfrak$}
\thicklines
\put(246,158){\vector(0,-1){60}}
\put(246,98){\vector(0,1){60}}
\put(243,83){$\bfrakh$}
\put(361,165){$\bfrak$}
\put(274,86){$ \schi^{-1}\bfrakh U\hc=\bh$}
\put(365,158){\vector(0,-1){60}}
\put(365,98){\vector(0,1){60}}
\put(361,83){$\bh$}

\put(67,70) {{\footnotesize Each vertical double arrow is a local isomorphism }}
\put(56,59) {$\foot{g}$}
\put(67,59) {{\footnotesize = Lie group elements: trivial monodromy, mixed under automorphisms}}
\put(56,47) {$\sgtiny$}
\put(67,47) {{\footnotesize = eigengroup elements: trivial monodromy, diagonal under automorphisms}}
\put(56,35) {$\foot{\hg}$}
\put(67,35) {{\footnotesize = group orbifold elements with (two-sided) diagonal monodromy}}
\put(56,23) {${\hat{\sgtiny}}$}
\put(67,23) {{\footnotesize = group orbifold elements with twisted boundary conditions}}
\put(98,4) {Fig.\,\ref{fig:goe}: Group and group orbifold elements}
\end{picture}
\myfig{fig:goe}
\vspace{.1in}

\noindent where we have included the commuting diagram for the corresponding tangent-space coordinates:
\begin{equation}
g = e^{i\b^a T_a}, \quad \sg = e^{i\bfrak^\nrm \st_\nrm}, \quad \sgh = e^{i\hat{\bfrak}^a T_a}, \quad 
   \hg = e^{i\bh^\nrm \st_\nrm} \, . \label{g=eibt}
\end{equation}
Here $T_a$ and $\st_\nrm (T,\s)$ are the untwisted and twisted representation matrices respectively, whose adjoints
are discussed in App.~D.

Again, the method of eigenfields and local isomorphisms \cite{deBoer:2001nw} follows the dashed line in Fig.~\ref{fig:goe}; the 
automorphic response of $g(T,\xi,t)$ is given in Eq.~\eqref{WZW-g-resp}, the eigengroup element $\sg(\st,\xi,t,\s)$ 
appears in Eqs.~\eqref{sg-Defn}, \eqref{sg-resp} of the text, and the (two-sided) diagonal monodromy of $\hg$ is
given in Eq.~\eqref{gMono}. One may also follow the path $g \rightarrow \hat{\sg} \rightarrow \hg$, where $\hat{\sg} 
(T,\xi,t,\s)$ is the group orbifold element with twisted boundary conditions
\begin{subequations}
\begin{gather}
\hat{\sg}(T,\xi+2\p,\s) = W(h_\s;T) \hat{\sg}(T,\xi,\s) W\hc(h_\s;T) \\
\hat{\sg}(T,\xi,\s) = U\hc(T,\s)\hg(\st,\xi,\s)U(T,\s) ,\quad \hg(\st,\xi,\s) = U(T,\s) \hat{\sg}(T,\xi,\s) U\hc(T,\s)
   \label{mono-dec-sg}
\end{gather}
\end{subequations}
where $W(h_\s;T)$ is the action of $h_\s \in H \subset Aut(g)$ in matrix representation $T$ and $U\hc (T,\s)$ is the 
eigenvector matrix of the extended $H$-eigenvalue problem \eqref{extHeig}. Again, we emphasize that $\hat{\sg}$ has matrix 
monodromy and the monodromy decomposition \eqref{mono-dec-sg} gives the group orbifold element $\hg$ with (two-sided) diagonal 
monodromy. Examples of the fields $\hat{\sg}$ were useful in the discussion \cite{Halpern:2002ab} of the charge conjugation 
orbifold on $\su(n)$. Except for the monodromies \cite{deBoer:2001nw}, the commuting diagram in Fig.~\ref{fig:goe} applies as well to the 
affine primary fields $g ,\sg$ and the corresponding twisted affine primary fields [11-14] $\hat{\sg} ,\hg$ of the WZW orbifolds.

For the tangent-space coordinates $\be ,\bfrak$ and their twisted counterparts $\bfrakh ,\bh$, the method of 
eigenfields and local isomorphisms begins with the transformation property \eqref{WZW-beta-resp} of the tangent-space 
coordinates $\be$ and gives
\begin{subequations}
\begin{align}
&\!\! \bfrak^\nrm (\xi,t,\s) \!\equiv \!\schisig_\nrm^{-1} \be^a(\xi,t) U\hc(\s)_a{}^\nrm ,\,\,\, \sg (\st(T,\s),\xi,t,\s) \!=\! 
   e^{i\bfrak^\nrm (\xi,t,\s) \st_\nrm (T,\s)} \label{C5a}
\end{align}\vsvs
\begin{gather}
\bfrak^\nrm (\xi,t,\s)' = \bfrak^\nrm (\xi,t,\s) E_{n(r)} (\s)^\ast = \bfrak^\nrm (\xi,t,\s) e^{\tp \nrrs} \\
\bfrak^\nrm (\xi,t,\s) \dual \bh^\nrm (\xi,t,\s) \\
\bh^\nrm (\xi +2\pi ,t,\s) = \bh^\nrm (\xi,t,\s) e^{\tp \nrrs}
\end{gather}
\end{subequations}
where $\bfrak$ is the tangent-space eigencoordinate and $\bh$ is the twisted tangent-space coordinate with diagonal
monodromy. Following the other path around the diagram, we encounter the tangent-space coordinates $\bfrakh$ with twisted 
boundary conditions
\begin{subequations}
\begin{gather}
\bfrakh^a(\xi+2\p,\s) = \bfrakh^b(\xi,\s)w\hc(h_\s)_b{}^{a} ,\quad
  \hat{\sg} (T,\xi,\s) = e^{i\smal{\bfrakh}^a (\xi,\s) T_a} \label{sbh-with-twisted-bc}\\
\b^a (\xi) \= \schisig_\nrm  \bfrak^\nrm(\xi,\s) U(\s)_\nrm{}^a \dual  \bfrakh^{a}(\xi,\s) \equivs \schisig_\nrm  \bh^\nrm(\xi,\s) 
   U(\s)_\nrm{}^a \label{beta-duals}
\end{gather}
\begin{gather}
\hg(\st,\xi,\s) = U(T,\s) \;g (T,\be(\xi)) \Big{|}_{\b \,\duals\, \bfrakh}  \;U\hc(T,\s) =e^{i\bh^\nrm(\xi,\s) \st_\nrm(T,\s) } \label{g-beta-duals} \\
\bh^\nrm(\xi,\s) \= \bfrakh^a(\xi,\s) U\hc(\s)_a{}^{\nrm} \schisig_\nrm^{-1} \label{mono-decomp-of-sg-sbh} 
\end{gather}
\end{subequations}
and the monodromy decomposition \eqref{mono-decomp-of-sg-sbh} of $\bfrakh$ gives the twisted tangent-space 
coordinates $\bh$ with diagonal monodromy.

For the Einstein coordinates and their twisted counterparts, we have the commuting diagram

\begin{picture}(328,178)(0,0)
\put(143,165){$x$}
\put(152,158){\line(1,0){5}}
\put(173,158){\line(1,0){5}}
\put(193,158){\line(1,0){5}}
\put(213,158){\line(1,0){5}}
\put(233,158){\line(1,0){5}}
\put(162,158){\line(1,0){5}}
\put(182,158){\line(1,0){5}}
\put(203,158){\line(1,0){5}}
\put(223,158){\line(1,0){5}}
\put(243,158){\line(1,0){5}}
\put(254,150){\oval(15,15)[tr]}
\put(261,147){\line(0,-1){5}}
\put(261,139){\line(0,-1){5}}
\put(261,131){\line(0,-1){5}}
\put(261,123){\line(0,-1){5}}
\put(261,114){\line(0,-1){5}}
\put(261,106){\vector(0,-1){10}}
\put(177,165){$\schi^{-1}xU\hc = \sx$}
\thicklines
\put(146,158){\vector(0,-1){60}}
\put(146,98){\vector(0,1){60}}
\put(143,86){$\hat{\sx}$}
\put(261,165){$\sx$}
\put(177,86){$\schi^{-1}\hat{\sx}U\hc = \hx$}
\put(265,158){\vector(0,-1){60}}
\put(265,98){\vector(0,1){60}}
\put(261,86){$\hx$}
\put(86,70) {{\footnotesize Each vertical double arrow is a local isomorphism }}
\put(75,59) {$\foot{x}$}
\put(86,59) {{\footnotesize = coordinates: trivial monodromy, mixed under automorphism}}
\put(75,47) {$\sxfoot$}
\put(86,47) {{\footnotesize = eigencoordinates: trivial monodromy, diagonal under automorphisms}}
\put(75,35) {$\foot{\hx}$}
\put(86,35) {{\footnotesize = twisted coordinates with diagonal monodromy}}
\put(75,23) {${\hat{\sxfoot}}$}
\put(86,23) {{\footnotesize = coordinates with twisted boundary conditions}}
\put(108,4) {Fig.\,\ref{fig:coords}: Coordinates and orbifold coordinates}
\end{picture}
\myfig{fig:coords}

\noindent where $x, \sx$ and $\hx$ were studied in Subsecs.~$4.1$ and $4.2$. As above, the Einstein coordinates $\sxh$ 
with twisted boundary conditions are locally isomorphic to the untwisted Einstein coordinates $x^i$ and satisfy
\begin{subequations}
\begin{gather}
\hat{\sx}_\s^i(\xi+2\p) = \hat{\sx}_\s^j(\xi)w\hc(h_\s)_j{}^i \label{sxh-Mono} \\
\hx_\s^\nrm(\xi) = \schisig_\nrm^{-1} \sxh_\s^i(\xi) U\hc(\s)_i{}^\nrm, \quad \sxh_\s^i(\xi) = 
   \schisig_\nrm \hx_\s^\nrm(\xi) U(\s)_\nrm{}^i \label{mono-decomp-of-sxh} 
\end{gather}
\end{subequations}
so that the twisted Einstein coordinates $\hx$ with diagonal monodromy are obtained by the monodromy decomposition of $\sxh$.
We have encountered the coordinates $\sxh$ in Eq.~\eqref{form-of-tw-G&B} and ubiquitously in Subsecs.~$4.3$,
$4.5$ and $4.7$.

For the special case of the WZW orbifolds, the preferred coordinate system of the text is 
\vspace{-0.1in}
\begin{equation}
x=\b ,\quad \sx=\bfrak ,\quad \sxh=\bfrakh ,\quad \hx=\bh 
\end{equation}
where $\b$, $\bfrak$, $\bfrakh$ and $\bh$ are the tangent-space coordinates in Fig.\,\ref{fig:goe}. With this 
identification, Eq.~\eqref{x-to-sx} is recognized as Eq.~\eqref{beta-duals}.

For the Einstein metrics and orbifold Einstein metrics, Eq.~\eqref{explicit-form-of-twisted-Einstein-tensor:G} and 
Fig.\,\ref{fig:metric} in Subsec.~$4.2$ give the corresponding diagram in a slightly different format, where
\begin{subequations}
\begin{gather}
\hG_{\nrm;\nsn} (\hx_\s) = \schisig_\nrm \schisig_\nsn U(\s)_\nrm{}^i U(\s)_\nsn{}^j \hat{\sG}_{ij} (\sxh_\s(\hx_\s)) \label{sGh-Mono-Dec} \\
\hat{\sG}_{ij} (\sxh_\s (\hx_\s)) \equiv G_{ij} (x \,\duals \,\sxh_\s (\hx_\s)) \\
\hat{\sG}_{ij} (\sxh_\s (\xi+2\pi)) = \w_i{}^k \w_j{}^l \hat{\sG}_{kl} (\sxh_\s (\xi)) \,.
\end{gather}
\end{subequations}
Here $\hat{\sG} (\sxh_\s)$ is the Einstein metric with twisted boundary condition, whose monodromy decomposition 
in \eqref{sGh-Mono-Dec} is the twisted Einstein metric $\hG (\hx)$ with diagonal monodromy. Similar commuting diagrams are 
easily constructed for all the other Einstein tensors of this paper and their orbifold counterparts. This tells us for example 
that Eqs.~($4.10$c,d) can also be understood as the monodromy decompositions of $\hat{\sb} \equiv B(x \duals \sxh)$ and 
$\hat{\sh} \equiv H(x \duals \sxh)$.

\section{The WZW Orbifold Matrix Adjoint Operation}

We consider here the action of complex conjugation and matrix adjoint in the tangent-space formulation of WZW orbifolds.

For the untwisted representation matrices $T$ and the untwisted tangent-space coordinates $\b$, we have
\vspace{-0.1in}
\begin{subequations}
\label{dagger}
\begin{gather}
 T_a\hc = \r_a{}^b T_b,\quad \b^{a\ast} = \b^b \r{}_b{}^a{}^\ast \\
 \r^\ast\r =1, \quad w(h_\s)^{\ast} \,\r\, w\hc(h_\s) = \r
\end{gather}
\end{subequations}
where dagger is matrix adjoint, star is complex conjugation and $\r$ is the complex conjugation matrix of Refs.~\cite{Halpern:1996js,
Halpern:2000vj}. It follows from \eqref{dagger} that the untwisted group elements are unitary as expected
\begin{gather}
 g(T,\xi) = e^{i\b^a(\xi)T_a} ,\quad g\hc(T,\xi) =  e^{-i\b^a{}^\ast T_a\hc} = g^{-1}(T,\xi) \label{g-dag}
\end{gather}
and that the linkage relation \eqref{Linkage} is preserved under the matrix adjoint operation.

Then using Eqs.~\eqref{st-Defn}, \eqref{C5a} we find as we move toward the orbifold that 
\begin{subequations}
\label{unitarity}
\begin{align}
 \st_\nrm(T,\s)\hc =& \sr_\nrm{}^\nsn(\s) \st_\nsn(T,\s) = \sr_\nrm{}^{-\nrn}(\s) \st_{-\nrn}(T,\s) \label{eq:tst}\\
 \bfrak^\nrm(\xi)^\ast =& \bfrak^\nsn(\xi) \sr_\nsn{}^{\nrm}(\s)^\ast = \bfrak^{-\nrn}(\xi) \sr_{-\nrn}{}^{\nrm}(\s)^\ast  \label{eq:script-bst}\\
 \bh^\nrm(\xi)^\ast =& \bh^\nsn(\xi) \sr_\nsn{}^{\nrm}(\s)^\ast = \bh^{-\nrn}(\xi)  \sr_{-\nrn}{}^{\nrm}(\s)^\ast  \label{eq:bst} \\
 \sr_\nrm{}^\nsn(\s) =& \schisig_\nrm{}^\ast \, \schisig_\nsn^{-1} \,U(\s)_\nrm{}^a{}^{\ast} \r_a{}^b U\hc(\s)_b{}^\nsn \nn\\ 
	=& \d_{n(r)+n(s),\, 0 \, mod\, \r(\s)} \sr_\nrm{}^{-n(r),\n}(\s),\quad \sr(\s)^\ast \sr(\s) = 1 \label{sr-sigma}
\end{align}
\end{subequations}
where $\bfrak^\nrm$ are the tangent-space eigencoordinates and $\bh^\nrm$ are the twisted tangent-space coordinates with diagonal
monodromy. Here Eq.~\eqref{eq:tst} defines the {\it orbifold matrix adjoint operation} on the twisted representation matrices, 
where $\sr(\s)$ in \eqref{sr-sigma} is the orbifold conjugation matrix of Ref.~\cite{Halpern:2000vj}. The orbifold conjugation matrix is dual 
to the conjugation matrix $\r$, and also controls the orbifold adjoint operation \cite{Halpern:2000vj,deBoer:2001nw} of the twisted current modes. 
Unitarity of the eigengroup elements and the group orbifold elements
\begin{subequations}
\begin{gather}
 \sg(\st,\xi,\s) = e^{i\smal{\bfrak}^\nrm(\xi)\st_\nrm(T,\s)} \goto \sg\hc(\st,\xi,\s) = \sg^{-1}(\st,\xi,\s) \label{eq:script-gst} \\
 \hg(\st,\xi,\s) = e^{i\bh^\nrm(\xi)\st_\nrm(T,\s)} \goto \hg\hc(\st,\xi,\s) = \hg^{-1}(\st,\xi,\s) \label{eq:gst} 
\end{gather}
\end{subequations}
then follows from \eqref{unitarity}, or directly from \eqref{sg-Defn} and local isomorphisms.

\section{Orbifolds on Abelian $g$}

Ref.~\cite{Halpern:2002ab} solved the general twisted vertex operator equations \cite{deBoer:2001nw} in an abelian limit of the WZW orbifolds to 
obtain the twisted vertex operators, correlators and group orbifold elements of a large class of orbifolds on abelian $g$
\begin{equation}
\frac{ A_{\text{Cartan } g} (H)}{H} ,\quad H \subset \text{Aut(Cartan } g) ,\quad \text{Cartan } g \subset g
\end{equation}
whose representation theory is provided by the ambient algebra $g$. Here, we use this development to discuss two simple 
examples of abelian inversion orbifolds, both of which are associated to the same $\Zint_2$ outer-automorphic inversion
\begin{equation}
\label{Invers}
J(\xi,t)' = -J(\xi,t) 
\end{equation}
of a single $\ugp (1)$ current $J$. Each of these examples has a single twisted sector $\s=1$, and although the twisted sectors differ 
in their representation theory, they share the same half-integral moded scalar field which appeared the first example \cite{Halpern:1971th} of a 
twisted sector of an orbifold. After this (essentially tangent-space) discussion we will also consider the Einstein geometry of the general 
orbifold on $g=\ugp (1)^n$.

As the first example, consider the inversion orbifold
\begin{equation}
\frac{ A_{\text{Cartan } \su (2)} (\Zint_2 )}{\Zint_2} \subset \frac{ A_{\text{Cartan } g} (H)}{H}
\end{equation}
which was constructed in Ref.~\cite{Halpern:2002ab} by embedding $J \rightarrow J_3$ as a component of the ambient $\su (2)$. At the 
classical level, the representation theory of the orbifold is provided by the choice of the classical group elements 
$g(T)$, and we have in this case \cite{Halpern:2002ab} 
\begin{equation}
g (T_3 (j),\xi,t) = e^{i x(\xi,t) T_3 (j)} ,\quad x(\xi,t)' = -x(\xi,t) 
\end{equation}
where $T_3 (j)$ is the third component of the spin $j$ representation of the ambient $\su (2)$. 
The corresponding group orbifold elements in the single twisted sector of this orbifold are \cite{Halpern:2002ab}
\begin{equation}
\label{Inv1-goe}
\hg (\st (j) ,\xi,t) = e^{i\hx (\xi,t) \st (j)} ,\quad \st (j) \equiv -T_2 (j) ,\quad \hx (\xi +2\pi,t) = - \hx(\xi,t)
\end{equation}
and it is easy to see from the abelian limit of Eq.~\eqref{Def-twVB}
\begin{equation}
\label{he-App}
\he_1 (\st(j)) = -i\hg^{-1} (\st(j)) \pl_\hx \hg (\st(j)) \equiv \he_1{}^1 \st (j)
\end{equation}
that $\he_1{}^1$ and hence the twisted Einstein metric $\hG_{11}$ of this orbifold are constant.

At the operator level, the twisted affine primary fields (twisted vertex operators) and correlators in the twisted sector 
of this orbifold are given in Ref.~\cite{Halpern:2002ab}
\begin{subequations}
\label{J3su2}
\begin{align}
\hg (\st (j),\bz,z) &= |z|^{-\frac{\st (j)^2}{k}} : e^{i \hbe^1 (\bz,z) \st (j)} :_M \\
i \hbe^1 (\bz,z) &= \frac{1}{k}\sum_{m \in \Zint} \frac{1}{m +\half}
\Big( \hjb_1^{\; \rm R} (m + \srac{1}{2}) \bz^{-(m+\half)}
-\hj_1 (m + \srac{1}{2}) z^{-(m+\half)} \Big) 
\end{align}\vsvsvs
\begin{equation}
[ \hj_1 (m+\srac{1}{2}) ,\hj_1 (n+\srac{1}{2})] = [\hjb_1^R (m+\srac{1}{2}) ,\hjb_1^R (n+\srac{1}{2})] = k(m+\srac{1}{2}) \de_{m+n+1,0}
\end{equation}
\begin{gather}
\bkspc \langle \hg (\st^{(1)},\bz_1,z_1)  \cdots \hg (\st^{(N)},\bz_N,z_N) \rangle_\s = \left( \prod_{\r} 
   |z_\r|^{- \srac{1}{k} \st^{(\r)} \st^{(\r)} } \right) \prod_{\r < \kappa}
   \left| \frac{ \sqrt{z_\r} -\sqrt{z_\kappa} }{\sqrt{z_\r} + \sqrt{z_\kappa}} \right|^{\srac{2}{k} \st^{(\k)}
   \st^{(\r)}} \label{npcornc} \\
\st^{(\r)} \equiv \st^{(\r)} (j_\r) = - T_2^{(\r)} (j_\r ) 
\end{gather}
\end{subequations}
where $T_2^{(\r)} (j_\r )$ is the second component of the spin $j=j_\r$ representation acting in the $\r$th subspace. The group orbifold
elements \eqref{Inv1-goe} are the classical (high-level) limit of these twisted vertex operators.

As the second example we consider another realization of the inversion \eqref{Invers}, which leads to the standard inversion
orbifold on compact $x$. In this case the classical group elements are 
\begin{equation}
g_n (\xi,t) = \left( \begin{array}{cc} e^{inx (\xi,t)} & 0 \\ 0 & e^{-inx (\xi,t)} \end{array} \right) = 
   e^{inx(\xi,t) \tau_3} ,\quad n \in \Zint ,\quad x(\xi,t)' = -x(\xi,t) 
\end{equation}
where $\tau_i ,\, i=1,2,3$ are the Pauli matrices. Then we may use the results above for $A_{\text{Cartan } \su(2)} 
(\Zint_2) /\Zint_2$ to read off the corresponding group orbifold elements $\hg_n$ from the map
\begin{subequations}
\begin{gather}
T_3 (j) \goto n \tau_3 ,\quad \st (j) \goto -n\tau_2 \label{AppD-Map} \\
\Rightarrow \hg_n (\xi,t) = e^{-i\hx (\xi,t) n\tau_2} ,\quad \hx(\xi +2\pi,t) = -\hx(\xi,t)
\end{gather}
\end{subequations}
and again we find that Eq.~\eqref{he-App} gives constant $\he_1{}^1$ and hence a constant twisted Einstein metric
$\hG_{11}$.

We may also use the map \eqref{AppD-Map} and $k=1$ to read off the operator results on the sphere for this case
\vspace{-0.1in}
\begin{subequations}
\begin{gather}
\hg_n (\bz ,z) = |z|^{-n^2} :e^{-i \hbe^1 (\bz,z) n\tau_2} :_M \\
i \hbe^1 (\bz,z) = \sum_{m \in \Zint} \frac{1}{m+ \half} \left( \hjb_1^R (m+\srac{1}{2}) \bz^{-(m+\half)} - \hj_1 (m+\srac{1}{2})
  z^{-(m+\half)} \right) \\
[ \hj_1 (m+\srac{1}{2}) ,\hj_1 (n+\srac{1}{2})] = [\hjb_1^R (m+\srac{1}{2}) ,\hjb_1^R (n+\srac{1}{2})] = (m+\srac{1}{2}) \de_{m+n+1,0}
\end{gather}
\begin{equation}
\langle \hg_{n_1} (\bz_1 ,z_1) \ldots \hg_{n_N} (\bz_N ,z_N) \rangle_\s = 
\left( \prod_\r |z_\r |^{-n^2_\r} \right) \prod_{\r < \kappa} \Big{|} \frac{ \sqrt{z_\r} - 
   \sqrt{z_\kappa}}{ \sqrt{z_\r} +\sqrt{z_\kappa}} \Big{|}^{2 n_\kappa n_\r \tau_2^{(\kappa)} \tau_2^{(\r)}} 
\end{equation}
\end{subequations}
from the relations in Eq.~\eqref{J3su2}. 

More generally (see e.g.~Ref.~\cite{Halpern:2002ab}), the group orbifold elements of any abelian orbifold on $g = \ugp(1)^n$ will have 
the form
\begin{subequations}
\label{E11}
\begin{gather}
\hg(\st,\xi,t,\s) = e^{i\hx_\s^\nrm (\xi,t) \st_\nrm (T,\s)} ,\quad \hx_\s^\nrm (\xi+2\pi ,t) = \hx_\s^\nrm (\xi,t) e^{\tp \nrrs} \\
[ \st_\nrm (T,\s) ,\st_\nsn (T,\s)] =0 ,\quad \srange \label{Ab-Orb-Lie}
\end{gather}
\end{subequations}
where the `momenta' $\st$ satisfy an abelian orbifold Lie algebra for each sector $\s$ as shown in \eqref{Ab-Orb-Lie}. (The monodromies 
of the twisted Einstein coordinates $\hx_\s$ of sector $\s$ follow as usual from the solution of the appropriate $H$-eigenvalue problem 
in the untwisted theory). Then Eqs.~\eqref{E11} and \eqref{Def-twVB} tell us that the twisted vielbein of sector $\s$ is always trivial
\begin{equation}
\he_\nrm (\st,\hx(\xi)) = \st_\nrm (T,\s) ,\quad \he (\hx_\s,\s)_\nrm {}^\nsn = \de_\nrm {}^\nsn 
\end{equation}
for orbifolds on abelian $g$. Moreover, it follows from the definition \eqref{G-hat-down} that the twisted Einstein metrics of each sector 
of all these abelian orbifolds
\begin{equation}
\hG_{\nrm;\nsn} (\hx_\s,\s) = \sG_{\nrm;\nsn} (\s) ,\quad \hpl_\nrm \hG_{\nsn;\ntd} (\hx) =0
\end{equation}
are {\it constant} because $\sG_\bullet (\s)$ is the twisted tangent-space metric. The diagonal monodromy \eqref{hG-Mono1} of the 
twisted Einstein metric
\begin{equation}
\hG_{\nrm;\nsn} (\hx (\xi+2\pi)) = e^{-\tp \srac{n(r)-n(s)}{\r(\s)}} \hG_{\nrm;\nsn} (\hx(\xi)) = \hG_{\nrm;\nsn} (\hx(\xi))
\end{equation}
reduces to trivial monodromy in all these cases because of the selection rule \eqref{sG-select} for $\sG_{\bullet} (\s)$.
Orbifolds on abelian $g$ are not a good laboratory for the study of non-trivial twisted Einstein metrics.

For the orbifolds on abelian $g$, the natural twisted torsion of sector $\s$, $\hh (\hx_\s) = \scf (\s) = 0$, also follows from 
Eq.~\eqref{hB-and-hH-defn}, but this can be modified by including other background $B,\hB$ fields.

\vskip .5cm 
\addcontentsline{toc}{section}{References} 
 
\renewcommand{\baselinestretch}{.4}\rm 
{\footnotesize 
 
\providecommand{\href}[2]{#2}\begingroup\raggedright\endgroup

\pagebreak

\end{document}